\documentclass[letterpaper]{JHEP3}
\usepackage{amsmath}
\usepackage{epsfig}

\newcommand{\roughly}[1]{\mathrel{\raise.3ex\hbox{$#1$\kern-0.85em
\lower1ex\hbox{$\sim$}}}}

\newcommand{\lsim}{\roughly<}
\newcommand{\gsim}{\roughly>}

\def\met{\slsh E_\ssT}

\def\la{{\langle}}
\def\ra{{\rangle}}
\def\cA{{\cal A}}
\def\cB{{\cal B}}

\def\cE{{\cal E}}

\def\cF{{\cal F}}
\def\cH{{\cal H}}

\def\cL{{\cal L}}
\def\cM{{\cal M}}
\def\cO{{\cal O}}
\def\cG{{\cal G}}

\def\cR{{\cal R}}

\def\cU{{\cal U}}
\def\cV{{\cal V}}

\def\cP{{\cal P}}
\def\cZ{{\cal Z}}

\newbox\charbox
\newbox\slabox
\def\slsh#1{{      
        \setbox\charbox=\hbox{$#1$}
        \setbox\slabox=\hbox{$/$}
        \dimen\charbox=\ht\slabox
        \advance\dimen\charbox by -\dp\slabox
        \advance\dimen\charbox by -\ht\charbox
        \advance\dimen\charbox by \dp\charbox
        \divide\dimen\charbox by 2
        \raise-\dimen\charbox\hbox to \wd\charbox{\hss/\hss}
        \llap{$#1$}
}}

\def\exd{{\hbox{d}}}
\def\d{\exd}

\def\cA{{\cal A}}
\def\cB{{\cal B}}

\def\cE{{\cal E}}
\def\cI{{\cal I}}

\def\cF{{\cal F}}
\def\cH{{\cal H}}

\def\cL{{\cal L}}
\def\cM{{\cal M}}
\def\cO{{\cal O}}
\def\cG{{\cal G}}

\def\cR{{\cal R}}

\def\cU{{\cal U}}

\def\cP{{\cal P}}
\def\cZ{{\cal Z}}

\def\bea{\begin{eqnarray}}
\def\eea{\end{eqnarray}}
\def\be{\begin{equation}}
\def\ee{\end{equation}}

\def\ol#1{\overline{#1}}

\def\ssB{{\scriptscriptstyle B}}
\def\ssC{{\scriptscriptstyle C}}
\def\ssD{{\scriptscriptstyle D}}
\def\ssF{{\scriptscriptstyle F}}
\def\ssG{{\scriptscriptstyle G}}
\def\ssH{{\scriptscriptstyle H}}

\def\ssL{{\scriptscriptstyle L}}
\def\ssM{{\scriptscriptstyle M}}
\def\ssN{{\scriptscriptstyle N}}

\def\ssQ{{\scriptscriptstyle Q}}
\def\ssR{{\scriptscriptstyle R}}
\def\ssS{{\scriptscriptstyle S}}
\def\ssT{{\scriptscriptstyle T}}

\def\ssW{{\scriptscriptstyle W}}
\def\ssX{{\scriptscriptstyle X}}
\def\ssY{{\scriptscriptstyle Y}}
\def\ssZ{{\scriptscriptstyle Z}}
\def\KK{{\scriptscriptstyle KK}}
\def\SM{{\scriptscriptstyle SM}}

\def\nn{\nonumber}

\def\d{\mathrm{d}}

\def\({\left(}
\def\){\right)}

\def\pref#1{(\ref{#1})}

\title{Bulk Stabilization, the Extra-Dimensional\\
Higgs Portal and Missing Energy in Higgs Events}

\author{Ross Diener and C.P.~Burgess \\
Department of Physics \& Astronomy, McMaster University\\ \qquad 1280 Main Street West, Hamilton ON, Canada.\\

Perimeter Institute for Theoretical Physics\\
\qquad 31 Caroline Street North, Waterloo ON, Canada.\\
}

\preprint{Preprint}

\date{\today}

\abstract { To solve the hierarchy problem, extra-dimensional models must explain why the new dimensions stabilize to the right size, and the known mechanisms for doing so require bulk scalars that couple to the branes. Because of these couplings the energetics of dimensional stabilization competes with the energetics of the Higgs vacuum, with potentially observable effects. These effects are particularly strong for one or two extra dimensions because the bulk-Higgs couplings can then be super-renormalizable or dimensionless. Experimental reach for such extra-dimensional Higgs `portals' are stronger than for gravitational couplings because they are less suppressed at low-energies. We compute how Higgs-bulk coupling through such a portal with two extra dimensions back-reacts onto properties of the Higgs boson. When the KK mass is smaller than the Higgs mass, mixing with KK modes results in an invisible Higgs decay width, missing-energy signals at high-energy colliders, and new mechanisms of energy loss in stars and supernovae. Astrophysical bounds turn out to be complementary to collider measurements, with observable LHC signals allowed by existing constraints. We comment on the changes to the Higgs mass-coupling relationship caused by Higgs-bulk mixing, and how the resulting modifications to the running of Higgs couplings alter vacuum-stability and triviality bounds.
}

\begin{document}

\section{Introduction}

In particle physics it is the best of times, and it is the worst of times. On one hand the recent discovery \cite{HiggsExp} of a new particle at the LHC moves us into the long-awaited study of the new particle's properties, after several decades spent exploring the physics of constraints. If the new particle's interpretation as a Higgs --- or {\em the} Higgs, if the Standard Model description continues to work --- survives, then we can anticipate an unprecedented new era probing vacuum physics.

On the other hand, the LHC has yet to produce compelling evidence for the kinds of physics widely expected to lie beyond the Standard Model. The hierarchy problem lies at the heart of these expectations, leading broadly to three main options\footnote{These need not be mutually exclusive, with some composite models potentially being equivalent to some extra-dimensional models \cite{AdSCFT}.} for LHC-observable new electroweak physics over the years: compositeness models \cite{Technicolor, Technirev}; supersymmetry \cite{SUSY, SUSYrev} (linearly realized\footnote{See, however, \cite{MSLED} for how supersymmetry could be present (but nonlinearly realized \cite{SSnonlin}) at electroweak energies and below, without requiring the existence of the superpartners that remain missing from experiments.}); and extra-dimensional scenarios (both warped \cite{RS} and unwarped \cite{ADD}). Absent compelling evidence for any of these three categories, it is crucial for theorists to seek new ways to distinguish the mechanisms underlying each.

The purpose of this paper is to identify new ways to use the properties of the Higgs to explore extra-dimensional models. Building on earlier work --- in 5D \cite{RSHiggs} and higher dimensional scenarios \cite{Giudice:2000av, 6DHiggs0, 6DHiggs1, 6DHiggsph} --- we track how the vacuum energetics of the Higgs potential interacts with the physics that stabilizes the extra dimensions, and show how this can open a new observable portal onto extra-dimensional dynamics.

At present most bounds on extra dimensions come from the kinematics of mixing and energy loss with the bulk gravitational degrees of freedom \cite{ADDgravem}. Yet a central part of solving the hierarchy problem using extra dimensions is understanding the vacuum physics that stabilizes their size at the required value, both for RS models (where the hierarchy between the electroweak and Planck scales comes from a size-dependent warp factor) and for ADD-type models (where it is the large extra-dimensional volume itself that provides the hierarchy). All of the known mechanisms for this stabilization involve introducing new bulk degrees of freedom (typically scalar fields), whose couplings to ordinary matter are only slightly less robust than those of the metric. It is these couplings to the Higgs that we aim to constrain. We identify two kinds of observable consequences for these couplings.
\begin{itemize}
\item {\em Modified Higgs mass-coupling relations:} due to the dependence of the Higgs potential on the new bulk fields. The interplay between these two fields changes the relationship between the Higgs mass and its couplings relative to Standard Model expectations;
\item {\em Contributions to the Higgs invisible `width':} due to mixing between the Higgs and bulk states. In particular, we find that the expected LHC bounds on this width are competitive with bounds from lower-energy observables, such as energy loss from astrophysical systems, anomalous magnetic moments and the like.
\end{itemize}

Two things are crucial about both of these effects. First, because the bulk fields involved are not the graviton, their couplings need not be precisely gravitational in strength. In particular (a point made earlier for ADD models in \cite{6DHiggsph}) depending on the number of extra dimensions present, they can involve dimensionless couplings, and so be less suppressed at low energies than are graviton interactions. (Dimensionless couplings can also arise for Higgs-curvature interactions in 6D, but unlike the Higgs-bulk portal they remain suppressed at low energies because of the derivative nature of the curvature couplings \cite{Giudice:2000av}.)

Second, the interplay between the (brane-localized) Higgs and extra-dimensional (bulk) stabilization mechanisms depends crucially on understanding how branes back-react on the bulk. Although this is understood relatively well for branes with one transverse dimension (such as arise in RS models) in terms of Israel junction conditions \cite{IJC}, it has only recently been systematically developed \cite{Cod2BR} for branes with two or more transverse dimensions, such as appear in the ADD picture. The understanding of codimension-2 back-reaction came comparatively late because of technical complications associated with the divergence of bulk fields near brane positions (which happens only with two or more transverse dimensions), and the need to absorb these into renormalizations of the brane couplings \cite{Cod2Renorm, claudia}.

\subsection{Higher-dimensional stabilization}

Until recently a big competitive advantage of RS models over ADD models was the existence of a simple and robust way to stabilize the extra dimensions: the Goldberger-Wise mechanism \cite{GW}. In this mechanism a bulk scalar field is introduced that couples to the branes situated at both ends of the RS scenario's one extra dimension, with couplings chosen to frustrate the scalar's ability to reach a constant vacuum configuration. (This can be achieved by having branes disagree with one another about the field value that minimizes the scalar potential.) Because branes are located at specific places in the extra dimension, the resulting frustration sets up gradients in the bulk scalar that make the minimum energy depend on the distance between the branes (and so also on the extra-dimensional size). An attractive feature of the RS model is that the warp factor then naturally exponentiates a modestly large extra-dimensional size into an enormous electroweak hierarchy. (Similar frustration can also be arranged with bulk scalars in more than one extra dimension, with sometimes intriguing implications for the Higgs vacuum \cite{6DHiggs0, 6DHiggs1}.)

A similarly robust mechanism for stabilizing large dimensions has been missing for standard ADD models, but an analogue was recently found \cite{6DExpStab} for their supersymmetric generalizations \cite{NS, SS, SLED} by applying to them a 6D cousin \cite{6DGW} of the Goldberger-Wise mechanism. In such theories the extra dimensions are stabilized classically through flux-stabilization, as is often possible for supersymmetric systems (and for which 6D systems provided the first examples \cite{SS}). In this mechanism the flux of a bulk magnetic field (which is typically required by anomaly cancellation to exist among the field content of the 6D supergravity \cite{NS, SS, SLED}) threads the two extra dimensions, that have the topology of a sphere. Dirac quantization of this flux makes it energetically costly to shrink the dimensions, providing a counterbalance against its gravitational collapse.

However complete stabilization purely within the bulk is never quite possible because of a classical scale invariance of the 6D supergravity action, which  leaves a flat direction parameterized by a bulk scalar field, $\chi$ (the `dilaton', which sits within the `extended' metric supermultiplet). Flux stabilization relates the extra-dimensional radius to this flat direction through the expression
\be \label{eq:rvsphi}
 r^2 = \ell^2 \, e^{-\chi} \,,
\ee
where $\ell$ is a length of order (but, in controlled calculations, parametrically moderately larger than) the 6D Planck scale, set by the flux stabilization.

Fixing $r$ completely requires breaking the classical scale invariance, and lifting the classical bulk flat direction. As shown in ref.~\cite{6DExpStab}, this can be achieved classically through its couplings to branes, whose interactions need not share the scale invariance of the bulk. In particular it is not difficult to arrange for moderately large negative values. Once this is done flux stabilization --- {\em via} eq.~\pref{eq:rvsphi} --- ensures the resulting radius is exponentially large in $\chi$, naturally ensuring an exponentially large hierarchy in these models as well. In the special case of supersymmetric ADD models \cite{SLED, SLEDrev} $\ell \sim (10 \, \hbox{TeV})^{-1}$, and so micron-sized dimensions can be achieved with $\chi \sim - 70$. But the stabilization mechanism itself doesn't rely on using an ADD framework, and could equally well apply if it were the Kaluza-Klein (KK) scale that were of electroweak size.

Of course quantum effects can modify eq.~\pref{eq:rvsphi} because they break the classical bulk scale invariance. But since each loop breaks scale invariance by a specific amount, these turn out to generate corrections as a series in $e^{2\chi}$ \cite{6DLoops}, and so do not ruin the exponentially large size of $r$.

\subsection{Relevance to the Higgs}

{}From the point of view of the Higgs, what is important about the above mechanisms (in both 5 and 6 dimensions) is that they require the presence of a coupling between a bulk scalar field and the brane on which the Higgs sits. For instance, in the 6D case the most general renormalizable interactions between a brane-localized Standard Model and a (canonically normalized) electroweak singlet bulk scalar, $\Phi$, have the form
\be
 S_{\rm int} = -\int \exd^4x \; \sqrt{- \gamma} \; U \left(H^\dagger H, \Phi \right) \,,
\ee
with
\bea \label{Ubintro}
  U \left( H^\dagger H, \Phi \right) &=& T_0 + \frac{\lambda_2}{2} \,  \left( \Phi_b + V^2 \right)^2 + g \, H^\dagger H \, \Phi_b + \lambda \left( H^\dagger H - \frac{v^2}{2} \right)^2 \nn\\
 &=& T + \mu_\Phi^2 \, \Phi_b + \frac{\lambda_2}{2} \,  \Phi_b^2 - \left( \mu_\ssH^2 - g \, \Phi_b \right) \, H^\dagger H + \lambda \left( H^\dagger H \right)^4 \,,
\eea
where
\be
  T := T_0 + \frac{\lambda_2 V^4}{2} + \frac{\lambda v^4}{4} \,, \qquad
 \mu_\Phi^2 := \lambda_2 V^2 \quad \hbox{and} \quad
 \mu_\ssH^2 := \lambda v^2 \,,
\ee
and $\Phi_b = \Phi(x,y = y_b)$, denotes the evaluation of the bulk scalar at the position of the brane. It is the dimensionless coupling $g$ that represents the unique Standard Model portal into extra dimensions within this six-dimensional context.

This means that the vacuum energetics of the Higgs field interacts with the physics that stabilizes the extra dimensions, and both Higgs and bulk fields must be varied to find the proper vacuum configuration. In particular, the bulk scalar couplings can act to help or hinder the propensity for electroweak symmetry breaking. For instance, to the extent that large volume requires $\Phi_b < 0$ we see that this acts to increase\footnote{For $g>0$.} the effective value $\mu^2_{\ssH\,{\rm eff}} = \mu_\ssH^2 - g \, \Phi_b$, and so assists the formation of a nonzero {\em v.e.v.}~for $H$. In what follows \S\ref{ssec:VacConf} fleshes this out more explicitly, with care being taken to handle properly the renormalizations required because $\Phi_b$ actually diverges at the brane position.

Similarly, using the replacement
\be
 H = \frac{1}{\sqrt 2} \left(
                         \begin{array}{c}
                           0 \\
                           v + h \\
                         \end{array}
                       \right) \,,
\ee
in the term $g \, H^\dagger H \, \Phi_b$ contributes to Higgs-bulk mixing, and so to invisible channels where energy leaks into the extra dimensions during Higgs-production processes. As we show below, such leakage looks like a Higgs invisible width, and so is subject to similar constraints. Furthermore, since the coupling $g$ is dimensionless, this loss rate is less suppressed at lower energies than would have been true for gravitational energy loss, and so allows better bounds and opportunities for detection \cite{6DHiggsph}. \S\ref{ssec:BBmix}\ computes this more carefully, extending the results of \cite{6DHiggsph} by taking full account of the Higgs-KK mixing brought about by brane-bulk back-reaction.

\begin{figure}[t]
\begin{center}
\epsfig{file=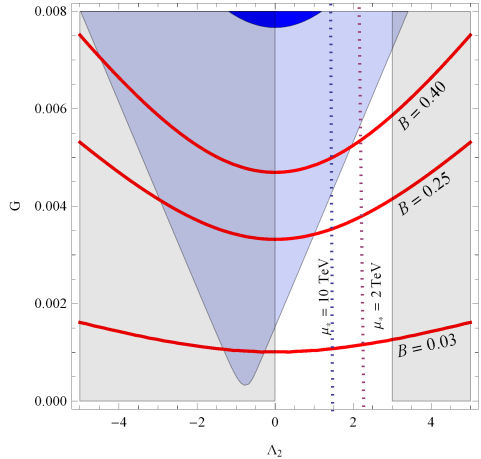,width=2.5in} \caption{A plot summarizing the various constraints and discovery potential of Higgs-bulk mixing in the $G=\bar g/\sqrt\alpha$ vs $\Lambda_2 = \bar\lambda_2/\alpha$ plane (renormalized at $\bar r = 1/m_h$), in the large-volume limit ($m_h \gg m_\KK$). The quantity $\alpha \sim 1$ is a measure of the defect angle near the brane, as defined in detail in \S\protect\ref{ssec:AcFE}. The dark (blue) shaded region is the region disfavoured by LHC global fits. The medium (blue) shaded region is the conservative bound from nucleon-bulk bremsstrahlung in SN1987a, assuming $T_{\ssS \ssN} = 20$ MeV. The lightest (gray) shade denotes regions excluded by demanding no Landau poles below $\mu_* = 1$~TeV, with the vertical dotted lines denoting how this bound changes with the choice of ultraviolet scale $\mu_*$. Also plotted are lines of constant invisible branching ratio $B$ that will be probed with additional data at the LHC or future experiments, all of which constrain this quantity.}
\label{resultgraph}
\end{center}
\end{figure}

The calculation in 6D in many ways resembles earlier work which considered Higgs-curvature mixing \cite{Giudice:2000av}, of the form $H^\dagger H \, R$, but with three differences. First, because the curvature couplings involve more derivatives than do the Higgs-scalar couplings, the curvature mixing remains suppressed at low energies (like other gravitational interactions). Secondly, unlike these earlier calculations, we are able to compute both the real and imaginary parts of the Higgs production amplitude and so can compute the full line-shape rather than just its effective width. We can do so because our treatment of back-reaction allows us to renormalize the divergences that complicate obtaining the real part, associated with the near-brane divergences of the bulk fields. This technology allows us to extend the study of mixing to invisible final states in astrophysics, and at colliders. Finally, we include {\em all} possible renormalizable interactions, including in particular the quadratic self-coupling, $\lambda_2$, for the bulk field on the brane. This inclusion has important consequences, since bounds on $g$ weaken with increasing $\lambda_2$, ultimately allowing a detectable invisible width at the LHC be consistent with strong constraints from low-energy astrophysics (see Fig.~\ref{resultgraph}).

Although our results apply both to the cases of large dimensions ($m_h \gg m_\KK$) and small ones ($m_h \ll m_\KK$), when discussing the phenomenology we focus on the case when the dimensions are large. We find, as did earlier authors \cite{Giudice:2000av, 6DHiggsph}, that a Higgs undergoing Higgs-bulk mixing in many ways resembles a Higgs that can decay into invisible channels. Indeed once both real and imaginary parts of the amplitude are computed, we find that the resemblance becomes perfect for processes with the Higgs resonantly produced in the narrow-width limit.

However, because the resonant, narrow-width limit is not always sufficient, there are also important differences between a bulk-mixed Higgs and one with access to invisible decays. Most important among these is the existence of strong bounds from astrophysical processes like SN1987a. These are not normally relevant for a Higgs with invisible decay channels (or for Higgs-bulk mixing through the $H^\dagger H \, R$ term), because the rate for producing the Higgs is too small at low energies to give an appreciable energy-loss channel. The same is not true for Higgs-bulk mixing in the scalar potential, however, since this is not suppressed at low energies, and is not dominated by resonant Higgs production. It is instead enhanced by the kinematic availability of a large number of very light states for which the couplings cannot be neglected. The resulting constraint is shown in Fig.~\ref{resultgraph}, together with the constraint coming from the successful Standard Model description of the observed Higgs, and contours indicating the size of the effective invisible Higgs width. Although astrophysics furnishes a very strong constraint, it does not exclude the range of interest to future LHC measurements. It does not do so because it is not a resonant process, and so involves a different combination of parameters than are measured at the LHC in the $g-\lambda_2$ plane.

The next sections present the details of this analysis as follows. First, \S\ref{sec:HBdyn} derives the main expressions for the propagation eigenstates when the Higgs mixes with bulk. Armed with these calculations we discuss some of the resulting phenomenology in \S\ref{sec:Phen}, including the lineshape for Higgs production at the LHC, and various bounds from lower energy phenomena. Our conclusions are summarized briefly in \S\ref{sec:Conc}.

\section{Higgs-Bulk Dynamics}
\label{sec:HBdyn}

In this section we compute in detail the implications of a Higgs-bulk interaction of the form given in eq.~\pref{Ubintro}. For concreteness we restrict from here on to the 6D case, for which back-reaction issues are much less well-explored.

We start, in \S\ref{ssec:AcFE} where the bulk and brane actions and field equations, including the conditions for back-reaction, are described. The next subsection, \S\ref{ssec:VacConf} then calculates how eq.~\pref{eq:rvsphi} changes the energy minimization for the Higgs and bulk scalar fields and so alters the expression for the Higgs {\em v.e.v.} in terms of the parameters in its potential. This is followed in \S\ref{ssec:BBmix} by a calculation of the spectrum of fluctuations, including a treatment of how the on-brane Higgs mixes with the bulk KK states. We consider two limits of interest in this mixing, depending on whether the KK mass is small or of the same order as the on-brane Higgs mass. The former is of most interest for ADD and supersymmetric large-dimension (SLED) scenarios, while the latter would be of interest for dimensions whose KK scale is of order the electroweak scale. We specialize to the case of large dimensions when examining phenomenology more explicitly.

\subsection{Field equations and back-reaction}
\label{ssec:AcFE}

We start by describing the 6D bulk and 4D brane systems of interest. For simplicity we focus purely on a single bulk scalar field, coupled to a Standard Model Higgs doublet on a space-filling codimension-2 brane situated at a specific spot in the two extra dimensions.

\subsubsection*{The Action}

Consider a massless, free, 6D bulk scalar, $\Phi$, with action
\be \label{bulkAction}
 S_\ssB = - \int \d^6 x \sqrt{-\cG} \left( \frac{1}{2} \, \cG^{\ssM \ssN} \partial_{\ssM} \Phi \, \partial_{\ssN} \Phi \right) \,.
\ee
We do not include a scalar potential in the bulk, and for ADD-type models this could be naturally enforced through a shift symmetry. The presence of scalars in the gravity supermultiplet and in the massless hypermultiplet representations of 6D supersymmetry also make it natural include light bulk scalars when the extra dimensions are supersymmetric. In the simplest case for bulk stabilization the supergravity of interest is gauged, chiral supergravity, and $\Phi = V^2 \chi$ represents the canonically normalized dilaton that transforms in the (extended) gravity multiplet \cite{NS, SS}. In this case there is a bulk scalar potential, $U_\ssB(\chi) \propto e^{\chi}$, which considerably complicates the treatment of fluctuations once the metric is included. However because the gravitational couplings are RG-irrelevant we omit them for simplicity of presentation, and expect our considerations explored here to apply at sufficiently low energies.

Next, consider a space-filling 4D brane that is located at a particular point, $y = y_b$, within the extra dimensions. With eq.~\pref{Ubintro} in mind we take the brane action to be
\be \label{braction}
 S_b = \int \d^4 x \sqrt{- \gamma}  \left(  \cL_{\SM} - T_0 - \frac{\lambda_2}{2} \, ( \Phi_b + V^2 )^2 - g \, H^\dagger H \, \Phi_b \right) \, ,
\ee
where $\Phi_b := \Phi(x,y=y_b)$ and $\gamma_{\mu\nu} = \cG_{\ssM \ssN}(x,y=y_b) \partial_\mu z^\ssM \partial_\nu z^\ssN$ is the induced metric on the brane, whose world-sheet is denoted $z^\ssM = \{x^\mu, y^m = y^m_b(x) \}$. $\cL_\SM$ denotes the Standard Model action, but for the present purposes we need only work with its Higgs part:
\be
 - \cL_{\SM} = \gamma^{\mu\nu} \partial_\mu H^\dagger \partial_\nu H + \lambda \left( H^\dagger H - \frac{\mu_\ssH^2}{2\lambda} \right)^2 \,.
\ee
Thus, the complete, on-brane scalar potential reads
\be \label{scalarPotential}
 U_b = T - \mu_\ssH^2 H^\dagger H + \lambda (H^\dagger H)^2 +  \mu_\Phi^2 \Phi_b + \frac{\lambda_2}{2} \, \Phi^2_b + g \, H^\dagger H \, \Phi_b \,,
\ee
as anticipated in eq.~\pref{eq:rvsphi}. This contains all possible terms involving only $H$ and the Standard Model that are local and involve only relevant or marginal couplings.

\subsubsection*{Background Geometry}

For the purposes of discussing Higgs energetics, consider the following unwarped, axisymmetric background geometry,
\be \label{geom}
 \d s^2 = \cG_{\ssM \ssN} \, \d x^\ssM \d x^\ssN = \eta_{\mu \nu} \, \d x^\mu \d x^\nu + f^{2}(r) \, \d \theta^2 + \d r^2 \, ,
\ee
where $r$ denotes proper distance away from the brane on which the Higgs resides. We allow for the possibility of a conical singularity at this brane by allowing a defect angle: $0 < \theta < 2 \pi \alpha$, with $0 < \alpha < 1$. Control of approximations usually requires a small defect angle, so $|\alpha - 1| \ll 1$. In real examples of interest the radial coordinate runs through a finite range, $0 < r < \pi R $, with $r =\pi R$ associated with another 4D brane at the opposite end of the extra dimensions.

For the present purposes we ask for simplicity that the singular behaviour of the extra-dimensional geometry be no worse than a conical singularity at the brane position, and so require $f(r) \approx r$ for $r \ll R$. This is not the most general case but is broad enough to include a variety of back-reacted examples, such as locally flat extra dimensions --- corresponding to $f(r) = r$ --- and spherical (or rugby-ball, for nonzero deficit angle) extra dimensions --- for which $f(r) = R \sin(r/R)$ --- as well as other potentially more exotic geometries.

We do not specify $f(r)$ explicitly other than this near-brane limit. This generality is possible because for collider applications to ADD-type models not much depends on $f(r)$. Physically, this is because it is only the enormous phase space associated with the large number of very high energy modes that allows observably large contributions to collider physics at all. But these modes have such short wavelengths that they are insensitive to the large-scale shape of the extra dimensions (see, for example, \cite{Leblond:2001ex} for explicit calculations that illustrate this point).

\subsection{Vacuum configurations}
\label{ssec:VacConf}

We now seek vacuum solutions to the coupled brane-bulk field equations, subject to the assumptions of 4D Lorentz invariance and axisymmetry in the extra dimensions.

\subsubsection*{Bulk field equations and vacuum solutions}

Using $\Phi = \Phi(r)$ in the bulk field scalar equation, $\Box \Phi = 0$, then gives
\be \label{bulkFE}
 \partial_r( f \partial_r \Phi )= 0 \,,
\ee
which integrates to give
\be \label{eq:rPhi'eq}
 \partial_r \Phi = \frac{\cA}{f(r)} \,,
\ee
for integration constant $\cA$. A second integration gives
\be \label{eq:Phisoln}
 \Phi(r) =  \cA \int\limits^r_{\hat{r}} \frac{\d u}{f(u)} := \cA F(r,\hat{r}) \,,
\ee
where we define a new coordinate, $F$, using the condition $\exd F := \exd r/f$.

In principle we also must satisfy the Einstein equations (and equations for any other bulk fields), but instead we use the fact that we do not require more than the near-brane form for $f(r)$ to side-step the effort of doing so. (See, however, \cite{6DSolns} for many explicit solutions to the 6D supergravity equations, including both those where the branes at $r = 0$ and $r = \pi R$ have different properties. Many among these solutions are consistent with the near-brane forms being assumed here.)

\subsubsection*{Boundary conditions and back-reaction}

We seek to eliminate the integration constants -- $\cA$, $\hat r$, {\em etc.} --- of the bulk solution in terms of the physical couplings of the brane action, and this is done using the near-brane boundary conditions that express how the branes back-react onto the bulk \cite{Cod2BR}. Specialized to the bulk scalar field considered here these state
\be \label{branebcI}
 -2 \pi \alpha \, f \,\Phi_b' - \frac{\delta S_b}{\delta \Phi} = - 2 \pi \alpha \, \cA + g H^\dagger H + \lambda_2 \Phi_b + \mu_\Phi^2 = 0 \,,
\ee
where $\Phi_b := \Phi(0)$, $\Phi_b' := (\partial_r \Phi)_{r=0}$ and the second equality in eq.~\pref{branebcI} uses the field equation, eq.~\pref{eq:rPhi'eq}, as well as the form, eq.~\pref{braction}, of the brane action. [One way of deriving this boundary condition -- for completeness, sketched in more detail in Appendix \ref{floating} -- is by excising the codimension-2 brane with a small regularizing codimension-1 cylinder (designed to dimensionally reduce to the above codimension-2 action when the cylinder's radius is very small), and using Israel junction conditions for the cylinder.]

There are similar equations governing the near-brane form of the metric and any other bulk fields, but for the present purposes these just dictate how the defect angle depends on the value of $U_b$ when evaluated at the classical solutions for $H$ and $\Phi$. Similar boundary conditions also apply for the brane at $r = \pi R$, and together with eq.~\pref{branebcI} these generically can be used to remove the two free integration constants in $\Phi(r)$ \cite{6DGW}.

The brane-localized fields must also satisfy their own classical field equations,
\be
 \frac{\delta S_b}{\delta H} = 0 \,,
\ee
and so for $x$-independent $H$ eq.~\pref{branebcI} should be supplemented with\footnote{See Appendix \ref{app:VEnerg} for the relative energetics of this solution compared with the solution $H = 0$.}
\be \label{higgsFE}
 H^\dagger H = \frac{1}{2 \lambda}(\mu_\ssH^2 - g \, \Phi_b ) \,.
\ee
This can be used to eliminate $H$ from (\ref{branebcI}), to give
\be \label{bulkbc}
  -2 \pi \alpha \, \cA + \lambda_{2\,{\rm eff}} \, \Phi_b + \mu^2_{\Phi\,{\rm eff}} = 0 \,,
\ee
where we define the `effective' couplings
\be \label{gammadeltaI}
 \lambda_{2\,{\rm eff}} := \lambda_2 - \frac{g^2}{ 2 \lambda}; \qquad \mu^2_{\Phi\,{\rm eff}} := \mu_\Phi^2 + \frac{g \mu_\ssH^2}{2 \lambda} \,.
\ee

\subsubsection*{Divergences and classical brane renormalization}

The complication of bulk divergences enters once eq.~\pref{eq:Phisoln} is used to eliminate $\Phi_b$. This diverges logarithmically near $r = 0$ due to the asymptotic limit $f \approx r$ there:
\be
 \Phi(r) = \cA \Bigl[ \log(r/\hat{r}) + \hbox{nonsingular} \Bigr]
 \qquad\qquad (\hbox{as $r \to 0$})\,.
\ee
Because $\Phi$ diverges logarithmically as $r \to 0$, we first regularize by taking $r \to \epsilon \ll R$, and then renormalize by allowing the brane couplings to be $\epsilon$-dependent in such a way that $\epsilon \to 0$ can be taken smoothly \cite{Cod2BR, Cod2Renorm}. Although unfamiliar in RS models, such classical divergences (and renormalizations) are generic to {\em any} theories with sources with two or more transverse dimensions (making RS models the exception, rather than the rule). Physically, these divergences arise from taking the source brane to be infinitely thin, and as such they can be lumped together with all of the other quantum ultraviolet (UV) effects that renormalizations of brane couplings would in any case have to encompass.

With this understanding the boundary condition (\ref{bulkbc}) becomes
\be \label{regBC}
 - 2 \pi \alpha \cA + \lambda_{2\,{\rm eff}}(\epsilon) \cA F(\epsilon, \hat r) + \mu^2_{\Phi\,{\rm eff}}(\epsilon) = 0 \,,
\ee
and we require the singular form of the couplings $\lambda_{2\,{\rm eff}}$ and $\mu^2_{\Phi\,{\rm eff}}$ in order to determine how the near-brane boundary condition relates the integration constants $\cA$ and $\hat r$. The one condition that eq.~\pref{regBC} remain finite is insufficient in itself to fix the $\epsilon$-dependence of all couplings, but these are easily determined by repeating the steps of \cite{Cod2BR, Cod2Renorm} and demanding the finiteness of a  few other quantities. For completeness, one way of doing this is described Appendix~\ref{floating}, which yields the same results as earlier authors when restricted to the couplings considered there.

The result for the $\epsilon$-dependence required of the brane couplings obtained in this way is simply summarized as follows,
\be \nn
 \ol{\mu}^2_\Phi(\bar{r}) = \frac{\mu_\Phi^2}{1 - \frac{{\lambda}_2}{2\pi \alpha}  \, F(\epsilon,\bar r)  } \,; \quad  \bar{g}(\bar{r}) = \frac{{g}}{1 - \frac{{\lambda}_2}{2\pi \alpha} \,F(\epsilon,\bar r) } \,; \quad \bar{\lambda}_2(\bar{r}) = \frac{{\lambda}_2}{1 - \frac{{\lambda}_2}{2\pi \alpha} \, F(\epsilon,\bar r) };
\ee
\be \label{rgsolved}
 \bar{\lambda}(\bar{r}) = \lambda + \frac12 \left( \frac{{g}^2}{2 \pi \alpha} \right) \frac{F(\epsilon,\bar r) }{1 - \frac{{\lambda}_2}{2\pi \alpha} \, F(\epsilon,\bar r) } \,; \qquad  \bar{\mu}_\ssH^2(\bar{r}) = \mu_\ssH^2 - \left( \frac{{g} {M}^2 }{2 \pi \alpha} \right) \frac{F(\epsilon,\bar r)}{1 - \frac{{\lambda}_2}{2\pi \alpha} \, F(\epsilon,\bar r) } \,,
\ee
where it is the renormalized (`barred') couplings that are held fixed as $\epsilon \to 0$. The associated RG equations can be found in eqs.~\pref{RGeqns}, below. Here $\bar{r}$ is an arbitrary renormalization scale, and the property $F(r,r) = 0$ ensures that the bare couplings may be interpreted as the renormalized couplings evaluated at $\bar r = \epsilon$. Given these expressions, the $\epsilon$-dependence of the effective coupling combinations appearing in eq.~\pref{regBC} are easily read off:
\be \label{gammadeltaII}
 {\bar{\lambda}}_{2\,{\rm eff}}(\bar{r}) = \frac{\lambda_{2\,{\rm eff}}}{1 - \frac{\lambda_{2\,{\rm eff}}}{2 \pi \alpha} \, F(\epsilon,\bar r) } \,; \qquad {\ol{\mu}}^{2}_{\Phi\,{\rm eff}}(\bar{r}) = \frac{\mu^2_{\Phi\,{\rm eff}}}{1 - \frac{\lambda_{2\,{\rm eff}}}{2 \pi \alpha} \, F(\epsilon,\bar r) } \,.
\ee

Using these expressions to eliminate $\mu^2_{\Phi\,{\rm eff}}$ and $\lambda_{2\,{\rm eff}}$ from eq.~(\ref{regBC}) gives a result for $\cA$ that is finite when $\epsilon \to 0$,
\be
 2\pi \alpha \, \cA = \frac{\mu^2_{\Phi\,{\rm eff}}}{1 - \frac{ \bar \lambda_{2\,{\rm eff}(\bar r)} }{2\pi \alpha} F(\epsilon,\hat r) } =  \frac{ \ol{\mu}^2_{\Phi\,{\rm eff}}(\bar r) } { 1 - \frac{ \bar \lambda_{2\,{\rm eff}(\bar r)} }{2\pi \alpha} F(\bar r, \hat r) } = \ol{\mu}^2_{\Phi\,{\rm eff}}(\hat r) \,,
\ee
where the second equality gives the required relation between $\cA$ and $\hat r$ in terms of the renormalized coupling $\ol{\mu}^2_{\Phi\,{\rm eff}}$ evaluated at an arbitrary scale $\bar r$. The third equality shows how this relation simplifies when expressed in terms of $\ol{\mu}^2_{\Phi\,{\rm eff}}$ defined at the renormalization scale $\hat{r}$. As these expressions make clear, the dependence of the right-side of the last equality on the arbitrary parameter $\bar r$ is illusory. In what follows we will often choose the arbitrary renormalization point for which the answer is most condensed, at the expense of making a logarithmic dependence implicit.

As mentioned earlier, the boundary condition at the distant brane at $r = \pi R$ imposes a second relation between $\cA$ and $\hat r$, in general fixing both and so completely fixing the bulk field configuration \cite{6DGW}. For the present purposes we leave $\hat r$ arbitrary, a placeholder for this faraway boundary condition. It is easy to track in what follows because it appears only through the function $F$, a dependence that is generically logarithmic and so quite weak.

With this understanding the bulk solution now becomes
\be \label{bulkSoln}
 \Phi(r) = \left( \frac{\ol{\mu}_{\Phi\,{\rm eff}}^2(\hat r)}{2 \pi \alpha} \right) F(r, \hat r) \,,
\ee
which, when substituted into eq.~(\ref{higgsFE}) gives the Higgs expectation as
\be \label{braneSoln}
 H^\dagger H = \frac{1 }{2 {\lambda}} \Bigl( \mu_\ssH^2 - g \, \Phi(\epsilon) \Bigr) = \frac{1 }{2 {\lambda}} \left[ \mu_\ssH^2 -  \left( \frac{g \ol{\mu}^2_{\Phi\,{\rm eff}}(\hat r)}{2 \pi \alpha} \right) F(\epsilon, \hat r) \right] = \frac{\bar{\mu}_\ssH^2(\hat r) }{2 \bar{\lambda}(\hat r)} \,,
\ee
which uses eqs.~\pref{rgsolved}. This is finite as $\epsilon \to 0$, as expected. The corresponding formula expressed in terms of renormalized couplings defined at a different scale is found simply by running them up or down according to (\ref{rgsolved}).

Eq.~\pref{braneSoln} shows that it is the renormalized combination $\bar \mu_\ssH^2 $ that must be positive for $H$ to become nonzero. Notice also that eq.~\pref{bulkSoln} shows that it is nonzero $\ol{\mu}^2_{\Phi\,{\rm eff}}(\hat r)$ that determines when $\cA \ne 0$, and so whether $\Phi$ has a nontrivial bulk profile. Physically, this profile arises because $\ol{\mu}^2_{\Phi\,{\rm eff}}$ controls the linear couplings of $\Phi$ to the brane, and having these nonzero precludes $\Phi$'s near-brane derivative from vanishing.

\subsection{Higgs-Bulk mixing}
\label{ssec:BBmix}

We next describe the fluctuations about this background solution, with a view towards identifying the extent to which the $H - \Phi$ couplings cause the Higgs particle to mix with KK $\Phi$-modes in the bulk. We only track here the mixing in the $H - \Phi$ sector, ignoring in particular potential mixing with other bulk fields that might arise within applications where the scalar interacts significantly with other fields (like the metric or fluxes) that are involved in extra-dimensional stabilization. This is known in particular to be a real complication when $\Phi$ is the dilaton that arises as part of the 6D bulk supergravity multiplet \cite{6DKKmixing}.

To study fluctuations we expand the regularized action in powers of the fluctuation fields,
\be
 \Phi = \left( \frac{\ol{\mu}_{\Phi\,{\rm eff}}^2(\hat r)}{2 \pi \alpha} \right) F(r,\hat r) + \phi(x, r, \theta) \,; \qquad
 H = \frac{1}{\sqrt{2}} \left( \begin{array}{c} 0 \\ v + h(x) \end{array} \right) \,,
\ee
where $x = \{ x^\mu \}$ and $v^2 = \bar{\mu}_\ssH^2(\hat r) / \bar{\lambda}(\hat r) = (246$ GeV$)^2$ is fixed from measurements of Fermi's constant, $G_\ssF$, in muon decay. This gives the bulk action
\be \label{eq:bulkActnap}
 S_\ssB = - \int \d^6 x \sqrt{-\cG} \left( \frac{1}{2} \, \cG^{\ssM \ssN} \partial_\ssM \phi \, \partial_\ssN \phi \right) \,,
\ee
while the on-brane potential of the scalar sector reads
\be \label{branePot}
 U_b = T + \lambda v^2  h^2 + \lambda v \, h^3 + \frac{\lambda}{4} \, h^4 + \frac{\lambda_2}{2} \, \phi^2(0) + gv \, h \, \phi(0) + \frac{g}{2} \, h^2 \, \phi(0) \, ,
\ee
where all of the above are `bare' couplings. The rest of the Standard Model looks like it usually does, with $h$ acting as the usual Higgs field.

Our goal is to compute the implications of the $gv \, h \, \phi(0)$ term that mixes the brane and bulk scalar degrees of freedom, and we do so in two ways. Although somewhat redundant, comparing both approaches provides insight and a check on our calculations.

\subsubsection{Perturbative method}
\label{greens}

A straightforward way to approach Higgs-bulk mixing \cite{Giudice:2000av} is to regard the terms $\frac{1}{2} \lambda_2 \, \phi^2(0)$ and $gv \, h \phi(0)$ as part of the interaction lagrangian so that the unperturbed system does not mix brane and bulk. The implications of mixing are then found by summing all possible types of insertions of the mixing interactions.

\begin{figure}[t]
\begin{center}
\epsfig{file=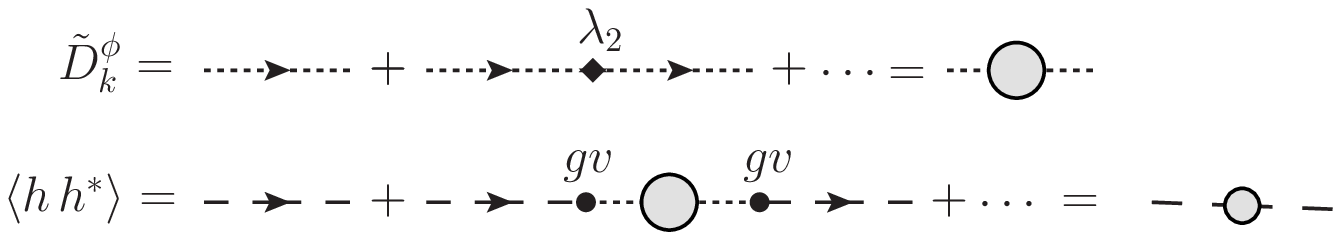,width=4.0in} \caption{Perturbative calculation of the Green's function for the $h$ operator. Dotted lines represent $D_k^\phi$ and dashed lines represent $D_k^h$ as defined in the text. Circles are used for the $gv h \phi(0)$ vertex, diamonds represent a $\frac{1}{2} \lambda_2 \phi^2(0)$ vertex and $\tilde D_k^\phi$ is the $\lambda_2$-resummed Green's function for the bulk field, Fourier transformed in the brane directions but not in the extra dimensions.}
\label{Fgraphs}
\end{center}
\end{figure}

For instance, at tree level a calculation of the correlator $\la hh^* \ra_k$ requires summing over the diagrams of Fig.~\ref{Fgraphs}.
\noindent Summing these graphs gives the following momentum-space result \cite{claudia},
\be \label{resum}
 \langle h h^* \rangle_k = \frac{D_k^h[1 + i \lambda_2 D_k^\phi(0,0)]}{1 + \left[ i \lambda_2 \  + (gv)^2 D_k^{h} \right] D_k^\phi(0,0)} \,,
\ee
This expression can alternatively be derived using a Schwinger-Dyson approach, as shown in Appendix~\ref{appsec:SchwDys}. The advantage of this approach is that it does not require $\lambda_2$ to be perturbatively small. In eq.~\pref{resum}, $D_k^h$ is the momentum-space propagator of $h$ in the unperturbed theory\footnote{Notice we use $\epsilon$ to denote the cutoff, while $\varepsilon$ governs the poles of the Feynman propagator.}
\be
 D_k^{h} = -\frac{i}{k^2 + 2 \lambda v^2 - i \varepsilon} \,,
\ee
and $D_k^\phi(0,0)$ is the unperturbed propagator for the bulk field, Fourier transformed only in the four brane directions and with both extra-dimensional positions evaluated at the Higgs-brane position ($r = 0$).

The unperturbed bulk propagator satisfies
\be
\label{propa}
 \left[ - f \, k^2 +  \frac{1}{f} \,\partial_{\theta}^2 + \partial_r \left( f \partial_r \right)  \right] D^\phi_k(r,\theta; r^\prime, \theta^\prime) = i \delta(r -r^\prime) \delta(\theta - \theta^\prime) \,,
\ee
and it is useful when solving this to expand in a basis of unperturbed eigenmodes for $\Box_2$:
\be
 \phi(k, r,\theta) = \sum_{n l} \phi_{n l}(k) Z_{n l}(r,\theta) \,,
\ee
where
\be \label{eq:boxZ}
 \Box_2 Z_{n l}(r,\theta) = \left[ \frac{1}{f^2} \, \partial_{\theta}^2 + \frac{1}{f} \, \partial_r ( f \, \partial_r ) \right] Z_{n l}(r, \theta) = -M^2_{n l} \; Z_{n l}(r,\theta) \,.
\ee

Rotational invariance of the background further allows the separation of variables
\be \label{eq:zseparation}
 Z_{n l}(r, \theta) = P_{n l}(r) e^{i n \theta / \alpha} \,,
\ee
where $n$ is an integer and the $P_{n l}$ are the corresponding set of radial mode functions.\footnote{Do not confuse these with Legendre functions, despite the notation.} Because these are unperturbed modes, they do not yet `know' about the Higgs-bulk couplings, and so satisfy the comparatively simple near-brane Neumann conditions corresponding to no brane couplings,
\be
 \left(f  \partial_r P_{n l} \right)_{r=0,\pi R} = 0 \,,
\ee
and satisfy the traditional Sturm-Liouville bulk normalization relations
\be \label{eq:Pnorm}
  2\pi \alpha \int\limits_0^{\pi R} dr \, f \, P_{n l}^* P_{n l'} = \delta_{l l'} \,.
\ee

The result that follows from using this expansion in (\ref{propa}) is
\be
  D^\phi_k(r,\theta;r^\prime,\theta^\prime) = - \frac{i}{2 \pi \alpha} \, \sum_{n l} \frac{P_{n l} (r) \, P^*_{n l }(r^\prime)}{k^2 +  M_{n l}^2- i \varepsilon}\; e^{in(\theta - \theta^\prime)/\alpha}  \,,
\ee
and, since only $n=0$ modes survive as $r, r^\prime \to 0$ (as is shown below), the required brane-to-brane propagator becomes
\be \label{eq:branetobrane}
  D_k^\phi(0,0) =  - \frac{i}{2 \pi \alpha} \sum_{l}  \frac{ P_{0 l} (0) \, P^*_{0 l }(0) }{k^2 +  M_{0 l}^2- i \varepsilon}  \,.
\ee

\subsubsection*{Continuum limit}

These expressions become particularly simple in the large-volume limit, where $|k^2 R^2| \gg 1$. In this case the discrete mode spacing is very small and the mode sum is well-approximated by a continuum momentum integral. This continuum limit is taken explicitly in Appendix \ref{app:toymodel}, starting from the discrete mode sum in a simple toy model, but the near-brane result can be obtained more simply by solving \pref{propa} directly for noncompact extra dimensions. This can be done explicitly near the branes, where $f \approx r$, since eq.~\pref{propa} becomes the equation for a free 2D field in cylindrical coordinates, whose solution are given in terms of Bessel functions.

Demanding normalizability near $r=0$ for the unperturbed functions, one finds
\be
 D^\phi_k(r,\theta; r^\prime, \theta^\prime) = -i \sum_{n = - \infty}^\infty  \int \left( \frac{q \, \d q}{2 \pi \alpha} \right) \frac{ e^{in(\theta -\theta^\prime)}}{k^2 + q^2 - i \varepsilon} \; J_{|n/\alpha|}(qr) J_{|n/\alpha|}(qr^\prime) \,.
\ee
Since only the $n = 0$ term contributes when evaluated at $r = r^\prime = 0$, the near-brane limit becomes
\be \label{eq:Dk00int}
 D_k^\phi(0,0) = \frac{-i}{4 \pi \alpha} \int\limits_0^\infty \frac{\d q^2}{k^2 + q^2 - i \varepsilon} \,.
\ee
The integral in eq.~\pref{eq:Dk00int} diverges logarithmically at large $q$, and once this is regularized with a cutoff $\Lambda = 1/\epsilon$ the result becomes
\be \label{eq:D0continuum}
 D_k^\phi(0,0) = \frac{i}{4\pi \alpha}\log\left(k^2 \epsilon^2 \right) \,.
\ee
The $i \varepsilon$ prescription tells us which branch of the logarithm should be used when $k^2$ is negative ({\em i.e.} $k^\mu$ is timelike), in which case
\be \label{eq:branches}
 \log \left( k^2 \epsilon^2 \right) = \log \left( - k^2 \epsilon^2 \right) -i \pi \,.
\ee

Using this in the Higgs two-point function, and expressing the result in terms of renormalized couplings gives the finite final (continuum) result
\be \label{dressed}
 \langle hh^* \rangle_k  = -i \left[ k^2 + 2 \bar{\lambda}(\bar{r}) v^2 + \left( \frac{\bar{g}^2(\bar{r}) v^2}{4 \pi \alpha} \right) \frac{\log(k^2 \bar r^2)}{1 - \left( \frac{\bar{\lambda}_2(\bar{r})}{4 \pi \alpha} \right) \, \log(k^2 \bar r^2) } \right]^{-1} \,,
\ee
where $\bar r$ is the same arbitrary renormalization energy scale at which the renormalized couplings are also evaluated, and in we have used $F(r,r^\prime) = \log(r/r^\prime)$ appropriate for the large $R$ limit. It is the implicit $\bar r$-dependence in these couplings that cancels the explicit dependence appearing in the logarithms, ensuring that $\bar r$ does not contribute to physical quantities computed from $\langle hh^* \rangle_k$.

\subsubsection{Direct mode diagonalization}
\label{sss:diagon}

We next provide an alternative derivation of the Higgs two-point function, which proceeds more directly by calculating $\langle hh^* \rangle_k$ explicitly by diagonalizing the KK and Higgs modes. This provides a more physical interpretation for the branch cut introduced into $\langle hh^* \rangle_k$ by the Higgs-bulk couplings, in terms of Higgs mixing with KK bulk states.

Propagation eigenstates are found by solving the field equations for $h$ and $\phi$, keeping track of the boundary conditions near the brane. In the present instance the relevant equations are the bulk scalar equation
\be \label{bulkeom}
 \left( \Box_4 + \Box_2 \right) \phi = \left[ \partial_\mu \partial^\mu + \frac{1}{f^2} \, \partial_{\theta}^2 + \frac{1}{f} \, \partial_r ( f \, \partial_r ) \right] \phi = 0 \,,
\ee
the (linearized) Higgs field equation on the brane,
\be \label{higgseom}
 -\partial_\mu \partial^\mu h + 2 \lambda v^2 \,  h + gv \, \phi(0) = 0 \,,
\ee
and the near-brane boundary condition for $\phi$,
\be \label{bceom}
 - 2 \pi \alpha \left( f \partial_r \phi \right)_{r = 0} + gv \, h + \lambda_2 \, \phi(0) = 0 \,.
\ee

As before we decompose the 6D scalar into a KK tower by expanding the solutions to eq.~\pref{bulkeom} in a basis of eigenfunctions of $\Box_2$,
\be
 \phi(k,r,\theta) = \sum_{n\ell} \varphi_{n\ell}(k) \, \cZ_{n\ell}(r,\theta) \,,
\ee
with
\be \label{eq:boxcurlyZ}
 \Box_2 \cZ_{n\ell}(r,\theta) = -M^2_{n\ell} \; \cZ_{n\ell}(r,\theta) \,.
\ee
We use the indices $(n \ell)$ rather than $(nl)$ here to emphasize that they run over a slightly different range, with $\ell$ including a value corresponding to $h$ in the special case $n=0$, in addition to the complete range of $l$ for the unperturbed $n=0$ KK modes. This change only happens for the $n=0$ modes because only these mix nontrivially with the brane.

As for the unperturbed case, we write
\be
 \cZ_{n\ell}(r, \theta) = \cP_{n\ell}(r) \, e^{i n \theta / \alpha} \,,
\ee
where $n$ is an integer and the $\cP_{n\ell}$ are the radial mode functions, satisfying
\be \label{bulkeom2}
 \left[ M^2_{n\ell} - \left( \frac{n}{\alpha f} \right)^2 + \frac{1}{f} \, \partial_r ( f \, \partial_r ) \right] \cP_{n\ell} = 0 \,.
\ee

In terms of the 4D modes $\varphi_{n\ell}$ and $h$, the $\phi$ and $h$ field equations, eqs.~\pref{bulkeom} and \pref{higgseom}, are as for the unperturbed case,
\bea \label{mixingeqsexplicit}
 \Bigl[ k^2 + M^2_{n\ell} \Bigr] \varphi_{n\ell} &=& 0 \nn\\
 \Bigl[ k^2 + 2 \lambda v^2 \Bigr]  h + \sum_{n\ell} \Bigl[ gv \,  \cP_{n\ell}(0) \Bigr] \varphi_{n\ell} &=& 0 \,,
\eea
but $\cP_{n\ell}$ differs from $P_{n\ell}$ by satisfying the near-brane boundary condition, eq.~\pref{bceom}, including the implications of the Higgs-bulk mixing,
\be \label{bcexplicit}
 gv \, h + \sum_{n\ell} \Bigl[ - 2 \pi \alpha \left( f \partial_r \cP_{n\ell} \right) + \lambda_2 \, \cP_{n\ell} \Bigr]_{r = 0} \varphi_{n\ell} = 0 \,.
\ee
Using eqs.~\pref{mixingeqsexplicit} to eliminate $h$ and $k^2$ gives
\be \label{eq:hvsphi}
 h = \sum_{n\ell} \left[ \frac{ gv \,  \cP_{n\ell}(0) }{M_{n\ell}^2 - 2 \lambda v^2} \right] \varphi_{n\ell} \, ,
\ee
and allows \pref{bcexplicit} to be rewritten as a boundary condition purely for $\cP_{n\ell}$:
\be \label{bcexplicit2}
 \left[ - 2 \pi \alpha \left( f \partial_r \cP_{n\ell} \right) + \left( \lambda_2 - \frac{(gv)^2}{M^2_{n\ell} - 2\lambda v^2} \right) \cP_{n\ell} \right]_{r = 0} = 0 \,.
\ee

What is unusual about this boundary condition is the presence of $M^2_{n\ell}$, which makes it mode-dependent, at least for those modes\footnote{The only normalizable modes for which $\cP_{n\ell}(0) \ne 0$ are those with $n = 0$, expressing conservation of angular momentum, as expected.} for which it is satisfied with $\cP_{n\ell}(0) \ne 0$. The presence of the unorthodox mode-dependent near-brane boundary conditions implies the eigenfunctions need not be orthogonal using the usual Sturm-Liouville (or Wronskian) inner product. Instead, as shown in detail in Appendix \ref{app:generalizedSturm}, the natural inner product adapted to this boundary-value problem also involves some boundary dependence.

The resulting generalized orthonormality condition for this new inner product is
\be \label{eq:cPnorm}
  2 \pi \alpha \int\limits_0^{\pi R} \d r \, f  \, \cP_s^*\, \cP_t + \frac{(gv)^2 \cP_s^*(0) \cP_t(0)}{(M_{s}^2 - 2 \lambda v^2 )(M_{t}^2 - 2 \lambda v^2 )} = \delta_{\,st} \,,
\ee
where we collectively denote $s,t = \{n \,\ell \}$. As shown in Appendix \ref{app:generalizedSturm}, it is this modified boundary condition that ensures that the KK expansion of the action gives a quadratic action that is diagonal in the $\varphi_{n\ell}$. This property would be ruined in the present instance for an expansion using the unperturbed mode functions, $\phi = \sum_{n l} \phi_{n l} \, P_{n l} $ by the term $gv \, h \, \phi(0) = gv \, h \sum_{n l}  \phi_{n l} P_{n l}(0)$ in the lagrangian, which causes bulk modes with $n = 0$ to mix with $h$. In Appendix~\ref{app:kkwfn} we make this discussion explicit by solving the perturbed wavefunctions subject to this boundary condition in an illustrative toy model.

The complete mass eigenstates of the theory obtained including this mixing are related to the unperturbed states discussed earlier by a linear rotation, as follows:
\be \label{redefs}
 h = \sum_s \cB_s \varphi_s \quad \phi_i = \sum_s \cU_{\,is} \varphi_s \,,
\ee
where the index $i$ denotes $\{n, l\}$ in the same manner as $s$ denotes $\{n, \ell\}$. Because only $n=0$ modes mix with the brane, in practice $\cU_{\,is} = \delta_{is}$ unless $n = 0$.

The quantity $\cB_s$ is of most interest, because it controls the two point function for the $h$ field, which can be written
\be \label{kkProp}
 \la hh^* \ra_k = \sum_s |\cB_s|^2 \; \la \varphi_s \varphi_s^* \ra_k =  -i \sum_s \frac{|\cB_s|^2}{k^2 + M_s^2 - i \varepsilon} \,.
\ee
Eq.~\pref{eq:hvsphi} gives $\cB_s$ as
\be
 \cB_s = \frac{gv \cP_s(0)}{M_{s}^2 - 2 \lambda v^2} \,,
\ee
and $\cU_{\,is}$ is found by using $\cP_{s}(r) = \sum_i \cU_{\,is} \, P_{i}(r) $ together with the normalization conditions, eqs.~\pref{eq:Pnorm} and \pref{eq:cPnorm}, respectively satisfied by $P_i$ and $\cP_s$. This gives
\be \label{unitary1}
 \sum_i \cU^*_{\,is} \, \cU_{\,it} + \cB_s^* \, \cB_t = \delta_{ st} \,,
\ee
as well as the remaining unitarity conditions
\be \label{unitary}
 \sum_s \cB_s^* \cB_s = 1 \,, \quad
  \sum_s \cU_{js} \cB_s^* = 0
  \quad \text{and} \quad \sum_s \cU_{i s} \, \cU^*_{j s} = \delta_{ij}  \,.
\ee

\subsubsection*{The continuum (large $R$) limit}

Although the formulae in \S\ref{sss:diagon} so far make no assumption about the relative size of $\lambda v^2$ and the KK scale, $m_\KK^2 \approx 1/R^2$, we pause here to display the simple result that obtains in the limit $\lambda v^2 \gg m_\KK^2$ (appropriate to large-volume models, say) for which KK mode sums are more usefully cast as integrals \cite{ADD, ADDgravem}.

In this limit it is a good approximation to write the sum appearing in the two-point function as an integral
\be \label{sumB}
 \la hh^* \ra_k = -i \sum_s \frac{|\cB_s|^2}{k^2 + M_s^2 - i \epsilon} \approx -i \int \d M^2  \frac{\rho_h(M^2)}{k^2 + M^2 - i\varepsilon} \,,
\ee
which defines the spectral function, $\rho_h(M^2)$. The simplest way to obtain an expression for $\rho_h$ is by using the explicit result for $\langle hh^* \rangle_k$ obtained above from the perturbative calculation. To read off $\rho_h(M^2)$ from this calculation we employ unitarity, in the form
\be \label{eq:unitaryrho}
 \pi \rho_h(M^2)  =  \text{Re} \, \langle hh^* \rangle_k \Bigr|_{k^2=-M^2} \,.
\ee
Using expression (\ref{dressed}), and separating the real and imaginary parts of the logarithms, gives the finite result
\bea \label{rhob}
  \pi \rho_h(M^2) &=& \left. \frac{ \bar{g}^2(\bar{r}) v^2 / 4 \alpha}{ \left[M^2 - 2 \bar{\lambda}(\bar{r}) v^2 \right]^2 + \pi^2 \left[\frac{\bar{\lambda}_2 (\bar{r})}{4 \pi \alpha} [M^2 - 2 \bar{\lambda}(\bar{r})v^2 ] + \frac{\bar{g}^2(\bar{r}) v^2}{4 \pi \alpha} \right]^2 } \, \right|_{\bar{r} = 1/M} \nn\\
  &=& \frac{v^2 \zeta(M^2) }{[M^2 - \Pi(M^2)]^2 + v^4 \zeta^2(M^2) } \,,
\eea
where
\be \label{eq:zeta}
 \zeta(M^2) := \frac{\bar{g}(\bar{r}) /4 \alpha }{\left[1 - \left( \frac{\bar{\lambda}_2(\bar{r})}{4 \pi \alpha} \right) \log( M^2 r^2)\right]^2 + \left(\frac{\bar{\lambda}_2(\bar r)}{4 \alpha} \right)^2} \,,
\ee
and
\be \label{eq:mstar}
 \Pi(M^2) := 2 \bar{\lambda}(\bar{r}) v^2 + \left( \frac{\bar{g}^2(\bar{r}) v^2}{4 \pi \alpha} \right) \frac{  \log(M^2 \bar{r}^2) \left[ 1 - \left( \frac{\bar{\lambda}_2(\bar r)}{4 \pi \alpha} \right) \log(M^2 \bar{r}^2)\right] - \frac{\pi \bar{\lambda}_2 (\bar r)}{4 \alpha} } { \left[ 1 - \left( \frac{\bar{\lambda}_2(\bar r)}{4 \pi \alpha} \right) \log(M^2 \bar{r}^2)\right]^2 + \left( \frac{\bar{\lambda}_2(\bar r)}{4 \alpha} \right)^2} \, .
\ee

For future reference we note here that many of the above expressions simplify if we make a specific choice, $\bar r = 1/M$, for the arbitrary renormalization scale, so that the $\log( M^2 \bar{r}^2 )$ terms disappear. In the present instance this leads to the simpler formulae
\be \label{zetadef}
  \zeta(M^2) = \left. \frac{\bar g^2(\bar{r}) /4\alpha}{1 + \left( { \bar\lambda_2 (\bar{r})/4 \alpha} \right)^2}  \right|_{\bar{r} = 1/M} \,,
\ee
and
\be \label{mstardef}
   \Pi(M^2) = 2 v^2 \left[ \bar\lambda(\bar{r}) - \frac{\zeta(M) \bar \lambda_2 (\bar r) }{8\,\alpha} \right]_{\bar{r} = 1/M}\,.
\ee

The weak-coupling limits of the spectral function are most easily seen using the second equality in eq.~\pref{rhob}, together with the representation $\pi \delta(x) = \lim_{\,\zeta \to \,0} \zeta/(x^2 + \zeta^2)$. This gives
\be
 \rho_h(M^2) \to \delta(M^2 - 2 \lambda v^2) \,,
\ee
both when $ g \to 0$ and when $\lambda_2 \to \infty$, illustrating how the Standard Model is obtained in both of these limits.\footnote{We describe these limits in terms of bare couplings, since complications associated with divergences vanish in these limits, and the renormalized couplings $\bar{\lambda}$, $\bar{g}$ and $\bar \lambda_2$  no longer run.} Although it is clear why this should hold when $g = 0$, it turns out also to hold when $\lambda_2 \to \infty$ because the boundary conditions, (\ref{bceom}), imply in this case that the KK modes all vanish on the brane.

It is the two functions $\zeta$ and $\Pi$ that control the phenomenology of Higgs-bulk mixing, as we now show.

\section{Phenomenological implications}
\label{sec:Phen}

In this section we discuss the leading sources of constraint on the Higgs-bulk mixing just described. We separate the effects into three types: virtual effects from exchanging KK modes; real effects from KK modes in the final state; and changes to the relation between the Higgs mass and couplings (together with associated changes to the constraints from vacuum stability).

\subsection{Virtual Higgs Exchange}
\begin{figure}[t]
\begin{center}
\epsfig{file=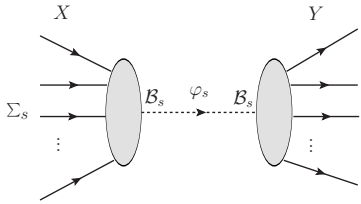,width=2.0in} \caption{Scattering via virtual exchange of the Higgs-bulk tower. A factor of $\cB_s$ comes from the vertex between $\varphi_s$ and both the initial and final state. All modes are summed over because any of the $\varphi_s$ can mediate this interaction.}
   \label{Fig:higgsExch}
\end{center}
\end{figure}

Consider first a tree-level Standard Model parton process of the form $X \to h \to Y$, where an initial state $X$ produces a virtual $h$ that subsequently produces state $Y$, as in Fig.~\ref{Fig:higgsExch}. The Standard Model amplitude for this process can be schematically written
\be
 \cM^\SM (X \to Y) = \cM(X \to h) \, \langle hh^* \rangle_k^\SM \, \cM(h \to Y) \,,
\ee
and so once $h$ mixes with the KK tower this becomes
\bea \nn
 \cM(X \to Y) &=& \cM(X \to h) \cM(h \to Y) \sum_s |\cB_s |^2 \langle \varphi_s \varphi_s \rangle_k  \\
 &=& \cM(X \to h) \cM(h \to Y) \langle h h^* \rangle_k \,,
\eea
where $k^\mu = p^\mu_\ssX = p^\mu_\ssY$ represents the 4-momentum flowing down the Higgs line. That is, virtual effects of Higgs-bulk mixing exchange processes are found by using $\la hh^* \ra_k$ in place of the Standard Model Higgs propagator.

\subsubsection{Lineshape}

Implications of this modification to Higgs exchange are easiest to study in the large-volume limit, for which the KK sums are well-approximated by integrals. In this case we may use eq.~(\ref{dressed}), which we rewrite in the Lorentzian form by separating the real and imaginary parts of the denominator
\be
 \la hh^* \ra_k  = -i \Bigl[ \Pi(-k^2) + k^2  -i v^2 \zeta(-k^2)   \Bigr]^{-1} \,,
\ee
where the functions are, not surprisingly, the same functions as in \S\ref{greens}.

\begin{figure}[t]
\begin{center}
\epsfig{file=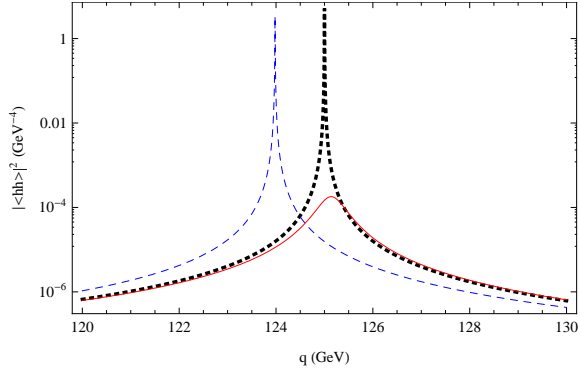,width=3.0in} \caption{The lineshape, $|\langle hh^* \rangle_k |^2$, for the KK tower exchange with various values of $(\bar{\lambda}, \bar{\lambda}_2, \bar{g},\alpha)$ evaluated at $\bar{r} = (125 \text{ GeV})^{-1}$. The dotted black line is the comparison lineshape for a Standard Model Higgs with mass, $m_h = 125$ GeV and a Standard Model width of $\Gamma_{\SM} = 4$ MeV. (In the Standard Model the Higgs coupling for this mass is $\bar \lambda = 0.1291$.) The dashed blue line shows a similarly narrow peak (chosen to lie near $124$ GeV rather than 125 GeV to avoid clutter in the figure), obtained using $(\bar{\lambda}, \bar{\lambda}_2, \bar{g},\alpha) = (0.127,0,0.0021,0.8)$. The solid red line shows instead a broad peak at $125$ GeV for an exaggerated choice of couplings $(\bar{\lambda}, \bar{\lambda}_2, \bar{g},\alpha) = (0.16,2,0.35,0.01).$}
   \label{Fig:lineshape}
\end{center}
\end{figure}

Physical processes depend on the squared magnitude,
\be
   |\la hh^* \ra_k|^2 = \frac{1}{\left[\Pi(-k^2) + k^2  \right]^2 + v^4 \zeta^2(-k^2)  } \, ,
\ee
which defines the resonant Higgs lineshape in the presence of mixing. A plot of this lineshape, and a comparison with its Standard Model counterpart, is given in Fig.~\ref{Fig:lineshape}, for several choices for the Higgs-bulk couplings.

The position of the resonant maximum occurs at
\be \label{eq:peak}
 m_h^2 = \Pi(m_h^2) = v^2 \left[ 2 \bar{\lambda}(\bar{r}) - \frac{ \bar{g}^2(\bar{r}) \, \bar{\lambda}_2(\bar{r}) }{(4 \alpha)^2 + \bar{\lambda}_2^2(\bar r)} \right]_{\bar{r} = 1/m_h} \, ,
\ee
where we have used the definitions of $\zeta$ and $\Pi$ in eqs.~\pref{zetadef} and \pref{mstardef}. This suggests the definition
\be
 m_h^2 := 2 \bar{\lambda}_{\rm eff}(\bar r) v^2 \Bigr|_{\bar{r} = 1/m_h} \quad \text{ with } \quad 2\bar{\lambda}_{\rm eff}(\bar r) := 2 \bar{\lambda}(\bar{r}) - \frac{ \bar{g}^2(\bar{r}) \, \bar{\lambda}_2(\bar{r}) }{(4 \alpha)^2 + \bar{\lambda}_2^2(\bar r)} \, .
\ee

\begin{figure}[t]
\begin{center}
\epsfig{file=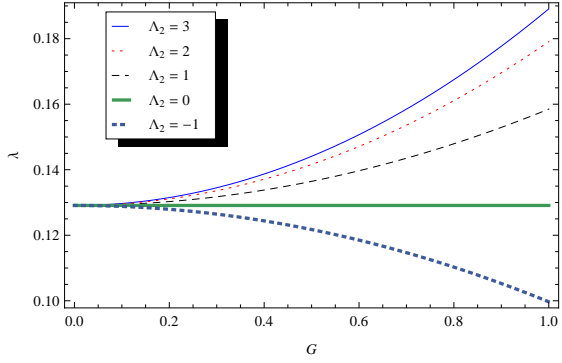,width=3.5in} \caption{The Higgs quartic coupling $\bar{\lambda}$ required to ensure $m_h = 125$ GeV, as a function of the brane bulk mixing parameter $G = \bar{g}/\sqrt{\alpha}$ for various choices of the coupling parameter $\Lambda_2 = \bar{\lambda}_2 / \alpha$ (as listed in the legend). The flat line $\bar{\Lambda}_2 = 0$ corresponds to the Standard Model value $\bar{\lambda} = 0.1291$. All couplings are renormalized and evaluated at a scale $\bar{r} = (125 \text{ GeV})^{-1}$.}
   \label{Fig:lambda}
\end{center}
\end{figure}

It is the value of the renormalized couplings at the scale $\bar r = 1/m_h$ that is relevant to many physical quantities, such as the condition $\bar\lambda_{\rm eff}(\bar r = 1/m_h) = 0.1291$ that is required to ensure a Higgs mass of 125 GeV. Because of this, in what follows a barred coupling without a specified renormalization scale is understood to be evaluated at $1/m_h$:
\be
 \bar{\lambda} = \bar{\lambda}(\bar r) \Bigr|_{\bar{r} = 1/m_h} \,, \quad \bar g = \bar g(\bar r) \Bigr|_{\bar r = 1/m_h} \,,
 \quad \hbox{and so on} \,.
\ee
Notice also that the condition $m_h = 125$ GeV imposes only a single relation amongst the couplings $\bar \lambda$, $\bar \lambda_2$ and $\bar g$, rather than fixing $\bar \lambda$ completely, as it would have done in the Standard Model. Fig.~\ref{Fig:lambda} plots the value predicted for $\bar \lambda$ as a function of $\bar g$ for various choices $\bar \lambda_2$, and we assume in what follows that $\bar \lambda$ is fixed in this way.

We see that the lineshape can resemble a single Higgs resonance despite its containing a sum over many KK states. Its width at its maximum is
\be
 m_h \Gamma_\ssB := \zeta(m_h^2) v^2 = \frac{\bar g^2v^2 /4\alpha}{1 + \left( \bar\lambda_2 /4 \alpha \right)^2}  \,.
\ee
This width is related by unitarity to the rate for invisible processes where the KK modes escape invisibly into the bulk, carrying with them missing energy. When $\Gamma_\ssB$ is sufficiently small, such as in the $g \to 0 $ limit, its role in the unitarity argument is eventually replaced by the Standard Model Higgs decay width, $\Gamma_\SM$, corresponding to the imaginary part of the usual Standard Model Higgs vacuum-polarization graphs. This suggests the definition $\Gamma_h = \Gamma_\ssB + \Gamma_\SM$. However, because the Standard Model contribution is a loop effect it should only be kept in the special case where it dominates $\Gamma_\ssB$ at the peak of a narrow resonance. Given these considerations, we write the corrected Green's function
\be
 \la hh^* \ra_k^\prime = -i \left[ \Pi(-k^2) + k^2 - iv^2 \zeta(-k^2) -i m_h \Gamma_\SM \right]^{-1} \, ,
\ee
so that
\be
| \la h h^*\ra_k^\prime |^2 = \frac{1}{[\Pi(-k^2) + k^2 ]^2 + m_h \left[ v^2 \zeta(-k^2) +  m_h \Gamma_\SM \right]} \, .
\ee

\subsubsection{Invisible width}
\label{sssec:invWidth}

In essence, mixing with the bulk introduces a new invisible channel into Higgs reactions while leaving unchanged the relative strength of all of the visible $h$ couplings to other Standard Model particles. The exchange of the KK tower (instead of just the Higgs) suppresses the overall rates for observable Higgs-mediated processes, while preserving their relative frequency. The success of the Standard Model description of the resonance at 125 GeV, as seen by both CMS and ATLAS, therefore provides an immediate constraint on such mixing, in much the same way as it constrains a more conventional Higgs invisible width.

In the present context the nature of this constraint is easiest to see within the narrow-width approximation, which applies in the phenomenologically most relevant case where $m_h \gg \Gamma_{\ssB} + \Gamma_{\ssS \ssM}$. In this case the resonant $h$ autocorrelation becomes
\be
 |\la hh^* \ra_k^\prime|^2 \approx \frac{\pi}{ m_h \left[\Gamma_{\ssB} + \Gamma_{\ssS \ssM}\right] } \; \delta(k^2+m_h^2) \,,
\ee
which neglects a factor $\left[ 1 - \Pi^\prime(m_h^2) \right]^{-1}$. This factor can be dropped because
\be \label{Wprimeexpr}
 \Pi{}^\prime (m_h^2) := \left( \frac{\partial \Pi}{\partial k^2} \right)_{k^2 = - m_h^2} = \frac{\zeta(m_h^2) \, v^2}{\pi m_h^2} \left[ \frac{1 - (\bar \lambda_2/4\alpha)^2}{1 + (\bar \lambda_2/4 \alpha)^2} \right] \approx \cO \left( \frac{\Gamma_\ssB}{m_h} \right) \,,
\ee
and so is similar in size to other contributions that have been neglected.

For comparison, the Standard Model Higgs distribution in the same narrow-width limit reads
\be
  |\la hh^* \ra_k|^2_\SM \approx \frac{\pi}{m_h \Gamma_{\ssS \ssM} } \; \delta(k^2+m_h^2) \,,
\ee
with $m_h$ being the physical mass of the Higgs. In this limit the momentum dependence of the two results is identical, and the exchange of the KK tower just provides an overall suppression to the rates for Higgs-mediated processes by the factor
\be
 \cR = \frac{\Gamma_{\ssS \ssM}}{\Gamma_{\ssB} + \Gamma_{\ssS \ssM}}
 \,,
\ee
relative to the Standard Model.

\begin{figure}[t]
\begin{center}
\epsfig{file=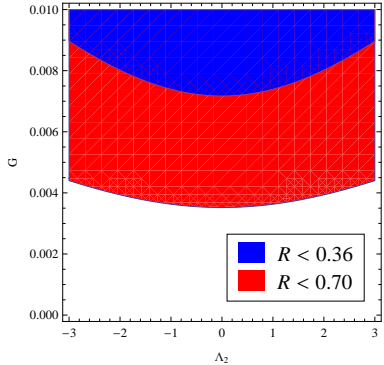,width=2.0in} \caption{Bounds on $\Lambda_2 = \bar{\lambda}_2 / \alpha$ and $G = \bar{g}/\sqrt{\alpha}$ from constraints on the invisible branching ratio of a Standard Model Higgs. The shaded region is excluded, for two sets of assumptions (described in the main text). Couplings are evaluated at a scale $\bar{r} = (125 \text{ GeV})^{-1}$.}
   \label{Fig:lineshapebounds}
\end{center}
\end{figure}

The bound that follows for $\cR$ can be inferred using the results of extant global fits to the LHC data that were performed to constrain the branching ratio into invisible decays of an otherwise Standard Model-like Higgs. The corresponding narrow-width suppression for a conventional invisible decay width would be $\cR_{\rm inv} = \Gamma_\SM/(\Gamma_\SM + \Gamma_{\rm inv}) = 1 - B_{\rm inv}$, where $B_{\rm inv}$ is the branching fraction into invisible decays. Recent fits give an upper bound $B_{\rm inv} \lsim (0.3 - 0.64)$ at 95\% CL \cite{brinv1, brinv2, brinv3, brinv4}, where the range depends on precisely the priors used when performing the fit. (Ref.~\cite{brinv2} finds $B_{\rm inv} < 0.64$ using 15 signals from the Tevatron, Atlas and CMS, while
ref.~\cite{brinv1} finds the slightly stronger limit of  $B_{\rm inv} < 0.4$ at 95\% CL using a different suite of 16 observables. The bound from ref.~\cite{brinv4} is similar. Ref.~\cite{brinv3} finds the strongest bound, $B_{\rm inv} < 0.2$ using a smaller set of Higgs signals that are argued to be more sensitive.)

Taking the most conservative of these limits, we infer the constraint
\be
 \cR \gsim 0.36 \,.
\ee
The corresponding constraint in the $\bar g - \bar \lambda_2$ plane is plotted in Fig.~\ref{Fig:lineshapebounds}, which also shows the result obtained from the more aggressive constraints. This plot shows that these global Higgs fits imply a conservative limit
\be
 \frac{\bar g}{\sqrt\alpha} \lsim 0.007 \quad \hbox{for $\bar \lambda_2 = 0$} \,;
\ee
with weaker constraints on $\bar g$ as $\bar \lambda_2$ increases. (This weakening of the $\bar g$ constraint with large $\bar\lambda_2$ is a general consequence of the decoupling of brane and bulk in the $\bar\lambda_2 \to \infty$ limit.) For comparison, Ref.~\cite{6DHiggsph} considered the phenomenological implications of the cubic vertex $\frac{1}{2} g h^2 \phi(0)$ but neglected Higgs-bulk mixing and $\lambda_2$. They found that $g = 0.18$ was accessible at the LHC with 100 fb${}^{-1}$ at 14 TeV in the $h \, \phi \to \gamma \, \gamma \, \phi$ channel.

Notice that $\Gamma_\ssB$ depends on the two variables $\bar g$ and $\bar \lambda_2$ only through the combination $\zeta(m_h^2)$, and this is generally true (once $m_h$ is fixed) for any observables for which the narrow-width approximation is justified. Whenever this is true it is more useful to quote the constraint directly on $\zeta(m_h^2)$, giving
\be
 \zeta(m_h^2) \lsim 5 \times 10^{-5}  \,.
\ee

\subsubsection{Low-energy bounds}

The effects of other virtual contributions of KK modes can be similarly computed given an expression for $\langle hh^* \rangle$. Important among these are constraints from low-energy precision measurements, such as the anomalous magnetic moment of the muon. By way of example we study this here for the large-volume limit where $\langle hh^* \rangle$ has a simple closed-form expression.

This section uses this calculation to conclude that these bounds are negligibly weak, in agreement with standard intuition for the large-volume case. This standard intuition starts from the observation that each KK mode couples with gravitational strength, and so it is only the enormous phase space for KK modes that can compensate for this suppression \cite{ADD}. Bounds on extra dimensions from low-energy observables are usually weaker than those from colliders because, for low-energy processes, only sub-TeV modes contribute, making the phase-space compensation incomplete and so insufficient to produce an observable result. Astrophysical energy-loss bounds on extra dimensions are an important exception to this intuition \cite{raffeltBook, HR, SNothers, SLEDph}, and we return to these below.

\begin{figure}[t]
\begin{center}
\epsfig{file=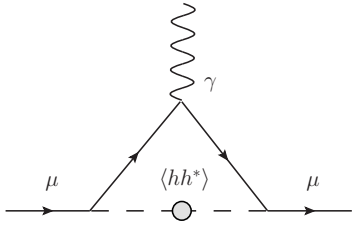,width=2.0in} \caption{The Feynman graph corresponding to the one-loop correction to the muon anomalous magnetic moment in the Higgs-bulk mixing scenario.}
   \label{Fig:muong2}
\end{center}
\end{figure}

Fig.~\ref{Fig:muong2} displays the Feynman graphs whose evaluation gives the KK mode contribution to a fermion's anomalous magnetic moment. We estimate that the difference between this graph and the corresponding graph for a Standard Model Higgs is quite small, as it should be if we consider perturbing in small $g$, and brane-bulk mixing effects are negligible.

To see this, consider first the Higgs contribution from the analogous graph in the Standard Model. This evaluates to an anomalous fermion magnetic moment of size
\be
 a_h   = \frac{y_f^2 m_f^2}{8 \pi^2 m_h^2} \int \limits_0^1  \d x \; \frac{( x-2) x^2}{  (1 - x) + x^2 (m_f^2/m_h^2) } \,,
\ee
where $m_f$ is the fermion's mass and $y_f := m_f/v$ is its Higgs Yukawa coupling. The regime of practical interest is $m_f \ll m_h$ and, because the integral diverges logarithmically near $x = 1$ when $m_f = 0$, the answer in this limit is dominated by
\bea
 a_h &\approx& \frac{y_f^2 m_f^2}{8 \pi^2 m_h^2} \left[ \log\left( \frac{m_f^2}{m_h^2} \right) + \frac{7}{6} + \cdots \right] \\ &\approx& -0.0021 \times 10^{-11} \quad \hbox{(muon)}\,, \nn
\eea
where the numerical values assume the fermion is the muon. This is negligible in comparison to both the electroweak boson contributions and the experimental precision, which are of order a few $100 \times 10^{-11}$ \cite{muong2}.

Repeating this exercise for the graph in Fig.~\ref{Fig:muong2} using the large-volume expression for $\la h h^*\ra_k$ allows us to estimate the difference between the Standard Model Higgs contribution $a_h$ and the contribution from the whole KK tower $a_\ssB$. We do not use $\la h h^*\ra^\prime_k$ with the inclusion of the Standard Model Higgs width because it is a higher-loop effect, and for simplicity we assume $\lambda_2 = 0$. Therefore, accounting for the new propagator in the graph is accomplished by making the replacement
\be
\frac{1}{k^2 + 2 \lambda v^2} \to \frac{1}{k^2 + 2 \bar{\lambda}(\bar r) v^2  + \left( \frac{\bar g^2 v^2}{4 \pi \alpha}\right) \log(k^2 \bar{r}^2) } \approx  \frac{1}{k^2 + 2 \bar{\lambda} (\bar r) v^2} + \left( \frac{\bar g^2 v^2}{4 \pi \alpha}\right) \frac{\log(k^2 \bar{r}^2)}{\left[ k^2 + 2 \bar{\lambda} (\bar r) v^2\right]^2} \, ,
\ee
where we write $\bar{g}(\bar r) = \bar{g}$ because it does not run in the limit of vanishing $\lambda_2$.

Although the above propagator is only one term in the graph, which also contains a loop integral over $k^2$, in order of magnitude, we estimate that the relative difference between the anomalous moment in the Higgs-bulk scenario and in the Standard Model is
\be
\left| \frac{a_\ssB - a_h}{a_h} \right| \approx \frac{ \bar{g}^2 }{ 8 \pi \alpha  \bar{\lambda} } \,,
\ee
and so can be ignored given the strong constraints already found for $\bar g$.

\subsection{Invisible final states}

This section considers the implications of Higgs-bulk mixing for observables with Higgs-bulk final states, rather than simply as intermediate states. We first argue that these states are invisible and then relate their production rates in various channels to the analogous Standard Model Higgs production rate. Invisible states appear as missing energy at high-energy colliders, and we discuss Higgs-bulk mixing signals at both LEP and the LHC. We then turn to low-energy constraints on missing energy in astrophysical systems. We conclude that LEP provides weak bounds on the Higgs-bulk couplings, but astrophysical bounds are considerably stronger. We also provide a preliminary estimate of the reach at the LHC, beyond the constraints already discussed coming from the suppression of the rates for producing Standard Model particles.

\subsubsection{Invisible-state production rates}
\label{sec:invfinal}

\begin{figure}[t]
\begin{center}
\epsfig{file=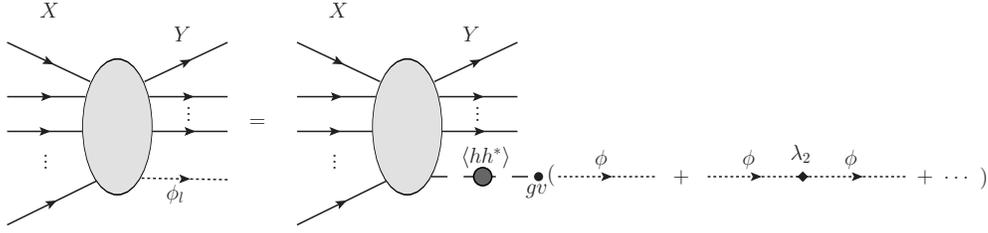,width=5.5in} \caption{The tree level contribution to vector boson fusion with an invisible final state.}
\label{vbf0}
\end{center}
\end{figure}

Consider a process whereby the tower of $\varphi_s$ states is produced through their overlap with $h$. For weak couplings the physics of this process resembles the physics of sterile neutrinos, in that production and detection of the produced state occurs only because of the mixing with a weakly-interacting Standard Model particle. For weak couplings invisible processes are therefore described by requiring the amplitude for producing a state orthogonal to $h$; that is, for being in one of the `flavour' eigenstates, $\phi_i$ rather than a propagation eigenstate, $\varphi_s$.

We therefore consider the squared amplitude to produce a $\phi_i$ final state
\bea \label{eq:metgeneral}
  |\cM(X \to Y \, \phi_i)|^2  \approx  | \cM_\SM(X \to Y \, h) |^2_{p_\phi} \, | \cM(h \to \phi_i) |^2_{p_\phi} \, ,
\eea
where the first factor is the Standard Model result for Higgs production, but evaluated using $p_\phi$, the momentum of the final state $\phi_i$, which in practice amounts to replacing $m_h \to M_i$. The second factor can be understood in two ways. Formally, $\cM(h \to \phi_i)$ is given in terms of $\langle h \, \phi^* \rangle_k$ by amputating the final unperturbed $\phi_i$ propagator and putting the correlation on shell (as usual) to obtain an amplitude with $\phi_i$ in the final state:
\be \label{hphiamp0}
 \cM(h \to \phi_i) = {\rm Amp} \, [\langle h \, \phi^*(0)  \rangle_{k=p_\phi}] := \langle h \, \phi^*(0)  \rangle_{k=p_\phi} \; [ D_{k=p_\phi}^\phi ]^{-1} \,.
\ee
As shown in Appendix~\ref{appsec:SchwDys}, the amputated $\langle h \, \phi^* \rangle_k$ correlation is given by
\be \label{hphiamp}
 {\rm Amp} \, [\langle h \, \phi^*(0)  \rangle_{k=p_\phi} ]= - \left. \frac{i \bar{g}(\bar{r}) v \, }{1 + i (\bar{\lambda}_2 (\bar r) / 4 \alpha) } \right|_{\bar r = 1/M_i} \la h \, h^* \ra_{k=p_\phi} \,.
\ee
where the renormalization point $\bar{r} = 1/M_i$ is chosen for notational convenience. Alternatively, the factorization in eq.~\pref{eq:metgeneral} and the expression in eq.~\pref{hphiamp} can be derived by directly summing the graphs of Fig.~\ref{vbf0}. However, the virtue of using a Schwinger-Dyson equation over the graphical methods is that it emphasizes that the result need not assume $\lambda_2$ is small.

Using this in eq.~\pref{eq:metgeneral} then gives the general result
\be \label{eq:metgeneralII}
 |\cM(X \to Y \, \phi_i)|^2 \approx 4 \alpha v^2 \zeta(M_i^2)   \, | \langle h \, h^* \rangle_{k=p_\phi} |^2 \, |\cM_\SM(X \to Y \, h)|^2_{p_\phi} \,,
\ee
where $\langle h \, h^* \rangle_{k=p_\phi}$ is given explicitly in terms of the unperturbed $h$ and $\phi$ propagators in earlier sections ({\em c.f.} eq.~\pref{resum}).

In practical applications, it is the sum over modes' squared amplitudes (possibly weighting some other function $O$) that is relevant to a physical observable. That is,
\be
 \cO = \sum_i |\cM(X \to Y \, \phi_i)|^2 O(M_i^2) \, .
\ee
This sum over bulk states, and the expression for $\langle h \, h^* \rangle_{k=p_\phi}$, both simplify considerably in the large-volume limit. In this limit we can also account for the on-resonance effects of decays into Standard Model particles, by using the large-volume expression for $\la h h^*\ra^\prime_{k=p_\phi}$ in place of $\la h \, h^* \ra_{k=p_\phi}$. Then, the observable can be written in the simple resonant form
\be
 \cO \approx \int \d M^2 \, \Upsilon(M^2) \, O(M^2) \,|\cM_\SM(X \to Y \, h)|^2_{p_\phi}\,,
\ee
where
\be \label{eq:upsilon}
\Upsilon(M^2) := \frac{v^2 \zeta(M^2) /\pi}{[M^2 - \Pi(M^2) ]^2 + [v^2 \zeta(M^2) + m_h \Gamma_\SM]^2} \,.
\ee
In taking this continuum limit we use the fact that each mode has only a gravitational-strength coupling to the brane, due to the proportionality of each normalized mode to $\cV_2^{-1/2}$, where $\cV_2$ is the extra-dimensional volume. Consequently, it is only the enormous phase space available at high energies that can compensate for the extremely feeble coupling of each mode, implying that it is only the density of states of the high-energy modes that is important. But the density of these modes does {\em not} depend on the details of the shape of the extra dimensions (unlike the density of states for the lowest-lying modes), which allows us to use the flat-space result appropriate to an extra-dimensional torus,
\be
  \frac{1}{\cV_2} \sum_i I(M_i^2) \approx \int \frac{\d^2p}{(2\pi \alpha)^2} \; I(p^2) = \int \frac{\d M^2}{4\pi \alpha} \; I(M^2) \,,
\ee
even for applications to more complicated geometries like spheres and rugby balls.

\subsubsection{Missing energy at LEP}
\label{sec:lep}

If the Higgs mixes significantly with invisible light states then these should have been produced at LEP, leading to a constraint on Higgs-bulk mixing. A convenient way to obtain this constraint is to use a particular search performed by the ALEPH, DELPHI, L3 and OPAL experiments at LEP II. These experiments have sought evidence for $Z$ boson production in association with a Higgs that decays 100\% invisibly while being produced with a Standard Model rate \cite{Searches:2001ab}. A combined analysis of each experiment's $\sqrt{s} = (200 - 209)$ GeV data has been used to place a lower bound on the mass of such a Higgs as $m_h \gsim 114.4$ GeV at 95\% CL.

This is a convenient search for the present purposes for two reasons. First, the resulting bounds usefully constrain the cross section at these energies for generic electron positron annihilation into a $Z$ boson plus missing energy, since it gives the same signal; that is
\be \label{expLEPvsHZ}
 \sigma_{\rm exp}(e^+ \, e^-  \to Z \, \met) < \sigma_\SM(e^+ \, e^-  \to Z \, h) \Bigr|_{m_h = 114 \, {\rm GeV}} \,,
\ee
where both sides are evaluated at $\sqrt s = 209$ GeV. This is true so long as the selection efficiencies are similar for the new process and the Higgs process. Second, as we expect from the discussion in \S\ref{sec:invfinal}, the invisible cross section predicted from Higgs-bulk mixing shares enough of the features of the Standard Model cross section to allow a simple inference of the constraints.

Any of the $\phi_i$ states produced at LEP would appear as missing energy, so that the total missing energy cross section is the sum of the individual cross sections
\be
 \sigma(e^+ \, e^-  \to Z \, \met) = \sum_i \sigma(e^+ \, e^-  \to Z \, \phi_i) \, .
\ee
The total missing energy cross section is therefore a weighted sum of squared $\phi_i$ production amplitudes. Using the results of \S\ref{sec:invfinal}, we can write the cross section for the missing energy process in the form of an integral
\be
 \sigma(e^+ e^- \to Z \, \met)  = \int\limits_0^{M^2_{\rm max}} \d M^2 \frac{\Upsilon(M^2)}{\cF } \int \frac{ \d^3 {\bf p}_\ssZ} {(2 \pi )^3 2 E_\ssZ}   \frac{ \d^3 {\bf p}_\phi} {(2 \pi )^3 2 E_\phi} |{\cM}_{\SM}|^2 _{p_\phi}(2 \pi)^4 \delta^4(p_\ssX - p_\ssZ + p_\phi) \, ,
\ee
where the upper bound on integration $M_{\rm max} = \sqrt{s} - m_\ssZ$ with $m_\ssZ$ as the mass of the $Z$ boson. This reflects the fact that only these modes are kinematically accessible. Additionally, $|\cM_{\SM}|_{p_\phi}^2 = |\cM_{\SM}(e^+ e^- \to Zh )|^2$ is the Standard Model Higgsstrahlung amplitude appropriately spin-summed/averaged, with the subscript, $p_\phi$, reminding us that it is evaluated using the $\phi_i$ final-state four-momentum $p_\phi = (E_\phi, {\bf p}_\phi )$ where $E_{\phi}^2 =  {\bf p}_{\phi}^2 + M^2$. Here and in the following, $\cF$ is the usual initial-state-dependent flux factor associated with a cross section, $p_\ssX$ is the four-momentum of this intial state and $p_\ssZ = (E_\ssZ, {\bf p}_\ssZ)$ is the four-momentum of the $Z$ boson.

\begin{figure}[t]
\begin{center}
\epsfig{file=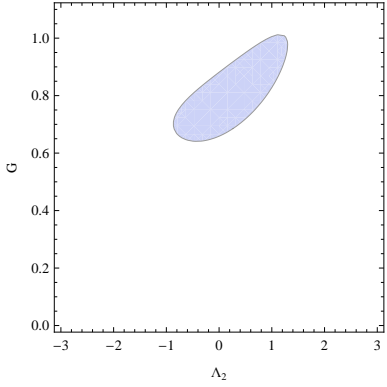,width=2.0in} \caption{Constraints from LEP expressed in the $\Lambda_2$--$G$ plane (with $G = \bar g/\sqrt{\alpha}$ and $\Lambda = \bar \lambda_2/\alpha$) from LEP constraints. The small shaded region is excluded. All couplings are evaluated at a scale $\bar r = (125 \text{ GeV})^{-1}$}
   \label{Fig:lepbounds}
\end{center}
\end{figure}

The Standard Model Higgsstrahlung cross section can be written
\be
 \sigma_\SM(e^+ e^- \to Z \, h) =   \frac{1}{\cF} \int \frac{ \d^3 {\bf p}_\ssZ} {(2 \pi )^3 2 E_\ssZ}   \frac{ \d^3 {\bf p}_h} {(2 \pi )^3 2 E_h} |{\cM}_{\SM}|^2_{p_h} (2 \pi)^4 \delta^4(p_\ssX - p_\ssZ + p_h) \,,
\ee
with $E_h^2 = {\bf p}_h^2 + m_f^2$ where $p_h = (E_h, {\bf p}_h)$, and $|\cM_{\SM}|^2$ is evaluated using the Higgs momentum $p_h$. We recognize the same expression in the missing energy cross section with the replacement $m_h \to M$. This allows the missing energy cross section to be rewritten as follows
\be \label{eq:metCS}
  \sigma(e^+ \, e^- \to Z \, \met)  = \int \limits_0^{M_{\rm max}^2}\d M^2 \, \Upsilon(M^2) \, \sigma_{\SM}(e^+ \, e^- \to Z \, h) \Bigr|_{m_h = M} \, .
\ee
This expresses our missing-energy prediction in terms of the well-known Standard Model Higgsstrahlung cross section $\sigma_{\SM}$, whose $m_h$- and $\sqrt s$-dependence is given by a standard result \cite{higgsphoton, higgsreviews},
\be
 \sigma_\SM(e^+ \, e^- \to Z \, h) \propto \sqrt{J(m_h^2)} \; \left[ J(m_h^2) + 12 M_\ssZ^2/s \right] \,,
\ee
where $J(m^2) := (1 - m^2/s - M_\ssZ^2/s)^2 - 4 m^2 M_\ssZ^2 / s^2$ is the two-body phase-space function. Using this, we may write the bound
\be
 \sigma(e^+ e^-  \to Z \slsh E_\ssT) < \sigma_{\rm exp}(e^+ e^-  \to Z \slsh E_\ssT) < \sigma_\SM(e^+ e^-  \to Z H) \Bigr|_{m_h = 114 \, {\rm GeV}}
\ee
in the form
\be
  \int \limits_0^{M_{\rm max}^2}\d M^2  \,  \Upsilon(M^2) \sqrt{J(M^2)} \left[ J(M^2) + 12 M_\ssZ^2/s \right] < \sqrt{J(m_h^2)} \left[ J(m_h^2) + 12 M_\ssZ^2/s \right] \Bigr|_{m_h = 114\, {\rm GeV}} \, ,
\ee
where all other pre-factors in the cross-section formula cancel. The integral can be evaluated numerically to determine the allowed region of parameter space, with the result plotted in the $\bar g - \bar \lambda_2$ plane in Fig.~\ref{Fig:lepbounds}. The result excludes an island in parameter space, and at first sight it might appear surprising that the bounds get weaker at large $\bar g$ as well as at small $\bar g$. The weakening for large $\bar g$ is a consequence of the resonant shape of $\Upsilon(M^2)$ together with the proximity of $\sqrt{s} - m_\ssZ = 118$ GeV to the resonance's maximum, because $\Upsilon \propto 1/(\zeta v^2)$ when $\zeta v^2$ is bigger than both $M^2 - \Pi(M^2)$ and $m_h \Gamma_\SM$. (A similar thing happens for bounds obtained for other observables dominated by the resonance, though the weakening of the constraints at large $\bar g$ for these occurs for $\bar g$ too large to justify our approximations, and so is not shown in the plots.)

The bounds obtained are clearly weaker than the LHC bounds on the Higgs invisible width discussed above (and the astrophysical bounds discussed below). So, although the issue of selection efficiency was ignored in arriving at these constraints, their weakness illustrates that a very large discepancy in the two signals' selection efficiencies, that furthermore favours Higgs-bulk events, would have to exist for LEP bounds to become significant.

\subsubsection{Missing energy at the LHC}
\label{sec:lhc}

The dominant production mechanisms for a $125$ GeV Higgs at the LHC are gluon fusion $gg \to h$ through a top-quark loop and vector-boson fusion  $qq \to q qh$. Additionally, the Higgsstrahlung process $q \bar{q} \to Vh$ provides a clean signal at the cost of a reduced cross section with respect to the other production channels.

These processes give rise to three different missing-energy signals for an invisible Higgs. At non-leading order, additional jets can radiate from coloured particle lines in gluon fusion, resulting in $j \, \met$ and $jj \, \met$ final states, the second of which is also attainable at leading order through vector boson fusion and Higgsstrahlung with a $Z$ decaying hadronically. Alternatively, a Higgsstrahlung $Z$ can decay leptonically, giving a signal $\ell^+ \ell^- \, \met$.

Missing energy rates from Higgs-bulk mixing at the LHC are also determined from the rate to produce $\phi_i$ final states. The arguments in \S\ref{sec:lep} up to eq.~\pref{eq:metCS} can be repeated for any of Higgs production processes at LHC, at least at parton-level, so that the missing energy, parton level cross section, $\hat \sigma$, is
\be \label{eq:lhcPartCS}
   \hat\sigma(X \to Y \, \met)  = \int \limits_0^{M < \sqrt{\hat s} - m_\ssY}\d M^2 \, \Upsilon(M^2)\, \hat\sigma_{\SM}(X  \to Y \, h) \Bigr|_{m_h = M} \,,
\ee
where $\hat{s}$ is the parton-level center of mass energy squared, and $m_\ssY$ is a placeholder for the region of $M$ phase space denied to $\phi_i$ by the presence of the other final state particles. For example, if the final state particles associated with $\phi_i$ are all massless, $m_\ssY = 0$, and for Higsstrahlung $m_\ssY = m_\ssZ$ as in \S\ref{sec:lep}.

An important difference between the LHC rate and the rate at LEP is that the LHC is sufficiently energetic to probe the peak of $\Upsilon(M^2)$ that lies at $m_h$. This allows us to employ the narrow width approximation for $m_h \gg \Gamma_\ssB + \Gamma_{\SM}$, in which
\be
 \Upsilon(M^2) \approx \left[ \frac{\Gamma_\ssB}{\Gamma_{\ssB} + \Gamma_{\SM}} \right] \delta(M^2 - m_h^2) \, ,
\ee
where we have used the fact that $v^2 \zeta(m_h) = m_h \Gamma_\ssB$. This simplifies the parton-level missing energy cross section
\be
 \hat\sigma(X \to Y \, \met) = \left[ \frac{\Gamma_\ssB}{\Gamma_\ssB + \Gamma_\SM}\right] \hat\sigma_{\SM}(X  \to Y \, h) \, .
\ee

One might worry that there are collisions where the initial state partons have small momentum fractions $x_{1,2}$, resulting in a small center of mass energy $\sqrt{\hat{s}} = \sqrt{x_1 x_2 s}$. If $\sqrt{\hat{s}}$ is too small, then the integral over $M^2$ will not saturate the delta function $\delta(M^2 - m_h^2)$. However, the resulting step function $\Theta(m_h + m_\ssY - \sqrt{\hat s})$ also appears in the Standard Model cross section where it encodes the threshold above which the Standard Model Higgs can be produced, and so we refrain from rewriting it. This calculation reveals that, in the narrow width limit, we expect the parton-level rate of producing an invisible Higgs to be suppressed relative to the Standard Model cross section by a factor of the invisible branching ratio $B = \Gamma_\ssB / (\Gamma_\ssB + \Gamma_\SM )$, as in other models with additional invisible Higgs decay channels.

The trivial mass and energy dependence of this factor allow us to convolve the parton distribution functions without complications. Furthermore, since it is only the modes near $M \approx m_h$ that contribute to the cross section in the narrow-width approximation, kinematic cuts will apply equally to single Higgs cross section and the Higgs-bulk cross section. Therefore, for a given production mechanism, the total missing energy proton-proton cross section at the LHC can be written
\be
 \sigma(pp \to Y \, \met) = B \times \sigma_{\SM}(pp \to Y \, h) \,,
\ee
where $B = \Gamma_\ssB/(\Gamma_\ssB + \Gamma_\SM)$ as before, and  $\sigma_{\SM}(pp \to Y \, h)$ is the Standard Model production cross section at the LHC.

New-physics constraints and discovery estimates for a Higgs with invisible decay channels, specific to a given production channel are cast in terms of the quantity
\be
 \xi^2 := \left( \frac{\sigma_{\ssB \ssS \ssM}}{\sigma_{\ssS \ssM}} \right) B_{\text{inv}}\,,
\ee
where $\sigma_{\ssB \ssS \ssM}$ is the production cross section of the proposed invisible Higgs, $\sigma_\SM$ is the Standard Model Higgs production cross section, and $B_{\text{inv}}$ is the invisible branching ratio predicted by new physics. Although these constraints envision a single Higgs state with new decay (and possibly production) modes, they apply equally well to Higgs-bulk production in the narrow width approximation, where we have $\sigma_{\ssB \ssS \ssM} = \sigma_{\SM}$ and $B_{\rm inv} = B$ and so $\xi^2 = B$. The constraints and reach of LHC invisible-Higgs searches are readily translated into constraints on Higgs-bulk mixing (as in \S\ref{sssec:invWidth}.)

Using the ATLAS monojet search at $\sqrt{s} = 7$ TeV with 1 fb${}^{-1}$ \cite{atlasMono}, the authors of Ref.~\cite{evasiveMono} were able to bound $\xi^2 \lsim 20$ for a 125 GeV Higgs-like particle. The updated CMS monjet data \cite{cmsMono} with 4.7 fb${}^{-1}$ at 7 TeV was subsequently used \cite{djouadiMono} to tighten this constraint to $\xi^2 < 1.3$ at 95\% C.L. and it was argued that the bound could be increased to $\xi^2 \lsim 0.9$ with 15 fb${}^{-1}$ at 8 TeV. It is only once $\xi^2 < 1$ becomes possible that measurements become sensitive to Higgs-bulk mixing. However, part of this region is already accessible (and ruled out) by constraints from the LHC global fits discussed earlier, since they require $B_{\text{inv}} \lsim 0.64$.

Nonetheless, higher integrated luminosity, and different channels, will allow Higgs-bulk mixing to be further probed. Ref.~\cite{baiMono} estimates that the LHC with 20 fb${}^{-1}$ at 7 and 8 TeV can exclude invisible rates down to $\xi^2 \approx 0.4$ at 95\% C.L. via the dijets plus missing energy signal. Finally, it was estimated that $\xi^2 \approx 0.25$ would be probed at $5 \sigma$ by the LHC with 300 fb${}^{-1}$ at 14 TeV \cite{ghoshInvis}, also via dijets plus missing energy. Both of these studies also considered the LHC sensitivity to Higgsstrahlung process, where the Higgs is invisible and the Z decays leptonically, but found it to be a weaker probe of $\xi^2$. Monojet searches were also concluded to be a weaker probe than vector boson fusion in ref.~\cite{baiMono}.

\subsubsection{Astrophysical constraints}

The new, invisible $\phi_i$ states can carry energy away from stars and supernovae. If the new emission process were too efficient, then it would conflict with the current understanding of stellar evolution. On this basis, it has been argued very generally \cite{raffeltBook} that any new energy loss channels in the sun must not exceed the solar energy loss rate $\dot{\cE}_{\text{sun}} \approx 2 \text{ erg} \, \text{s}^{-1} \text{g}^{-1} $ and new channels in red giants and horizontal-branch stars cannot exceed $\dot{\cE}_{\ssR \ssG} \approx 10 \text{ erg} \, \text{s}^{-1} \text{g}^{-1}$. Additionally, neutrino observations from SN1987a suggest that a new channel must not release energy at a rate exceeding the neutrino rate $\dot{\cE}_{\ssS \ssN} \approx 10^{19} \text{ erg} \, \text{s}^{-1} \text{g}^{-1} $ during core collapse.

In this section, we consider a variety of $\phi$ emission processes in these settings and determine the strength of the corresponding bounds on Higgs-bulk mixing. Since the goal of this section is to estimate constraints, we make various simplifying assumptions and neglect the effects of dense media, interference effects from multiple scattering \cite{raffeltBook, multScatter} and the possibilities of trapping the KK modes inside of an astrophysical object, or of their decaying before exiting the astrophysical medium.

\subsubsection*{Electron-positron annihilation in supernovae}

The temperature in the core of SN1987a $T_{\ssS \ssN} = (20 - 60)$ MeV was high enough to produce electron-positron pairs. These pairs could have subsequently annihilated into an $h$ state that then mixed over to a $\phi_i$ mode that carried energy away from the core. We calculate the associated energy loss rate for this simple process as an example to guide our discussion of astrophysics constraints on Higgs-bulk mixing.

The energy loss rate for this process is given by the sum of the individual $\phi_i$ emission rates
\be \label{eq:epsSum}
 \dot{\cE} = \sum_i \dot{\cE}_i \, .
\ee
where the emission rate to a single state $\phi_i$ is given by
\be \label{eq:emsRate}
 \dot \cE_i = \frac{1}{4 \rho} \int \frac{ \d^3 {\bf p}_1} {(2 \pi )^3 2 E_1}  \frac{ \d^3 {\bf p}_2} {(2 \pi )^3 2 E_2} \frac{ \d^3 {\bf p}_\phi} {(2 \pi )^3 2 E_\phi}  (f_1 f_2 E_\phi) |\cM|^2_{p_\phi} (2\pi)^4 \delta^4(p_1 + p_2 - p_\phi) \, ,
\ee
In the above equation $\rho = 3 \times 10^{14} \text{g cm}^{-3} \approx \Lambda_{\ssQ \ssC \ssD}^4$ is the density of the supernova core, $|\cM|^2= | \cM (e^+ e^- \to \phi_i) |^2$ is the spin-summed amplitude for the annihilation process, $p_{(1,2)}$ is the momentum of the initial state electron (positron) with energy $E_{(1,2)}$ and ${\bf p}_\phi$ is the momentum of the outgoing KK mode with energy $E_\phi$. The occupation numbers $f_{(1,2)}$ are given by the Fermi-Dirac distribution for a relativistic electron (positron)
\be
 f_{(1,2)} = \frac{1}{e^{\left(E_{(1,2)} \mp \mu_e \right)/T} + 1} \, ,
\ee
where $\mu_e \approx 345 \text{ MeV}$ is the chemical potential for the electron in the supernova core. Because the final state KK mode is assumed to escape, there is also no need for a Bose-Einstein final-state factor for it in the energy-loss rate.
Eqs.~\pref{eq:epsSum} and \pref{eq:emsRate} show that the total energy loss rate is a weighted sum of squared $\phi_i$ production amplitudes, so the results of \S\ref{sec:invfinal} allow us to write the total energy loss rate in the form of an integral over $M^2$
\be \label{eq:fullEps}
 \dot \cE = \int\limits_0^\infty \d M^2 \frac{\Upsilon(M^2)}{4 \rho}\int \frac{ \d^3 {\bf p}_1} {(2 \pi )^3 2 E_1}  \frac{ \d^3 {\bf p}_2} {(2 \pi )^3 2 E_2} \frac{ \d^3 {\bf p}_\phi} {(2 \pi )^3 2 E_\phi}  (f_1 f_2 E_\phi) |\cM_{\SM}|^2_{p_\phi} (2\pi)^4 \delta^4(p_1 + p_2 - p_\phi) \, ,
\ee
where $|\cM_{\SM}|^2 = |\cM_{\SM}(e^+ e^- \to h)|^2$ is the spin-summed Standard Model Higgs amplitude squared.

As in \S\ref{sec:lep} and \S\ref{sec:lhc}, we could proceed by noting that the energy-loss rate can be written
\be \label{eq:epsForm}
 \dot \cE = \int\limits_0^\infty  \d M^2 \, \Upsilon(M^2) \dot \cE_{\SM}(e^+ e^- \to h) \Bigr|_{m_h = M} \, ,
\ee
where $\dot \cE_{\SM}$ is the Standard Model Higgs emission rate, and it should be evaluated for a Higgs with mass $m_h = M$. However, since $\dot \cE_{\SM}$ is unknown, it is more straightforward to determine the total energy-loss rate via eq.~\pref{eq:fullEps}, a task to which we now turn.

The spin-summed, squared, Standard Model amplitude is
\be
 |\cM_{\SM}(e^+ e^- \to h)|^2 = 4 (y_e^h)^2 E_1 E_2 (1 - \cos \theta_{12}) \, ,
\ee
where $y_e^h $ is the Standard Model Higgs-electron Yukawa coupling, $\theta_{12}$ is the angle between the momentum vectors of the electron and positron, and the mass of the electron is neglected because $T_{\ssS \ssN} \gg m_e$. Using standard techniques to integrate phase space, we find
\be
 \dot \cE = \frac{ (y_e^h)^2 }{64 \pi^3 \rho} \int\limits_0^\infty  \d E_1 \, \d E_2 \, \d M^2 \, \Upsilon(M^2) \,  M^2 \, f_1 f_2 (E_1 + E_2) \, \Theta(4E_1 E_2 - M^2) \, ,
\ee
where $\Theta$ is a step function encoding the threshold above which a $\phi_i$ mode with mass $M$ can be produced. This integral must be integrated numerically using the expression for $\Upsilon(M^2)$ in eq.~\pref{eq:upsilon}.

The low-energy form of $\Upsilon(M^2)$ simplifies greatly if we specialize to $\lambda_2 = 0$, and consider the $\bar g \ll 1$ limit justified by existing constraints. First, eq.~\pref{eq:zeta} gives $\zeta(M^2) = \bar{g}^2 / 4 \alpha$ as constant. Additionally, assuming $\bar g \ll 1$ allow us to approximate eq.~\pref{eq:mstar} as $\Pi(M^2) = m_h^2$. Finally, since $T_{\ssS \ssN}^2 \ll m_h^2$  we find that in these three limits $\Upsilon(M^2)$ is a constant
\be \label{eq:simpleUps}
 \Upsilon(M^2) \approx \frac{\bar{g}^2 v^2}{4 \pi \alpha m_h^4}\, ,
\ee
where this approximation is better than 1\% for $M < 1 {\text{ GeV}}$ and $\bar{g}/\alpha < 0.007$. This simplifies the total energy-loss rate to
\be \label{eq:annihilationrate}
 \dot \cE \approx \frac{(y_e^h)^2 T^5}{8 \pi^3 \rho} \left( \frac{\bar{g}^2 v^2 T^2 }{4 \pi \alpha  m_h^4 } \right) \int\limits_0^\infty \d x \, \d y  \, \frac{ (x + y) (xy)^2 }{(e^{x - \mu_e/T}+ 1)(e^{y + \mu_e/T}+ 1)}\, ,
\ee
where $x = E_1/T$ and $y=E_2/T$. This gives
\be
 \dot \cE \approx \left( \frac{\bar{g}^2}{\alpha} \right)  \left( 4.4 \times 10^{17} \text{ erg} \text{ s}^{-1} \text{g}^{-1} \right) T_{20}^7 \, \cI(\mu_e / T) \, ,
\ee
where $T_{20} = T/(20 \text{ MeV})$ and $\cI$ is the integral factor in (\ref{eq:annihilationrate}). It is exponentially sensitive to $\mu_e/T$ and ranges from $2\times10^{-3}$ to $4.4$ as temperature ranges from 20 to 60 MeV. However, even the most severe bound on Higgs-bulk mixing from electron-positron annihilation in SN1987a, which is found by assuming $T_{\ssS \ssN} = 60$ MeV, only constrains $\bar{g}/\sqrt{\alpha} \lsim 0.05$, which is already ruled out (for $\bar\lambda_2=0$) by the LHC global-fit constraints in Fig.~\ref{Fig:lineshapebounds}.

\subsubsection*{Stellar processes}

Rather than calculating additional subdominant constraints on Higgs-bulk mixing, we next use existing constraints on the Yukawa coupling of a single light scalar $\psi$ to estimate whether a given emission process will result in a significant bound. For example, the strongest stellar bound on the Yukawa coupling of a light scalar to nucleons, $y_\ssN^\psi < 4.3 \times 10^{-11}$, comes from the Compton process $A \, \gamma \to A \, \psi$ in red giants, where $A = \{ {}^4\text{He}, p \}$ and the temperature of red giants is $T_{\ssR \ssG} \sim 10 \text{ keV}$ \cite{raffeltBook, yukawabounds}. The bound is derived by assuming that the light scalar couples only to to nucleons, which is a good approximation to the Higgs-bulk mixing scenario since the $\phi_i$ states couple to the stellar constituents through the $h$ state, and the Standard Model Higgs-nucleon Yukawa coupling is much larger than the electron Yukawa coupling $y_\ssN^h \approx (340 \text{ MeV}/v) \gg y_e^h$ \cite{higgsnucleon}.

To estimate the energy loss rate for the Compton process $A \, \gamma \to A \, \phi$ in Higgs-bulk mixing, we write it the same form as eq.~\pref{eq:epsForm}
\be
 \dot \cE = \int\limits_0^\infty  \d M^2 \, \Upsilon(M^2) \dot \cE_{\SM}(A \, \gamma \to A \, h) \Bigr|_{m_h = M} \, ,
\ee
where $\dot\cE_{\SM}$ is the Standard Model Higgs emission rate. In order of magnitude, we assume the Standard Model Higgs emission rate is related to the emission rate of a single light scalar $\psi$ as follows
\be
\dot \cE_{\SM}(A \, \gamma \to A \, h) \Bigr|_{m_h = M} \approx \left( \frac{y_\ssN^h}{y_\ssN^\psi}\right)^2 \dot \cE_\psi(A \, \gamma \to A \, \psi) \Theta(T_{\ssR \ssG} - M) \, ,
\ee
where $\dot\cE_{\psi}$ is the emission rate of a light scalar. We include a step function, $\Theta$, to account for the fact that only Higgses of mass $M \lsim T$ will have appreciable production rates. This follows from the fact that $A$ is much heavier than the energy of the photon, so we can neglect its recoil and assume that the outgoing $\phi_i$ mode has the energy of the incoming photon $E_\phi  \approx E_\gamma$. The emission rate for heavier $\phi_i$ modes is therefore suppressed by a Boltzmann factor $e^{-E_\gamma/T} \approx e^{-M/T} \ll 1$.

Using this approximation, and taking the $\lambda_2 = 0$, small $\bar g$, low-energy limits for $\Upsilon(M^2)$ in eq.~\pref{eq:simpleUps} then gives an order of magnitude estimate for the Higgs-bulk energy loss rate in terms of the $\psi$ emission rate
\be \label{eq:boundEstimate}
 \dot \cE = \left( \frac{\bar g^2 v^2 T^2_{\ssR \ssG}}{4 \pi \alpha m_h^4} \right)  \left( \frac{y_\ssN^h}{y_\ssN^\psi}\right)^2 \dot \cE_\psi(A \, \gamma \to A \, \psi) \, .
\ee
This allows the bound on $y_\ssN^\psi$ to be translated into a bound on Higgs-bulk mixing (with $\lambda_2 = 0$)
\be
 \frac{ \bar{g}^2 v^2  T_{\ssR \ssG}^2 (y_\ssN^h)^2 }{4 \pi \alpha m_h^4 } \lsim \left( 4.3 \times 10^{-11} \right)^2 \, ,
\ee
from which it is estimated that even the tightest constraint $g/\sqrt{\alpha} \lsim 0.8$ from stellar physics is subdominant to constraints from LHC global fits in Fig~\ref{Fig:lineshapebounds}. For completeness, we note that there are similar constraints on the Yukawa coupling of a light scalar to nucleons from the Compton process in the sun, and the bremsstrahlung process $A \, e \to A \, e \, \psi$ in the sun and red giants \cite{yukawabounds}, but none of these are estimated to give improved bounds on Higgs-bulk mixing.

\subsubsection*{Photon annihilation}

Photon annihilation is a relevant process in both both stars and supernovae. However, the Standard Model Higgs couples to photons through $W$ and heavy quark loops, and so must the KK modes. For processes like photon annihilation $\gamma \gamma \to h$, in which the photons and Higgs are all on shell, the effect of these loops is well captured by the following effective Lagrangian \cite{higgsphoton, higgsreviews}
\be
 \cL = - \sum_\ell c_{\gamma} F_{\mu \nu} F^{\mu \nu} h \quad {\rm with } \quad c_\gamma \approx \frac{\alpha_{em}}{6 \pi v} \,.
\ee
We estimate that
\be
 |\cM_{\SM}(\gamma \, \gamma \to h)|^2 \approx \left( \frac{\alpha^2_{em} T}{6 \pi v \, y_e^h}\right)^2 |\cM_\SM(e^+ \, e^- \to h) |^2 \, ,
\ee
from which it follows that energy loss rate from photon annihilation is less than the rate from electron-positron annihilation, in stars and supernovae, and therefore gives a negligible bound.

\subsubsection*{Nucleon-bulk bremsstrahlung in supernovae}

In addition to the bounds from the sun and red giants, there is a similar bound on the nuleon Yukawa coupling of a light scalar from SN1987a data $y_\ssN^\psi \le 4 \times 10^{-11}$ that follows from considering the nucleon bremsstrahlung process $N \, N \to N \, N \, \psi$ \cite{snYukawa} and assuming $T_{\ssS \ssN} = 60$ MeV. This can be used to estimate the constraint coming from the energy loss process $N \, N \to N \, N \, \phi$. Although this upper limit is similar to the stellar bounds on the same coupling, it provides a much more stringent constraint on Higgs-bulk mixing because the supernova core is much hotter than red giants, thereby enhancing the Higgs-bulk energy loss rate by a factor $T^2_{\ssS \ssN}/T^2_{\ssR \ssG} \sim 10^6$. Generalizing eq.~\pref{eq:boundEstimate} to bulk-nucleon bremsstrahlung, we expected a constraint $\bar{g}/\sqrt{\alpha} \lsim 10^{-4}$ if the supernova temperature is assumed to be $T_{\ssS \ssN} = 60 \text{ MeV}$, which is dominant to all other constraints discussed so far.

In order to confirm this estimate, we calculate the energy loss rate in the one pion exchange approximation, using the following interactions
\be
 \cL = -ig_{\pi \ssN \ssN} \ol{N}\gamma_5 N \, \pi^0 - y_n^h  h \ol{N} N \, ,
\ee
where $N$ represents the neutron, $\pi^0$ the neutral pion, and $g_{\pi \ssN \ssN} \approx 13$ phenomenologically. This approximation is known to overestimate the rate of axion and KK graviton bremsstrahlung, where it can be tested against model-independent calculations \cite{omegapole}. Since model-independent methods are not applicable to scalar radiation via a Yukawa coupling \cite{saxion}, the emission rate and associated bound are just crude estimates. We are content to further simplify the calculation by neglecting bremsstrahlung from protons, because the proton fraction in the supernova core is small, and by neglecting the Higgs-pion coupling, which is a dimension-5 operator. We also neglect a suppression due to multiple scattering effects \cite{multScatter} that is expected to reduce the emission rates of scalars in nucleon bremsstrahlung by as much as a factor of 5 \cite{saxion}.

The total energy loss rate for this process is given by
\bea \nn
 \dot \cE &=& \int\limits_0^\infty   \d M^2 \frac{\Upsilon(M^2)}{4 \rho} \int \prod_{j=1,...4} \left( \frac{\d^3 {\bf p}_j} {(2 \pi )^3 2 E_j} \right) \frac{\d^3 {\bf p}_\phi} {(2 \pi )^3 2 E_\phi}  \left[ f_1 f_2 (1 - f_3) (1-f_4) E_\phi\right]  \\
 &&  \qquad\qquad\qquad\qquad\qquad \times |\cM_\SM|^2_{p_\phi} \,  (2\pi)^4 \delta^4(p_1 + p_2 - p_3 - p_4 - k) \,,
\eea
where $|\cM_{\SM}|^2 = |\cM_\SM(N \, N \to N \, N \, h)|^2$ is the spin-summed Standard Model Higgs amplitude squared, $p_j$ represent the neutron four-momenta, and everything else maintains its old definition, except that it is now appropriate to use Maxwell-Boltzman distributions for the non-relativistic neutrons with number density $n_\ssN$
\be
 f_j = \frac{n_\ssN}{2} \left( \frac{2 \pi}{m_\ssN T} \right)^{3/2} \exp\left({-\frac{{\bf p}_j^2}{2 m_\ssN T}}\right) \,,
\ee
and in the phase space measures $E_j \to m_\ssN$, where $m_\ssN$ is the nucleon mass.

\begin{figure}[t]
\begin{center}
\epsfig{file=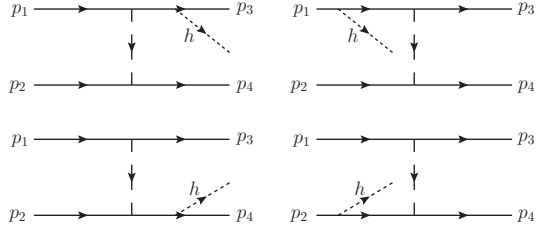,width=3.0in} \caption{The four uncrossed graphs for Higgs-nucleon bremsstrahlung.}
   \label{Fig:brems}
\end{center}
\end{figure}

We proceed by calculating the spin-summed Standard Model Higgs amplitude $|\cM_{\SM}|^2$. There are 8 relevant diagrams for this process, four of which can be found in Fig.~\ref{Fig:brems}. The other four correspond to the crossed analogues of those listed. In the nonrelativistic limit $|{\bf p}_j|\ll m_\ssN$, and we assume the relevant KK modes are much lighter than the neutron $M \ll m_\ssN$. In this limit, the squared, summed matrix element reads
\be
 |\cM_\SM|^2 = \frac{8 (y_\ssN^h)^2 g_{\pi \ssN \ssN}^4}{m_\ssN^2} \left[ \left( \frac{|\bf{k}|^2}{|{\bf k}|^2 + m_\pi^2 } \right)^2 +  \left( \frac{|\bf{l}|^2}{|{\bf l}|^2 + m_\pi^2 } \right)^2  +  \frac{ |{\bf k}|^2 |{\bf l}|^2 - 2({\bf k} \cdot {\bf l})^2}{\left( |{\bf k}|^2 + m_\pi^2 \right) \left( |{\bf l}|^2 + m_\pi^2 \right)}  \right] \,,
\ee
where ${\bf k} = {\bf p}_2 - {\bf p}_4$, ${\bf l} = {\bf p}_2 - {\bf p}_3$. The individual terms in the square brackets are $\cO(1)$ since $|{\bf k}|,|{\bf l}| \sim \sqrt{m_\ssN T} \approx m_\pi$ so that we approximate the matrix element as a constant \cite{raffeltBook}
\be
 |\cM_\SM|^2 \approx \frac{8 (y_\ssN^h)^2 g_{\pi \ssN \ssN}^4}{m_\ssN^2} \,.
\ee

\begin{figure}[t]
\begin{center}
$\begin{array}{cc}
\epsfig{file=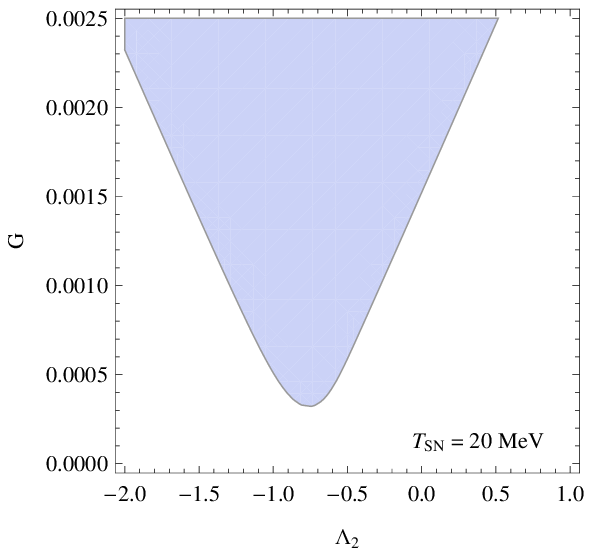,width=2.1in} \quad &
\epsfig{file=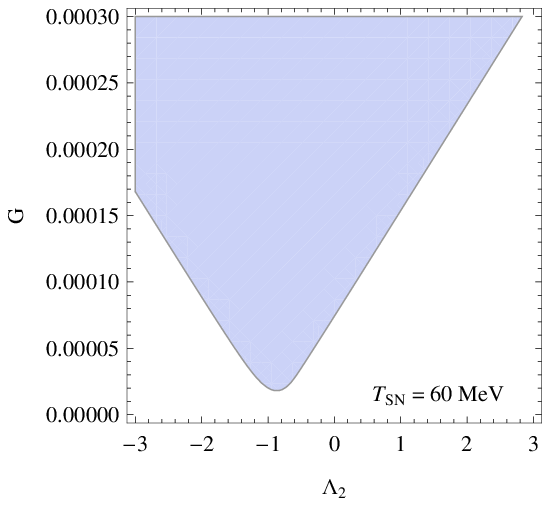,width=2.1in}
\end{array}$
\caption{Constraints on Higgs-bulk mixing from bulk-nucleon bremsstrahlung in SN1987a. The left panel shows the constraints obtained assuming $T_{\ssS \ssN} = 20$ MeV and the right panel shows the analogous constraints assuming $T_{\ssS \ssN} = 60$ MeV. Couplings are evaluated at $1/\bar{r} = m_h = 125$ GeV.}  \label{Fig:bremBounds}
\end{center}
\end{figure}

We evaluate the phase space integral in the non-degenerate limit. Although the neutrons in a supernova core are somewhat degenerate, in more detailed calculations of scalar-nucleon bremsstrahlung \cite{saxion}, the non-degenerate limit approximated the full emission rate well for moderate supernova temperatures $T_{\ssS \ssN} \gsim 20$ MeV. In this limit we neglect the blocking factors $1- f_{3,4} \approx 1$, and after integrating over phase space we find
\bea \nn
\dot{\cE} &=& \frac{ n_\ssN^2 \, \, (y_n^h)^2 \, g_{\pi \ssN \ssN}^4  \, T^{7/2}}{128 \, \pi^{7/2}  M_n^{9/2} \rho} \int\limits_0^\infty \d x  \, \d y \, \d z \, \Upsilon(zT^2) \, xe^{-y-x}   \left( y^2 + xy \right)^{1/2} \left(x^2 - z \right)^{1/2} \theta(x^2 - z)\\
&\approx& \frac{ n_\ssN^2 \, (y_n^h)^2 \, g_{\pi \ssN \ssN}^4  \, T^{7/2}}{128 \, \pi^{7/2}  M_n^{9/2} \rho} \int\limits_0^\infty   \d x  \int\limits_0^{x^2} \d z \, \Upsilon(zT^2) \, x e^{-x}   \left( 1+ x\pi/4 \right)^{1/2} \left(x^2 - z \right)^{1/2} \,,
\eea
where $x = E_\phi/T$, $z = M^2 / T^2$ and the approximation made in the second line is good to within 2.2\% \cite{raffeltBook}. This integral can be evaluated numerically, and the allowed region of parameter space from SN1987a constraints is plotted in Fig.~\ref{Fig:bremBounds}.

This somewhat crude calculation and the associated bounds verify that bulk-nucleon bremsstrahlung in SN1987a is likely the strongest constraint on Higgs-bulk mixing and, for $T_{\ssS \ssN} = 60$ MeV, gives $\bar{g}/\sqrt{\alpha} \lsim 10^{-4}$ when $\lambda_2 = 0$, as estimated. However, since we neglected many effects that might suppress energy loss rate, and SN1987a bounds from nucleon bremsstrahlung are at best order of magnitude estimates, we assume the $T_{\ssS \ssN} = 20$ MeV bounds in order to be conservative.

\subsection{Mass-coupling relations and vacuum stability}

A great virtue of the Standard Model Higgs is the tight connection between the strength of its coupling to a particle and that particle's mass. In particular, knowledge of the mass of the Higgs itself reveals the strength of its self-coupling. Although it is not yet possible to directly measure this self-coupling, it plays a role in how couplings run at higher energies and so indirectly constrains the possibilities for UV physics.

These constraints differ in the presence of Higgs-bulk mixing, because this mixing changes the relation between the Higgs mass and its self-coupling. Furthermore, brane-bulk interactions quite generally introduce a new source of running for brane-localized interactions. Both of these observations work to change the nature of the constraints on UV physics.

There are generically two kinds of UV constraints that arise for Higgs couplings: `vacuum stability' \cite{HiggsMassBound} and `triviality' \cite{Triviality}. The first of these demands that the relevant quartic coupling of the Higgs potential not run to negative values. The second demands that any Landau poles (where the running couplings diverge) not arise at too low an energy. The depth of one's worry about these bounds depends fairly strongly on the depth of one's convictions as to how far this running can be trusted before some at-present-unknown UV physics intervenes. In the present instance triviality turns out to provide a fairly strong constraint on the brane-bulk coupling $\bar\lambda_2$.

To see how Higgs-bulk mixing changes things, we briefly restate the two sources of running. For the Standard Model Higgs self-coupling, one-loop renormalization --- including the Higgs, top quark and the gauge bosons in the loop --- gives the following beta function for $\bar\lambda$ \cite{sally}
\be
 \left( \mu \, {\d \bar \lambda \over \d \mu} \right)_\SM := \beta_\SM \approx {3\over 4 \pi^2}\biggl[
 \bar\lambda^2 + \bar\lambda y_t^2 - y_t^4 - {\bar\lambda \over 8}  (3 g_2^2
 + g_1^2 ) + {1\over 64} (3 g_2^4 +2  g_2^2 g_1^2 + g_1^4) \biggr] \,,
\label{renorm}
\ee
where $y_t = m_t / v$ is the top-quark Yukawa coupling, and $g_1$ and $g_2$ are (respectively) the $U(1)_\ssY$ and $SU(2)_\ssL$ gauge couplings, with the mass of the $W$ and $Z$ bosons given by $m_\ssW^2 = g_2^2 v^2$ and $m_\ssZ^2 = \frac{1}{4} \left( g_2^2 + g_1^2 \right)v^2$ at tree-level, as usual. To these must be added the new Higgs-bulk contributions to the UV running of brane couplings, and writing $\mu \d/\d \mu = - f(\bar r) \, \d/\d \, \bar r$ these are given by
\bea \nn
 \left( \mu \, \frac{\d \bar{\mu}_\Phi^2}{\d \mu} \right)_\ssB = \frac{\bar{\lambda}_2 \bar{\mu}_\Phi^2}{2 \pi \alpha}; \qquad \left( \mu \, \frac{\d \bar{g}}{\d \mu} \right)_\ssB = \frac{\bar{g} \bar{\lambda}_2 }{2 \pi \alpha}; \qquad \left( \mu \, \frac{\d \bar{\lambda}_2}{\d \mu} \right)_\ssB = \frac{\bar{\lambda}_2 ^2}{2 \pi \alpha} \\
 \left( \mu \, \frac{\d \bar{\lambda}}{\d \mu} \right)_\ssB = \frac{\bar{g}^2 }{4 \pi \alpha}; \qquad \left( \mu \, \frac{\d \bar{\mu}_\ssH^2}{\d \mu} \right)_\ssB = \frac{\bar{g} \bar{\mu}_\Phi^2}{2 \pi \alpha}; \qquad \left( \mu \, \frac{\d\ol{T}}{\d \mu} \right)_\ssB = \frac{ \bar{\mu}_\Phi^4}{4 \pi \alpha} \,. \label{RGeqns}
\eea
where a derivation can be found in Appendix~\ref{floating}. 

We've seen that $\zeta$ and $\bar \lambda_{\rm eff}$ are two combinations of these couplings appear quite frequently in Higgs observables. Eqs.~\pref{RGeqns} imply these couplings satisfy
\be
 \left( \mu \, \frac{\d \zeta}{\d \mu} \right)_\ssB = \left( \frac{\zeta \bar{\lambda}_2 }{ \pi \alpha} \right) \frac{1}{ 1 + (\bar{\lambda}_2 / 4 \alpha)^2 } \,,
\ee
and
\be \label{lambdaeffbetafn}
 \left( \mu \, \frac{\d \bar\lambda_{\rm eff}}{\d \mu} \right)_\ssB = \left( \frac{\zeta }{ \pi } \right) \frac{ 1 - (\bar{\lambda}_2 / 4 \alpha)^2  }{ 1 + (\bar{\lambda}_2 / 4 \alpha)^2 } \,.
\ee
Notice that there is no requirement that $\bar \lambda_2$ be small, so we need not expand the denominator in these expressions.

\subsubsection*{Triviality}

An example of a solution to eqs.~\pref{RGeqns} is
\be \label{eq:grunning}
 \bar g(\mu) = \frac{ \bar g(m_h) }{1 - \frac{\bar \lambda_2(m_h)}{2\pi \alpha} \, \log(\mu/m_h)} \,,
\ee
where we show the running relative to the Higgs mass scale, $\mu = m_h$. This shows how strongly the running depends on the coupling $\bar \lambda_2$. The triviality bound comes from demanding that couplings like $\bar g$ remain within the perturbative regime throughout the energy ranges of interest. In particular, we require that the Landau pole (where $\bar g \to \infty$ and so $\bar \lambda_2(m_h) \log(\mu/m_h) = 2\pi \alpha$), not occur within this energy range.

Notice that the Landau pole arises for scales $\mu < m_h$ whenever $\bar\lambda_2 < 0$, but occurs for $\mu > m_h$ when $\bar \lambda_2 > 0$. If we demand no such pole at energies below 1 TeV then we must therefore require $\bar \lambda_2$ to lie in the range
\be \label{Trivialityeq}
 0 \le \Lambda_2 = \frac{\bar\lambda_2}{\alpha} \lsim 3 \,,
\ee
as indicated in Fig.~\ref{resultgraph}. The upper limit becomes smaller if no Landau pole is allowed for energies above 1 TeV. This represents a significant constraint since none of our results required perturbing in $\bar\lambda_2$, and so were not restricted {\em a priori} to small $\bar \lambda_2$.

\subsubsection*{Vacuum stability}

The vacuum-stability bound demands that $\bar \lambda$ remains positive as it is extrapolated into the UV (at least up to the point where any new UV physics intervenes to change how things run). In the Standard Model this provides the strongest constraint for light Higgs masses, for which $\bar\lambda$ must be small. In this case, neglecting $\bar \lambda$ in its RG equation gives
\be
 \left( \mu \, \frac{\d \bar{\lambda}}{\d \mu} \right)_\SM \approx {3\over 4 \pi^2} \biggl[
 - y_t^4 + {1\over 64} (3 g_2^4 +2 g_2^2 g_1^2 + g_1^4) \biggr] \,,
\label{renormsmall}
\ee
and the constraint arises because the dominant term is negative, eventually driving $\bar \lambda$ negative.

The presence of Higgs-bulk mixing can change this constraint in at least three ways, two of which act to weaken the vacuum stability constraint. It first does so by moving the starting point for $\bar \lambda$ to more positive values. That is, the condition $m_h = 125$ GeV implies $\bar \lambda_{\rm eff} (m_h) = 0.1291$ and so
\be \label{eq:point13}
 \bar \lambda(m_h) = 0.1291 + \frac{\bar g^2(m_h) \bar \lambda_2(m_h)}{(4\alpha)^2 + \bar \lambda_2^2(m_h)} \,,
\ee
rather than simply $\bar \lambda(m_h) = 0.1291$, as would have been the case for the Standard Model. Because the triviality bound requires $\bar \lambda_2 > 0$ we see that the presence of bulk couplings moves the initial condition, $\bar\lambda(m_h)$, to more positive values.

The second change is to the RG equations governing the running of $\bar \lambda$. Including both the contributions from the bulk and from Standard Model loops, we have
\be \label{eq:betalambda}
 \left( \mu \, \frac{\d \bar{\lambda}}{\d \mu} \right)_\ssB = \beta_\SM + \frac{\bar{g}^2 }{4 \pi \alpha} \,,
\ee
where $\beta_\SM$ represents the contribution of eq.~\pref{renorm}. This shows that the bulk couplings always make $\d \bar \lambda/\d \mu$ more positive, and make it more difficult for $\bar \lambda$ to become negative at higher energies.

Finally, the third way Higgs-bulk mixing changes the logic of these bounds is by providing new UV physics, beyond which a naive extrapolation using the renormalization group need not apply. Perhaps the most dramatic way this might happen can be seen in the large-volume case, for which the KK scale is much smaller than the Higgs mass. In this kind of scenario the extra-dimensional Newton constant is much smaller than the 4D Planck mass, and all extrapolations must break down at the mass scale associated with the extra-dimensional Newton constant due to the intervention of the UV physics (perhaps string theory) that is required to unitarize gravitational interactions.

\subsubsection*{Ultraviolet Fixed point for $\lambda$}

The new, positive bulk contribution to the beta function also allows $\bar \lambda$ to reach an ultraviolet fixed point. Demanding the vanishing of eq.~(\ref{eq:betalambda}) at some UV scale $\mu$ gives
\be \label{eq:fxpt}
 \frac{\bar g^2(\mu) }{\alpha} \approx -{3\over \pi}\biggl[
 \bar\lambda^2 + \bar\lambda y_t^2 - y_t^4 - {\bar\lambda  \over 8}  (3 g_2^2
 + g_1^2 ) + {1\over 64} (3 g_2^4 +2  g_2^2 g_1^2 + g_1^4) \biggr] \,.
\ee
To approximate the numerical value of this expression we evaluate the gauge couplings, top Yukawa coupling and Higgs quartic coupling at the weak scale, which is justified by their weak running and the proximity of $\mu$ to the electroweak scale. (A scale much higher than the weak scale would exceed the extra-dimensional gravity scale, as discussed above.) We also approximate $\bar{\lambda}(m_h) = 0.1291 + \cO (\bar g^2(m_h) ) \approx 0.1291$ and so we find that $\bar{g}(\mu)/\sqrt{\alpha} \approx 0.4$ is required to obtain a UV fixed point.

This condition can be run down to the Higgs mass scale using eq.~(\ref{eq:grunning}), which gives
\be \label{eq:fixpointcondition}
 \frac{\bar g (m_h) }{\sqrt{\alpha}} = \frac{\bar g (\mu) }{\sqrt{\alpha}}  \left[1 - \frac{\bar \lambda_2(m_h) }{2 \pi \alpha} \log(\mu/m_h ) \right] \,.
\ee
Since constraints on Higgs bulk mixing restrict $\bar{g} (m_h) /\sqrt{\alpha} \ll \bar{g}(\mu) / \sqrt{\alpha} \approx 0.4$, the value of $\bar{\lambda}_2 (m_h)$ required to attain a UV fixed point is given approximately by the vanishing of eq.~\pref{eq:fixpointcondition}. Choosing $\bar \lambda_2 (m_h)$  in this way also ensures a Landau pole at $\mu$, and so we conclude that the UV fixed point for $\bar \lambda$ and Landau pole are approximately coincident.

\section{Conclusion}
\label{sec:Conc}

In summary, this paper examines the phenomenological implications of Higgs mixing with a bulk scalar field within an extra-dimensional brane-world scenario with the Standard Model localized on a brane. We focus in particular on the {\em Higgs portal}: the dimensionless couplings that can exist in such a scenario between a brane-localized Standard Model Higgs and a bulk scalar field if there are precisely two extra dimensions. We have a variety of motivations for studying this problem, with the main one being the requirement for such couplings in the recently discovered mechanism \cite{6DExpStab} for stabilizing two dimensions at naturally large values (through a manner similar to the Goldberger-Wise mechanism \cite{GW} with one extra dimension).

In a nutshell, we find that the Higgs portal causes the Higgs to mix with KK modes of the bulk scalar, generically leading to new channels for emitting missing energy in processes that can lead to Higgs emission. This can give observable signals at the LHC that strongly resemble the phenomenology of a Higgs with a branching ratio to an invisible decay channel. Unlike Higgs-curvature mixing, or bulk emission by a Higgs (through the trilinear $hh\phi$ coupling), Higgs-bulk mixing through the extra-dimensional portal can give rise to an appreciable energy-loss rate in supernovae, leading to significant constraints on their couplings. As shown in Fig.~\ref{resultgraph}, although strong, these constraints need not preclude an observable signal for an invisible Higgs `decay' channel at the LHC.

In more detail, we find:

\begin{itemize}
\item {\em The strongest constraints come from nucleon bremsstrahlung in SN1987a:} the phase space made available by the high temperature of the supernova, the appreciable Higgs coupling to nucleons, and the strongly coupled nature of nucleon interactions makes this a very strong bound, as we expect from experience with graviton emission in extra dimensional models. However, the one pion exchange approximation we employ is known to overstimate the emission rate, the temperature of the supernova core is uncertain and our calculation assumes the non-degenerate limit, which overestimates the rate at small temperatures \cite{saxion}. Furthermore, the neglected effect of multiple scatterings will decrease the emission rate. Therefore we conservatively estimate bounds assuming $T_{\ssS \ssN} = 20$ MeV, which gives the dominant bound $\bar{g}/\sqrt{\alpha} \lsim 1.5 \times 10^{-3}$ (or $\zeta \lsim 5.6 \times 10^{-7}$) when $\bar{\lambda}_2 = 0$.
\item {\em LHC global fits:} the LHC can indirectly probe an invisible Higgs. An additional, invisible width supresses Higgs signals, and too large a suppression would be in tension with the Standard-Model-like strength of signals currently being observed at the LHC. Global fits to the Tevatron and LHC Higgs data currently imply bounds $B_{\text{inv}} < (0.3 - 0.64)$ the most conservative of which imposes $\bar{g}/\sqrt{\alpha} \lsim 0.007$ (or $\zeta \lsim 5 \times 10^{-5}$), a bound subdominant to the SN1987a bound. If more LHC data were to begin favouring a universal suppression to Higgs signals, then this would suggest an invisible Higgs width, possibly from Higgs-bulk mixing.
\item {\em Invisible final states at the LHC:} the LHC can also directly search for invisible Higgs decays. The LHC will be most sensitive to the $2j + \met$ signal from vector boson fusion into an invisible Higgs. Current searches are only sensitive to invisible Higgs cross sections roughly equal to the Standard Model cross section. However, at design energy and 300 fb${}^{-1}$ of integrated luminosity, it was estimated that this channel would allow the $5\sigma$ discovery of an invisible Higgs with Standard Model production cross section and an invisible branching ratio as small as $B_{\text{inv}} \approx 0.25$. With a more modest 20 fb${}^{-1}$ of luminosity at 7 and 8 TeV, the LHC should be able to rule out $B_{\text{inv}} \lsim 0.4$ at 95\% C.L. In both cases, there are regions of allowed parameter space that predict these branching ratios.
\item {\em Additional cubic interaction:} Higgs-bulk mixing also predicts a cubic Higgs-Higgs-bulk interaction $\frac{1}{2} g h^2 \phi(0)$. This interaction was studied in \cite{6DHiggsph} in the context of the $h \, \phi \to \gamma \, \gamma \, \phi$ final state before the discovery of the new 125 GeV Higgs-like resonance, and without $\lambda_2$ and Higgs-bulk mixing effects. It was concluded that the LHC can probe down to $g = 0.18$ with 100 fb${}^{-1}$ at 14 TeV. The possibility of large $\lambda_2$ and a $h \, \phi \to b \bar{b} \, \phi$ final state through this interaction make the $\gamma \gamma \, \met$ and $b \bar{b} \, \met$ signals worth investigating as a probe of Higgs-bulk physics. The observation of an invisible Higgs width consistent with an increased rate in one or both of these channels would provide strong evidence of Higgs-bulk mixing.
\item {\em Beyond the large-volume limit:} we present constraints and signals in the large-volume limit, but expect the the phenomenology to change a great deal if the volume were not large, so that $m_h \lsim m_\KK$. A single state in the diagonal KK tower would be identified as the 125 GeV resonance, and the spacing of nearby states would be governed by $R^{-1}$. There is no reason to believe that the main signals of large $R$ Higgs-bulk mixing -- invisible final states and suppressed Higgs production rates -- would persist, and the astrophysical bounds would no longer apply to the much heavier states of this scenario. This scenario might resemble Higgs-radion mixing in Randall-Sundrum models, and we regard its exploration to be worth pursuing.
\item {\em Future colliders:} although we refrained from discussing these in any detail, a future muon collider running at $\sqrt{s} = m_h$ with 0.5 fb${}^{-1}$ of data and beam energy resolution of 0.01\% (0.003\%) could directly probe the Higgs width to within 0.85 MeV (0.30 MeV) \cite{muonWidth}. A future linear collider with an integrated luminosity of 250 fb${}^{-1}$ at 250 GeV would be able to constrain the Higgs invisible branching ratio to less than a few \% \cite{ilc}. This would correspond to the $B=0.03$ line in Fig.~\ref{resultgraph}.

\end{itemize}

In short, mixing through the Higgs-bulk portal provides a particular example of what detailed Higgs studies might ultimately tell us about the nature of vacuum energetics. We hope that this is the path Nature chooses, and that the Higgs is shown to have exotic invisible properties, rather than persisting in having invisible exotic properties.

\section*{Acknowledgements}

We thank Hugo Beauchemin, Leo van Nierop, Brian Shuve, Matt Williams and Itay Yavin for useful discussions. Research was supported in part by funds from the Natrual Sciences and Engineering Research Council (NSERC) of Canada. Research at the Perimeter Institute is supported in part by the Government of Canada through Industry Canada, and by the Province of Ontario through the Ministry of Research and Information (MRI).

\appendix

\appendix

\section{Renormalization of brane couplings}
\label{floating}

\subsubsection*{Codimension-1 Regularization}

In this Appendix we compute the near-brane divergences and how they are renormalized, following the analysis of Ref.~\cite{Cod2Renorm}.

A more formal means of deriving the regulated boundary condition in (\ref{regBC}) and the subsequent renormalized brane couplings (\ref{rgsolved}) is to model the codimension-2 brane as the dimensional reduction of an arbitrarily small codimension-1 brane. We introduce a codimension-1 brane action at a small distance $r = \epsilon$ in the extra dimensions, such that its dimensional reduction matches the codimension-2 brane action as $\epsilon \to 0$. This might not seem like a palpable addition to the theory. However, since new brane has codimension 1, it resolves the singularities typically associated with codimension-2 sources, at the expense of an introducing arbitrarily short distance $\epsilon$. It is in this sense that the codimension-1 brane is analogous to introducting cutoff in typical quantum field theories.

Consider the following codimension-1 action
\be \label{codim1}
 S_\epsilon = - f(\epsilon)  \int \d^4 x \, \d \theta \left( T_{\epsilon} - (\mu_\ssH^{\epsilon})^2 {H}^\dagger {H} + \lambda_{\epsilon} ({H}^\dagger {H})^2 +  (\mu_\Phi^{\epsilon})^2 \Phi_\epsilon + \frac{1}{2} \lambda_2^{\epsilon} \Phi^2_\epsilon + g_{\epsilon} {H}^\dagger {H} \Phi_{\epsilon} \right) \,,
\ee
where the factor $ f(\epsilon)$ comes from the determinant of the induced metric, $\Phi_\epsilon = \Phi(x,\epsilon,\theta)$ is a function of $x$ and $\theta$ evaluated at $r = \epsilon$, and the rest of the Standard Model terms are neglected. Couplings with an $\epsilon$ script are new couplings defined on the codimension-1 brane, in analogy with the quantities in (\ref{scalarPotential}). The field $H$ is not a function of the extra dimensional coordinates.

We can match the codimension-1 action to the codimension-2 action in (\ref{scalarPotential}) by dimensional reduction. The KK modes of $\Phi_{\epsilon}$ in the $\theta$ direction acquire a mass $n/\epsilon$ so all but the $n=0$ modes decouple for arbitrarily small $\epsilon$. The surviving $n=0$ modes have constant profiles in the coordinate $\theta$ that we can choose to be unity, giving
\be
 S_\epsilon = - 2 \pi \alpha f(\epsilon) \int \d^4 x   \left(T_{\epsilon} - (\mu_\ssH^\epsilon)^2 H^\dagger H + \lambda_\epsilon (H^\dagger H)^2 +  (\mu_\Phi^\epsilon)^2 \Phi_b + \frac{1}{2} {\lambda}_2^\epsilon \Phi_b^2 + {g_{\epsilon}} H^\dagger H \Phi_b \right) \,,
\ee
where we have identified $\Phi_b = \Phi(x,0)$ as the zero mode of $\Phi_\epsilon$. To reproduce the action (\ref{scalarPotential}) requires
\be
 2 \pi \alpha f(\epsilon) c_{\epsilon} = c \,,
\ee
where $c_\epsilon$ is any of the couplings in the codimension-1 action (\ref{codim1}) and $c$ is the associated coupling in the codimension-2 action (\ref{scalarPotential}).

Having matched the codimension-1 action to the original theory, we use it to derive a boundary condition for $\Phi$. The variation of the bulk action (\ref{bulkAction}) gives a contribution on the boundary at $r = \epsilon$
\be
 \delta S_\ssB \supset -\int \d^4 x \, \d \theta \, \delta \Phi_\epsilon \left( -f \partial_r \Phi \right)_{r = \epsilon} \,,
\ee
which must vanish when combined with the variation of the codimension-1 action (\ref{codim1}), giving the boundary condition
\be
 -( f \partial_r \Phi)_{r= \epsilon} + f(\epsilon) \left(  (\mu_\Phi^{\epsilon})^2 + \lambda_2^{\epsilon} \Phi_\epsilon + g_{\epsilon} H^\dagger H \right) = 0\,.
\ee
This can be rewritten in terms of the couplings in the codimension-2 action
\be
 -2 \pi \alpha \left( f \partial_r \Phi \right)_{r= \epsilon} +  \mu_\Phi^2 + \lambda_2 \Phi_\epsilon + g H^\dagger H  = 0 \,,
\ee
which is a regularized version of eq. (\ref{branebcI}). Varying (\ref{codim1}) with respect to $H$ yields the regularized version of (\ref{higgsFE}), which can be used to derive (\ref{regBC}).

\subsubsection*{Floating Brane Renormalization}

As in typical field theories, the cutoff $\epsilon$ is not a physical quantity, and as $\epsilon \to 0$ the boundary condition still diverge. To remedy this, we consider a theory that is sourced by a ``floating'' codimension-1 brane at $r = \bar{r}$, that can be dimensionally reduced to a codimension-2 brane. The properties of the floating brane are fixed by demanding that the solution to the field equations for $r \geq \bar{r}$ matches the regularized solution. The dependence on $\epsilon$ will be eliminated when the bare couplings $c$ are traded for the coupling constants of the floating brane $\bar{c}$. It is in this sense that $\bar r$ can be thought of as a subtraction scale, and $\bar{c}$ can be thought of as the renormalized couplings.

After dimensional reduction to codimension-2, we assume the floating brane has the form
\be
 \bar{S}_b = -\int d^4 x \left( \bar{T}-\bar{\mu}_\ssH^2 H^\dagger H + \bar{\lambda} (H^\dagger H)^2 +  \ol{\mu}_\Phi^2 \Phi_{\bar{r}} + \frac{1}{2} \bar{\lambda}_2 \Phi^2_{\bar{r}} + \bar{g} H^\dagger H \Phi_{\bar{r}} \right) \,,
\ee
where $\Phi_{\bar{r}} = \Phi(x,\bar{r})$ and the theta dependence has been integrated out. Varying the action gives two boundary field equations
\be \label{renormCond}
-2 \pi \alpha \left( f \partial_{r} \Phi \right)_{r= \bar{r}} + \bar{g} H^\dagger H + \bar{\lambda}_2 \Phi_{\bar{r}} + \ol{\mu}_\Phi^2 = 0; \qquad H^\dagger H = \frac{1}{2 \bar{\lambda}}\left(\bar{\mu}_\ssH^2- \bar{g} \Phi_{\bar{r}}\right) \,,
\ee
in direct analogy with the boundary field equations in \S\ref{ssec:VacConf}. These equations should be read as fixing the floating brane couplings, since we have demanded that $\Phi$ and $H$ solve the regularized field equations and so their functional form is already fixed. For a given value of $\bar{r}$, the floating brane couplings will have to be chosen appropriately, and will change with a change in $\bar{r}$. However, since $\bar{r}$ is arbitrary, this change in the couplings cannot have any effect on physical quantities, such as $H^\dagger H$. For example, changes in (\ref{renormCond}) under a change in $\bar r$ should vanish, giving
\bea \label{longRGs} \nn
 H^\dagger H \partial_{\bar r} \bar{g} + \Phi({\bar r}) \partial_{\bar r} \bar{\lambda}_2 +   \bar{\lambda}_2 \partial_{\bar r} \Phi({\bar r}) +  \partial_{\bar r} \ol{\mu}_\Phi^2 &=& 0\\
 2 H^\dagger H \partial_{\bar r} \bar{\lambda} - \partial_{\bar r} \bar{\mu}_\ssH^2  + \Phi({\bar r}) \partial_{\bar r} \bar{g} + \bar{g}  \partial_{\bar r} \Phi({\bar r}) &=& 0 \, ,
\eea
where we have used the fact that $\partial_{\bar r}(f \partial_{r} \Phi )_{r = \bar r} = (\partial_r f \partial_{r} \Phi)_{r = \bar r} = 0 $ by the bulk equation of motion (\ref{bulkFE}), and $\partial_{\bar r} (H^\dagger H)= 0$ because it is a physical quantity that should not depend on $\bar r$. We premultipy both equations by $f(\bar r)$ to facilitate the use of the relation
\be
 f({\bar r}) \partial_{\bar r} \Phi({\bar r}) = (f \partial_r \Phi)_{r = \bar r} = \frac{1}{2 \pi \alpha} \left( \ol{\mu}_\Phi^2 + \bar{\lambda}_2 \Phi({\bar r}) + \bar{g} H^{\dagger} H \right).
\ee
Substituting this into (\ref{longRGs}) and equating powers of $\Phi$ and $H^\dagger H$ yields the following RG equations
\be \nn
 \partial_{\bar F} \ol{\mu}_\Phi^2 = -\frac{\bar{\lambda}_2 \ol{\mu}_\Phi^2}{2 \pi \alpha}; \qquad \partial_{\bar F} \bar{g} = -\frac{\bar{g} \bar{\lambda}_2 }{2 \pi \alpha}; \qquad  \partial_{\bar F}\bar{\lambda}_2 = -\frac{\bar{\lambda}_2 ^2}{2 \pi \alpha};
\ee
\be
\partial_{\bar F} \bar{\lambda} = -\frac{\bar{g}^2 }{4 \pi \alpha}; \qquad \partial_{\bar F}\bar{\mu}_\ssH^2 = \frac{\bar{g}\, \ol{\mu}_\Phi^2}{2 \pi \alpha}; \qquad  \partial_{\bar F}\bar{T} = -\frac{ \ol{\mu}_\Phi^4}{4 \pi \alpha} \,,
\ee
where we have used $f(\bar r)\partial_{\bar r} = \partial_{\bar F}$. The solutions can be found in (\ref{rgsolved}). Although not explicitly derived in this section, for completness the RG equation for $\bar T$ is listed here and can be derived from (\ref{tensionRenorm}) in the following appendix.

\section{Probe-brane energetics}
\label{app:VEnerg}

In this section we evaluate the energy density to identify when the solution with nonzero $H$ is energetically preferable to the solution $H = 0$. For illustrative purposes (following the discussion of \cite{6DHiggs1}) we do so here ignoring the energetics of the gravitational back-reaction of the branes. This is known not to be a good approximation in general for codimension-2 objects since in many cases it is the classical bulk back-reaction that cancels the brane tensions to allow the classical brane geometries to be flat \cite{SS, 6DSolns, 6DMagic}. (Indeed, it is this property --- together with bulk supersymmetry --- that underlies their use as a starting point for seeking solutions to the cosmological constant problem \cite{SLED, SLEDrev, TNCC}.)

The regularized energy density for $H$ and $\Phi$ is
\be
 \cH = \cH_\ssB + \sum_b U_b \,,
\ee
where the bulk contribution is
\be
 \cH_\ssB = \pi \alpha \int \limits_{\epsilon}^{\pi R} \d r f(r)  (\partial_r \Phi)^2 \,,
\ee
and the `Higgs brane' gives
\be
  U = T - \mu_\ssH^2 \, H^\dagger H + \lambda (H^\dagger H)^2 + \mu_\Phi^2 \, \Phi_\epsilon + \frac{\lambda_2}{2} \, \Phi^2_\epsilon + g \, H^\dagger H \, \Phi_\epsilon \,.
\ee

Evaluating using the bulk and brane solutions, eqs.~\pref{bulkSoln} and \pref{braneSoln}, and renormalizing as before gives
\be
 \cH_\ssB + U = - \left( \frac{\bar{\mu}^4_{\Phi\,{\rm eff}}(\hat r)   }{4 \pi \alpha} \right) F(\hat r, \pi R) + T + \left( \frac{\ol{\mu}_\Phi^4(\hat r)}{4 \pi \alpha} \right) \frac{ F(\epsilon, \hat r)}{1
 + \frac{\bar{\lambda}_2
 (\hat r) }{2 \pi \alpha} \, F(\epsilon,\hat r)} - \frac{\bar{\mu}_\ssH^4(\hat r)}{4 \bar{\lambda}(\hat r)} \,,
\ee
which suggests we renormalize the brane tension using
\be \label{tensionRenorm}
 \ol{T}(\bar r) = T + \left( \frac{\mu_\Phi^4}{4 \pi \alpha} \right) \frac{F(\epsilon,\bar r) }{1 - \frac{{\lambda}_2}{2\pi \alpha} \, F(\epsilon,\bar r) } \,,
\ee
so that the vacuum energy density coming from the bulk and the Higgs brane is
\be \label{vacuumE}
 \cH_\ssB + U = - \left( \frac{\ol{\mu}^4_{\Phi\,{\rm eff}}(\hat r)   }{4 \pi \alpha} \right)  F(\hat r, \pi R) + \ol{T}(\hat r) - \frac{\bar{\mu}_\ssH^4(\hat r)}{4 \bar{\lambda}(\hat r)} \,.
\ee

Eq.~\pref{vacuumE} allows two simple conclusions to be drawn. First, since the result with $H=0$ only omits the last term, this shows that nonzero $H$ is a preferred vacuum to the extent that $\bar\mu_\ssH^2(\hat r) > 0$. (Boundedness of $\cH$ from below precludes $\bar\lambda$ from being negative.) Second, the sign of the $\bar \mu_{\Phi\,{\rm eff}}^4$ term depends on the sign of $F(\hat r, \pi R)$, which in turn depends on the boundary condition at $\pi R$ . We can imagine imposing a general boundary condition
\be
 2 \pi \alpha f \Phi^\prime({\pi R}) + \lambda_3 \Phi({\pi R}) = 0 \, ,
\ee
which fixes $\hat r$ as follows
\be
 F(\hat r, \pi R) = \frac{\lambda_3}{2 \pi \alpha} \, .
\ee
This illustrates that the bulk field contribution to the vacuum energy density can be positive or negative depending on the sign of $\lambda_3$, so that the preferred vacuum will depend on the details of the faraway brane. We assume that the faraway brane is such that the general solution we found in the main text is preferred.

\section{Schwinger-Dyson equation}
\label{appsec:SchwDys}

In this Appendix we derive the relation of eq.~\pref{hphiamp},
\be \label{appyyy}
 [{\rm Amp}\langle h \, \phi^*(0) \rangle_k]^* = {\rm Amp}\langle \phi(0) \, h^* \rangle_k = \left. -\frac{i\bar g(\bar r) v \, \langle h \, h^* \rangle_k}{1 - i (\bar \lambda_2(\bar r)/4\alpha)} \right|_{\bar r^2 = -1/k^2} \,,
\ee
relating the amputated mixed $h-\phi$ propagator to the $h-h$ autocorrelation, and the relation of eq.~\pref{dressed}
\be \label{appzzz}
 \langle h h^* \rangle_k = \frac{D_k^h[1 + i \lambda_2 D_k^\phi(0,0)]}{1 + \left[ i \lambda_2 \  + (gv)^2 D_k^{h} \right] D_k^\phi(0,0)} \,,
\ee
that gives the dressed two point function $h$, and largely controls the phenomenology of Higgs-bulk mixing.

Our goal is to compute relations amongst the four correlation functions of interest, given by
\bea
 &&\langle h \, h^* \rangle_k := G_{hh}(k) \,, \quad
 \langle h \, \phi^*(y) \rangle_k := G_{h \phi}(k; y) \,, \nn\\
 &&\langle \phi(y) \, h^* \rangle_k := G_{\phi h}(k; y) \quad
 \hbox{and} \quad
 \langle \phi(y) \, \phi^*(y') \rangle_k := G_{\phi\phi}(k; y,y') \,,
\eea
where $G_{h\phi}(k;y) = G^*_{\phi h}(k;y)$. In these expressions $y^m$ denotes the spatial coordinates in the two extra dimensions while $k^\mu$ is the Fourier transform variable in the four on-brane directions.

The most direct way to obtain the desired relations is to express the Higgs-bulk interactions as delta-function localized terms in the lagrangian density, following arguments made in the appendix of ref.~\cite{6DHiggs1}.\footnote{A disadvantage of the delta-function formulation is the requirement to deal with expressions like $f(x) \, \delta(x)$, with $f(x) \to \infty$ as $x \to 0$. This requires a more careful treatment of regularization and renormalization, along the lines of the codimension-one formulation used in the main text, but in the present instance leads to the same conclusions.} The starting point is the field equations for linearized fluctuations
\bea
 \sqrt{\cG_2}(\Box_4 + \Box_2) \phi - \Bigl[ \lambda_2 \, \phi + gv \, h \Bigr] \, \delta^2(y) &=& 0 \nn\\
 \Box_4 h - 2 \lambda v^2 \, h - gv \, \phi(0) &=& 0 \,.
\eea
which imply the following equations for the propagators
\bea \label{app:Geqns}
 \sqrt{\cG_2}(-k^2 + \Box_2) G_{\phi\phi}(k; y,y') - \Bigl[ \lambda_2 \, G_{\phi \phi}(k;0,y') + gv \, G_{h\phi}(k; y') \Bigr] \delta^2(y) &=& i \delta^2(y-y') \nn\\
 \sqrt{\cG_2}(-k^2 + \Box_2) G_{\phi h}(k; y) - \Bigl[ \lambda_2 \, G_{\phi h}(k;0) + gv \, G_{hh}(k) \Bigr] \delta^2(y)  &=& 0 \nn\\
 (k^2 + 2 \lambda v^2 )G_{hh}(k) + gv \, G_{\phi h}(k; 0) &=& -i \nn\\
  (k^2+2 \lambda v^2) G_{h\phi}(k;y) + gv \, G_{\phi\phi}(k; 0,y) &=& 0 \,.
\eea
By contrast, the unperturbed propagators in the absence of Higgs-bulk couplings satisfy
\bea \label{app:Deqns}
 \sqrt{\cG_2}(-k^2 + \Box_2) D_k^{\phi}(y,y') &=& i \delta^2(y-y') \nn\\
 (k^2 + 2 \lambda v^2 )D_k^h &=& -i \,.
\eea

We use the first of eqs.~\pref{app:Deqns} to solve the second of eqs.~\pref{app:Geqns}, leading to
\bea
 G_{\phi h}(k; y) &=& -i \int \d^2y' D_k^{\phi}(y,y')\Bigl[ \lambda_2 \, G_{\phi h}(k;0) + gv \, G_{hh}(k) \Bigr] \delta^2(y') \nn\\
 &=& -i D_k^{\phi}(y,0)\Bigl[ \lambda_2 \, G_{\phi h}(k;0) + gv \, G_{hh}(k) \Bigr]  \,,
\eea
and this, when specialized to $y=0$, in turn implies
\be
 G_{\phi h}(k; 0) = -i D_k^{\phi}(0,0)\Bigl[ \lambda_2 \, G_{\phi h}(k;0) + gv \, G_{hh}(k) \Bigr]  \,,
\ee
which may be solved to give
\be \label{eq:gphih}
 G_{\phi h}(k; 0) = -i \left[ \frac{gv \, G_{hh}(k)}{1 + i \lambda_2 D_k^{\phi}(0,0)} \right] \, D_k^{\phi}(0,0)\,.
\ee

The overall factor of $D_k^\phi(0,0)$ is removed when the external $\phi$-line is amputated, and for the denominator we use the continuum result, eq.~\pref{eq:D0continuum}, to evaluate $D_k^\phi(0,0)$,
\be \label{app:branetobrane}
  D_k^\phi(0,0) = \frac{i}{4\pi \alpha} \Bigl[ \log(-k^2 \epsilon^2) -i \pi \Bigr] \,,
\ee
and renormalize the divergence into the brane couplings, $\bar g$ and $\bar \lambda_2$, using eqs.~\pref{rgsolved}. Eq.~\pref{appyyy} then follows by choosing the renormalization point so that $k^2 \bar r^2 = 1$ and the logarithms vanish.

We can also use the second of eqs.~\pref{app:Deqns} to solve the third of eqs.~\pref{app:Geqns}, giving
\be
 G_{hh}(k) + i gv \, G_{\phi h}(k; 0) D_k^h  = D_k^h
\ee
To solve for $G_{hh}(k)$ we substitute eq.~(\ref{eq:gphih}) into the above expression, which can be rearranged to give eq.~\pref{appzzz} as desired.

\section{Toy model: unperturbed modes}
\label{app:toymodel}

In this appendix we explicitly take the continuum limit of (\ref{eq:branetobrane}) to arrive at eq. (\ref{eq:Dk00int}). This is accomplished in a toy model in which the extra dimensions are a flat disc: $f(r) = r$ for $0 \le r \le \pi R$. We can explicitly solve the wavefunctions on this background, which allows for a straightforward move to the large $R$ limit, although the results are true for all $R$.

Using eq. (\ref{eq:zseparation}) in eq. (\ref{eq:boxZ}) on the disc geometry gives the field equation for the $n=0$ wavefunctions
\be \label{eq:bulkEOM}
 \left[ M^2_{0 l} + \frac{1}{r} \, \partial_r ( r \, \partial_r ) \right] P_{0 l} = 0 \,,
\ee
with the following boundary conditions
\be
 \left(r  \partial_r P_{0 l} \right)_{r=0,\pi R} = 0 \, ,
\ee
and normalization conditions
\be
 2 \pi \alpha \int\limits_0^{\pi R} dr \, r \, P_{0 l}^* P_{0 l'} = \delta_{l l'} \, .
\ee
The properly normalized solutions and eigenvalue conditions read
\be
 P_{0 l}(r) = \frac{1}{\sqrt{\pi^3 \alpha R^2 }} \left( \frac{J_0(M_{0 l} \, r) }{ J_0(M_{0 l} \pi R ) }\right)  \qquad \text{with}  \qquad J_1(\pi R M_{0 l}  ) = 0 \, ,
\ee
where $J_0$ is the zeroth Bessel function of the first kind. The brane-to-brane propagator is given by
\be
   D_k(\epsilon,0) =  \sum_{l}\left(  \frac{-i}{k^2 +  M_{0 l}^2- i \epsilon} \right) \frac{ J_0( \epsilon M_{0l})   }{J_0^2( \pi R M_{0l}) \pi^3 \alpha R^2} \, ,
\ee
since $J_0(0) = 1$. Using the fact that $J_0(x) \to \sqrt{\frac{2}{\pi x}}$ for large $x$ gives
\be
 D_k(\epsilon,0) = \sum_{l}\left(  \frac{-i}{k^2 +  M_{0 l}^2- i \epsilon} \right) \frac{ J_0( \epsilon M_{0l}) M_{0 l}}{2 \pi \alpha R} \, .
\ee
Sums over closely spaced modes in $d$ dimensions can be replaced by integrals as follows
\be
 \frac{1}{\Omega^d} \sum_{\vec{n}} f_{\vec n} \to \int\frac{  d^d M }{(2 \pi)^d} \, f(M) \, ,
\ee
where, in this case, the sum is over the radial index, so the conversion is one-dimensional. Using $\Omega^d = 2 \pi R$ for the diameter of the disc is gives
\be \label{eq:integralap}
 D_k^\phi(\epsilon,0) = \frac{-i}{2 \pi \alpha} \int\limits_0^\infty \d q \frac{ q J_0( \epsilon M) } {k^2 + q^2 - i \varepsilon} \, .
\ee
The integral can be computed
\be
 D_k^\phi(\epsilon,0) = \frac{-i}{2 \pi \alpha} K_0 \left( \sqrt{k^2} \, \epsilon \right) \,,
\ee
where $K_0$ is the zeroth modified bessel function. For small arguments $K_0(x) \to - \log(x/2) + \gamma$ so the divergent part of the brane-to-brane propagator reads
\be
 D_k^\phi(\epsilon,0) = \frac{i}{4 \pi \alpha} \log(k^2 \epsilon^2 ) \,,
\ee
in agreement with eq.~(\ref{eq:D0continuum}).

\section{Beyond Sturm Liouville}
\label{app:generalizedSturm}

In this Appendix we describe how the Sturm-Liouville orthogonality conditions generalize to the case of interest in the main text, for which the boundary conditions differ for different modes.

For the present purposes the eigenvalue condition for the mode functions $\xi_n(x)$ has the general form
\be \label{stlv}
 \partial_x \left[ p(x) \partial_x \xi_n \right] - q(x) \xi_n  + \lambda_n w(x) \xi_n  = 0 \,,
\ee
in an interval $x_0 \le x \le x_1$, with $p,q,w$ known real functions and $\lambda_n$ the corresponding eigenvalue. The unusual part relative to Sturm-Liouville problems of childhood days is that they satisfy $n$-dependent boundary conditions at the edges of the domain of interest:
\be \label{genbc}
 \Bigl[ J_b (\lambda_n - K_b) p \, \partial_x \xi_n + (\lambda_n - L_b) \xi_n \Bigr]_{x=x_b} = 0 \,,
\ee
where $J_b,K_b,L_b$ are again known coefficients. These boundary conditions ruin the orthonormality of the mode functions under the usual inner product,
\be
 \int\limits_{x_0}^{x_1} \d x \; w(x) \, \xi_m^* \xi_n \neq \delta_{mn} \,,
\ee
which in turn ruins the diagonalization of the 4D action once decomposed in terms of these modes.

To identify how the inner product must generalize in order to maintain orthogonality with the new boundary conditions we follow standard steps. First multiply eq.~(\ref{stlv}) by $\xi_m$ then subtract the complex conjugate of the same equation with $(m \leftrightarrow n)$ and integrate the result over $x$. This yields
\be
 (\lambda_m - \lambda_n) \int\limits_{x_0}^{x_1} \d x \; w(x) \, \xi_m^* \xi_n = \Bigl[ p \left(  \xi_m^* \, \partial_x \xi_n - \xi_n \, \partial_x \xi_m^*\right) \Bigr]_{x_0}^{x_1} \,,
\ee
which would vanish for the usual Stum-Liouville boundary conditions. However, with the $n$-dependent boundary conditions of the form (\ref{genbc}) we instead have
\be \label{eq:intstep}
 (\lambda_m- \lambda_n ) \int\limits_{x_0}^{x_1} \d x \; w(x) \, \xi_m^* \xi_n = (\lambda_m - \lambda_n) \sum_b (-1)^{1-b} \left( \frac{ L_b - K_b}{J_b} \right) \frac{ \xi_m^*(x_b) \xi_n(x_b)   }{(\lambda_n - K_b)(\lambda_m - K_b)}   \neq 0 \,.
\ee

What allows us to devise an inner product with respect to which the modes are automatically orthogonal is the property that the $n$-dependence of the boundary conditions is linear in $\lambda_n$, since this ensures both sides of eq.~\pref{eq:intstep} depend on $n$ through their common factor of $(\lambda_m - \lambda_n)$. This suggests defining the following inner product
\be \nn
 \langle \xi_m, \xi_n \rangle = \int\limits_{x_0}^{x_1} \d x \, w(x) \xi_m^* \xi_n + \sum_b (-1)^{1-b} \left( \frac{ L_b - K_b}{J_b} \right) \frac{ \xi_m^*(x_b) \xi_n(x_b) }{(\lambda_m - K_b)(\lambda_n - K_b)} \,,
\ee
since eq.~\pref{eq:intstep} then shows that the boundary conditions imply
modes with different eigenvalues are automatically orthogonal, and so a basis of eigenmodes can be chosen to be orthonormal: $\langle \xi_m, \xi_n \rangle = \delta_{mn}$.

In the dimensional-reduction problem the constants $J_b,K_b,L_b$ are read from the brane action, and so are the quantities that appear in the quadratic lagrangian once bulk fields are decomposed in terms of these mode functions. This ensures that the action diagonalizes as it would have done for a standard KK decomposition without endpoints.

For example, for the zero modes in the brane bulk mixing scenario we send $x \to r$ and $n,m \to s,t$ and use $p(r) = 2\pi \alpha f(r)$, $q(r) = 0$ and $w(r) = 2 \pi \alpha f(r)$. We replace the eigenvalues with the KK masses $\lambda_\ell = M_{0 \ell}^2$. There is only one brane at $r_0 = 0$ and (neglecting subscripts) it gives $J = - 1/\lambda_2$, $K = 2v^2 \lambda$ and $L = 2v^2 \lambda + (gv)^2 / \lambda_2$ so that the inner product reads
\be
 \la \cP_{s} ,\cP_{t}\ra = 2 \pi \alpha \int\limits_0^{\pi R} \d r f \, \cP_{s}^* \cP_{t} + \frac{(gv)^2 \cP_{s}^*(0) \cP_{t}(0)}{(M_{s}^2 - 2 \lambda v^2 )(M_{t}^2 - 2 \lambda v^2)} \,,
\ee
and the orthonormality relationship (\ref{eq:cPnorm}) in the text follows.

\subsection*{Diagonalization of the quadratic action}

We now show that in the case of interest in the main text, this modified inner product is just what is required to diagonalize the quadratic action, including the Higgs-bulk mixing terms. For simplicity, we only include the $n = 0$ modes, but the extension to any $n \neq 0$ level of the KK tower follows readily. We still use $s = \{n,\ell\}$ with the understanding that $n=0$.

In terms of KK modes the the bulk action (\ref{eq:bulkActnap}) reads
\bea \nn
 S_\ssB &=& - 2 \pi \alpha \int d^4 x \int dr \sum_{s ,t} \left( f \cP_{s}^* \, \cP_{t} \right) \left[ \frac{1}{2} \, \partial_\mu \varphi_{s} \, \partial^\mu \varphi_{t} \right] \\
 &&-   2 \pi \alpha \int d^4 x \int dr \sum_{s,t} \left( f \partial_r \cP_{s}^* \partial_r \cP_{t}\right) \left[ \frac{1}{2} \, \varphi_{s} \varphi_{t}  \right] \,,
\eea
where terms have been organized into their $r$-dependent parts, which are in round brackets, and their $x$-dependent parts, which are in square brackets. They have also been written on separate lines for organizational purposes. Integrating the second term by parts gives
\bea \nn
 S_{\ssB} &=& - 2 \pi \alpha \int d^4 x \int dr \sum_{s,t}  \left(f  \cP_{s}^* \, \cP_{t} \right) \left[ \frac{1}{2} \, \partial_\mu \varphi_{s} \, \partial^\mu \varphi_{t}\right] \\ \nn
&&+ 2 \pi \alpha \int d^4 x \int dr \sum_{s,t} \left( \cP_{t} \, \partial_r f  \partial_r \, \cP_{s}^* \right) \left[ \frac{1}{2} \varphi_{s} \varphi_{t} \right] \\ &&+ 2\pi \alpha \int d^4 x \sum_{s,t}  \left( f \cP_{t} \, \partial_r \cP_{s}^*  \right)_{r=0} \left[ \frac{1}{2}\, \varphi_{s} \varphi_{t} \right] \,,
\eea
where the term on the bottom line is a boundary term, and it is assumed that the other boundary term for the faraway brane vanishes by the boundary conditions, or is cancelled by the faraway brane's action. The bulk equation of motion (\ref{bulkeom2}) allows the second line to be combined with the first as follows
\bea \nn
  S_{\ssB} &=& - 2 \pi \alpha \int d^4 x \int dr \sum_{s,t}   \left( f \cP_{s}^* \, \cP_{t} \right) \left[ \frac{1}{2} \, \partial_\mu \varphi_{s} \, \partial^\mu \varphi_{t} + \frac{1}{2} \, M_{s}^2 \varphi_{s} \varphi_{t} \right] \\ &&+ 2\pi \alpha \int d^4 x \sum_{s,t}   \left( f \cP_{t} \, \partial_r \cP_{s}^*  \right)_{r=0} \left[ \frac{1}{2}\, \varphi_{s} \varphi_{t}  \right] \, .
\eea
The integration over the radial coordinate can be completed using the orthonormality relationship (\ref{eq:cPnorm}) so that the bulk action contributes three terms to the dimensionally reduced Lagrangian that will be called $\cL_{1,2,3}$
\bea \nn
 \cL_1 &=& - \sum_s  \left[ \frac{1}{2} \partial_\mu \varphi_{s} \, \partial^\mu \varphi_{s} + \frac{1}{2} M_{s}^2 \varphi_{s}^2 \right] \\ \nn
 \cL_2 &=&   \sum_{s,t}  \frac{(gv)^2 \cP_{s}(0) \cP_{t}(0)}{(M_{s}^2 - 2 \lambda v^2 )(M_{t}^2 - 2 \lambda v^2 )}  \left[ \frac{1}{2} \, \partial_\mu \varphi_{s} \, \partial^\mu \varphi_{t} + \frac{1}{2} \, M_{s}^2 \varphi_{s}  \varphi_{t} \right]  \\
\cL_3 &=&  2\pi \alpha  \sum_{s,t}  \left( f \cP_{t} \, \partial_r \cP_{s}  \right)_{r=0}  \left[ \frac{1}{2}\, \varphi_{s} \varphi_{t} \right] \, .
\eea
The first line is a canonically normalized KK tower of scalar fields with masses $M_{s}$, which is the desired final result. The second and third term are cancelled by terms in the brane action as will be shown explicitly. For example, writing the $h$ kinetic and mass term in terms of eigenstates $\varphi_s$ and then combining them with $\cL_2$ gives
\be \label{eq:halfgvap}
\cL_{2} + \cL_h = \cL_2 - \frac{1}{2} \partial_\mu h \partial^\mu h - \lambda v^2 h^2 =  \sum_{s,t}  \frac{(gv)^2 \cP_{s}(0) \cP_{t}(0)}{(M_{t}^2 - 2 \lambda v^2 )}  \left[ \frac{1}{2} \, \varphi_{s}  \varphi_{t} \right] \,,
\ee
while $\cL_3$ and the brane mass term for $\phi$ give
\be
 \cL_{3} + \cL_\phi =  \cL_3 - \frac{1}{2} \lambda_2 \phi^2(0) = \sum_{s,t} \cP_{t}(0) (2 \pi \alpha f \partial_r \cP_{s}(0) - \lambda_2 \cP_{s}(0) ) \left[ \frac{1}{2} \, \varphi_{s}  \varphi_{t} \right] \,,
\ee
which is identical to (\ref{eq:halfgvap}) once the boundary condition (\ref{bcexplicit2}) is employed. Finally the mixing term gives
\be
 \cL_{h\phi} =  - gv h \phi(0) = - 2 \sum_{s,t}  \frac{(gv)^2 \cP_{s}(0) \cP_{t}(0)}{(M_{t}^2 - 2 \lambda v^2 )}  \left[ \frac{1}{2} \, \varphi_{s}  \varphi_{t} \right] \,,
\ee
so that $\cL_2 + \cL_3 + \cL_\phi + \cL_h + \cL_{h\phi} = 0$ and the dimensionally reduced theory is the KK tower of massive scalar fields found in $\cL_1$.

\section{Solving for the KK mode functions}
\label{app:kkwfn}

In this appendix we solve solve the $n=0$ perturbed wavefunctions $\cP_{0 \ell}$ for $f(r) = r$, and $0 < r < \pi R$ with Dirichlet boundary conditions at $r = \pi R$. In this geometry the general solution can be written in terms of Bessel functions
\be
 \cP_{0 \ell}(r) = N_\ell \left[ \frac{\pi}{2} Y_0(r M_{0\ell}) + D_\ell J_0(r M_{0\ell}) \right] \, ,
\ee
where $N_\ell$ are normalization constants, $D_\ell$ are integration constants and the factor of $\pi/2$ is chosen for convenience. It is straightforward to impose Dirichlet BCs at $r =\pi R$, which imply
\be \label{DBcs}
D_\ell = -\frac{\pi}{2} \frac{ Y_0 (\pi RM_{0\ell})}{J_0(\pi RM_{0\ell})} \, .
\ee
Imposing the UV boundary condition, on the other hand, is more complicated. This is because of the UV divergences we expect in this theory. Near the origin, the relevant Bessel functions behave like
\bea \nn
  Y_0(x) &\approx& \frac{2}{\pi} \left[\ln\left(x/2\right)  + \gamma \right] \\ \nn
  Y_1(x) &\approx& -\frac{2}{\pi} \frac{1}{x} \\ \nn
  J_0(0) &=& 1 \\
  J_1(0) &=& 0 \, .
\eea
So, as in the vacuum solutions, the boundary condition near the brane diverges, and must be regulated and renormalized. We cut off the boundary condition at $r=\epsilon$ and find
\be
  D_\ell = \frac{2 \pi \alpha}{\beta_\ell} - \log(\epsilon M_{0\ell}/2) - \gamma \quad \text{with} \quad \beta_\ell = \lambda_2 + \frac{(gv)^2}{M_{0\ell}^2 - 2v^2 \lambda} \, .
\ee
We can rewrite this boundary condition in terms of the renormalized quantities of (\ref{rgsolved}), rendering it finite and cutoff-independent
\be \label{renBC}
  D_\ell= \frac{2 \pi \alpha}{\bar{\beta}_\ell(\bar{r})} - \log(\bar{r} M_{0\ell}/2) - \gamma \quad \text{with} \quad \bar{\beta}_\ell(\bar{r}) = \bar{\lambda}_2(\bar{r}) + \frac{\bar{g}^2(\bar{r})v^2}{M_{0\ell}^2 - 2v^2 \bar{\lambda}(\bar{r})} \, .
\ee
Equating the two expressions for $D_i$ yields an eigenvalue equation for the $M_\ell^2$ masses
\be
-\frac{\pi}{2}  \frac{ Y_0 (\pi RM_{0\ell})}{J_0(\pi RM_{0\ell})} = \frac{2 \pi \alpha}{\bar{\beta}_\ell(\bar{r})} - \log(\bar{r} M_{0\ell}/2) - \gamma \, ,
\ee
which is, unfortunately, quite difficult to solve.

The normalization condition for the perturbed wavefunctions reads
\be \label{normCond}
2 \pi \alpha \int\limits_\epsilon^{\pi R} r dr \, \cP_{0 \ell}^2(r) + \frac{(gv)^2 \cP_{0\ell}^2(0)}{(M_{0\ell}^2 - 2v^2 \lambda )^2} = 1 \, .
\ee
We break this calculation into parts. First, we calculate the integral
\be
\int\limits_\epsilon^{\pi R} r dr \, \cP_{0 \ell}^2(r) = N_\ell^2 \int\limits_\epsilon^{ \pi R} r dr \left[\frac{\pi}{2} Y_0(rM_{0 \ell}) + D_\ell J_0(rM_{0 \ell})\right]^2 := N_\ell^2 I_\ell \,.
\ee
Using the identity
\be
 \frac{d}{dx} \left[\frac{1}{2} x^2 (Z_0^2(x) + Z_1^2(x)) \right] = x Z_0^2(x) \, ,
\ee
for any function $Z_0$ that satisfies Bessel's equation, we can write
\bea \nn
 I_\ell &=& \frac{1}{M_{0\ell}^2}  \left[\frac{1}{2} x^2 \left( \frac{\pi}{2} Y_0(x) + D_\ell J_0(x) \right)^2 + \frac{1}{2} x^2 \left( \frac{\pi}{2} Y_1(x) + D_i J_1(x) \right)^2 \right]_{\epsilon M_{0\ell}}^{\pi R M_{0\ell}} \\ \nn
&=& \frac{1}{2}  \pi^2 R^2 \left[ \left( \frac{\pi}{2} Y_0(\pi RM_{0\ell}) + D_i J_0(\pi RM_{0\ell}) \right)^2 + \left( \frac{\pi}{2} Y_1(\pi RM_{0\ell}) + D_i J_1(\pi RM_{0\ell}) \right)^2 \right] - \frac{1}{2 M_{0 \ell}^2} \, ,
\eea
where the second term follows from taking the $\epsilon \to 0$ limit. Using the boundary condition (\ref{DBcs}) gives
\be
 I_{\ell} = \frac{1}{2}  \frac{\pi^2 R^2}{J_0^2(\pi RM_{0 \ell})} \left( \frac{\pi}{2} Y_1(\pi RM_{0 \ell})J_0(\pi RM_{0 \ell}) -  \frac{\pi}{2} Y_0(\pi RM_{0 \ell}) J_1(\pi RM_{0 \ell}) \right)^2  - \frac{1}{2 M_{0 \ell}^2} \, .
\ee
Bessel functions obey the following identity
\be
\frac{\pi}{2} Y_1(\pi RM_i)J_0(\pi RM_i) -  \frac{\pi}{2} Y_0(\pi RM_i) J_1(\pi RM_i) = -\frac{1}{\pi RM_i} \, ,
\ee
so that
\be
 I_\ell = \frac{1}{2 M_{0\ell}^2} \left( \frac{1}{J_0^2(\pi RM_{0\ell})} - 1 \right) \, .
\ee
Now we move to the second term in (\ref{normCond}), which is equal to $\cB_\ell^2$. Using the $r=0$ boundary condition allows us to write
\be
 \cB_\ell =   \frac{gv \cP_{0 \ell}(0) }{M_{0\ell}^2 - 2v^2 \lambda} = \frac{2 \pi \alpha gv \left[r \partial_r \cP_{0 \ell} \right]_{r=0} }{(M_{0 \ell}^2 - 2v^2 \lambda) \lambda_2 + (gv)^2 } = \frac{2 \pi \alpha gv N_\ell }{(M_{0 \ell}^2 - 2v^2 \lambda) \lambda_2 + (gv)^2 } \, ,
\ee
which can be inserted into (\ref{normCond}) to give the following equation
\be \label{normm}
 N_\ell^{-2} = 2 \pi \alpha \left[\frac{1}{2 M_{0 \ell}^2}\left( \frac{1}{J_0^2(\pi RM_{0 \ell})} -1 \right) + \frac{ 2\pi \alpha (gv)^2 }{\left[ \lambda_2 \left(M_{0 \ell}^2 - 2v^2 \lambda \right) + (gv)^2 \right]^2} \right].
\ee
From the above normalization we find the mixing coefficients
\be \label{overlap}
 \cB_\ell^{-2} = 1 + \frac{1}{2 M_{0 \ell}^2}\left( \frac{1}{J_0^2(\pi RM_{0 \ell})} - 1 \right) \frac{\left[ \lambda_2 \left(M_{0 \ell}^2 - 2v^2 \lambda \right) + (gv)^2\right]^2}{2 \pi \alpha (gv)^2 } .
\ee
Note that the mixing coefficients vanish as $g \to 0$ or $\lambda_2 \to \infty$ unless $M_i^2 = 2v^2 \lambda$.


\begin{thebibliography}{99}



\bibitem{HiggsExp}
 G.~Aad {\it et al.}  [ATLAS Collaboration],
  ``Observation of a new particle in the search for the Standard Model Higgs boson with the ATLAS detector at the LHC,''
  Phys.\ Lett.\ B {\bf 716} (2012) 1
  [arXiv:1207.7214 [hep-ex]];

 S.~Chatrchyan {\it et al.}  [CMS Collaboration],
  ``Observation of a new boson at a mass of 125 GeV with the CMS experiment at the LHC,''
  Phys.\ Lett.\ B {\bf 716} (2012) 30
  [arXiv:1207.7235 [hep-ex]].

\bibitem{AdSCFT}
 J.~M.~Maldacena,
  ``The large N limit of superconformal field theories and supergravity,''
  Adv.\ Theor.\ Math.\ Phys.\  {\bf 2} (1998) 231
  [Int.\ J.\ Theor.\ Phys.\  {\bf 38} (1999) 1113]
  [hep-th/9711200];

   S.~S.~Gubser, I.~R.~Klebanov and A.~M.~Polyakov,
  ``Gauge theory correlators from non-critical string theory,''
  Phys.\ Lett.\  B {\bf 428} (1998) 105
  [hep-th/9802109];

  E.~Witten,
  ``Anti-de Sitter space and holography,''
  Adv.\ Theor.\ Math.\ Phys.\  {\bf 2} (1998) 253
  [hep-th/9802150].

\bibitem{Technicolor}
 S.~Weinberg,
  ``Implications of Dynamical Symmetry Breaking,''
  Phys.\ Rev.\ D {\bf 13} (1976) 974;

  L.~Susskind,
  ``Dynamics of Spontaneous Symmetry Breaking in the Weinberg-Salam Theory,''
  Phys.\ Rev.\ D {\bf 20} (1979) 2619.

\bibitem{Technirev}
  E.~Farhi and L.~Susskind,
  ``Technicolor,''
  Phys.\ Rept.\  {\bf 74} (1981) 277.

\bibitem{SUSY}
 J. Wess and B. Zumino, Nucl. Phys. B70 (1974) 39;

  E.~Witten,
  ``Dynamical Breaking of Supersymmetry,''
  Nucl.\ Phys.\ B {\bf 188} (1981) 513;

  S.~Dimopoulos and H.~Georgi,
  ``Softly Broken Supersymmetry and SU(5),''
  Nucl.\ Phys.\ B {\bf 193} (1981) 150.

\bibitem{SUSYrev}
 H.~P.~Nilles,
  ``Supersymmetry, Supergravity and Particle Physics,''
  Phys.\ Rept.\  {\bf 110} (1984) 1.\;

H.~E.~Haber and G.~L.~Kane,
  ``The Search for Supersymmetry: Probing Physics Beyond the Standard Model,''
  Phys.\ Rept.\  {\bf 117} (1985) 75.

\bibitem{MSLED}
   C.~P.~Burgess, J.~Matias and F.~Quevedo,
  ``MSLED: A Minimal supersymmetric large extra dimensions scenario,''
  Nucl.\ Phys.\  B {\bf 706} (2005) 71
  [arXiv:hep-ph/0404135];

 J.~Matias and C.~P.~Burgess,
  ``MSLED, neutrino oscillations and the cosmological constant,''
  JHEP {\bf 0509} (2005) 052
  [arXiv:hep-ph/0508156].

  \bibitem{SSnonlin}
 D.~V.~Volkov and V.~P.~Akulov,
  ``Is the Neutrino a Goldstone Particle?,''
  Phys.\ Lett.\ B {\bf 46} (1973) 109;

  E.~A.~Ivanov and A.~A.~Kapustnikov,
  ``General Relationship Between Linear And Nonlinear Realizations Of Supersymmetry,''
  J.\ Phys.\ A {\bf 11} (1978) 2375;

  E.~A.~Ivanov and A.~A.~Kapustnikov,
  ``The Nonlinear Realization Structure Of Models With Spontaneously Broken Supersymmetry,''
  J.\ Phys.\ G {\bf 8} (1982) 167;

   S.~Samuel and J.~Wess,
  ``A Superfield Formulation Of The Nonlinear Realization Of Supersymmetry And Its Coupling To Supergravity,''
  Nucl.\ Phys.\ B {\bf 221} (1983) 153.

  J.~Bagger and J.~Wess,
  ``Partial Breaking Of Extended Supersymmetry,''
  Phys.\ Lett.\ B {\bf 138} (1984) 105;

  J.~Hughes and J.~Polchinski,
  ``Partially Broken Global Supersymmetry and the Superstring,''
  Nucl.\ Phys.\ B {\bf 278} (1986) 147.

\bibitem{RS}
  L. Randall, R. Sundrum,
  ``A Large Mass Hierarchy from a Small Extra Dimension''
  { Phys.\ Rev.\ Lett.} {\bf 83}
  (1999) 3370 [hep-ph/9905221];
   ``An Alternative to Compactification''
  Phys.\ Rev.\ Lett.\ {\bf 83} (1999)
  4690 [hep-th/9906064].

\bibitem{ADD}
  N.~Arkani-Hamed, S.~Dimopoulos and G.~R.~Dvali,
  ``The hierarchy problem and new dimensions at a millimeter,''
  Phys.\ Lett.\ B {\bf 429} (1998) 263
  [arXiv:hep-ph/9803315];

  I.~Antoniadis, N.~Arkani-Hamed, S.~Dimopoulos and G.~R.~Dvali,
  ``New dimensions at a millimeter to a Fermi and superstrings at a TeV,''
  Phys.\ Lett.\ B {\bf 436} (1998) 257
  [arXiv:hep-ph/9804398].

\bibitem{RSHiggs}
  F.~Coradeschi, S.~De Curtis, D.~Dominici, J.~R.~Pelaez,
  ``Modified spontaneous symmetry breaking pattern by brane-bulk interaction terms,''
  JHEP {\bf 0804}, 048 (2008).
  [arXiv:0712.0537 [hep-th]].

\bibitem{Giudice:2000av}
  G.~F.~Giudice, R.~Rattazzi and J.~D.~Wells,
  ``Graviscalars from higher dimensional metrics and curvature Higgs mixing,''
  Nucl.\ Phys.\ B {\bf 595}, 250 (2001)
  [hep-ph/0002178];

  D.~Dominici and J.~F.~Gunion,
  ``Invisible Higgs Decays from Higgs Graviscalar Mixing,''
  Phys.\ Rev.\ D {\bf 80}, 115006 (2009)
  [arXiv:0902.1512 [hep-ph]].

\bibitem{6DHiggs0}
   E.~Dudas, C.~Papineau and V.~A.~Rubakov,
  ``Flowing to four dimensions,''
  JHEP {\bf 0603} (2006) 085
  [hep-th/0512276];

\bibitem{6DHiggs1}
 C.~P.~Burgess, C.~de Rham and L.~van Nierop,
  ``The Hierarchy Problem and the Self-Localized Higgs,''
  JHEP {\bf 0808} (2008) 061
  [arXiv:0802.4221 [hep-ph]].

\bibitem{6DHiggsph}
  P.~H.~Beauchemin, G.~Azuelos and C.~P.~Burgess,
  ``Dimensionless coupling of bulk scalars at the LHC,''
  J.\ Phys.\ G G {\bf 30}, N17 (2004)
  [hep-ph/0407196];

  G.~Azuelos, P.~H.~Beauchemin and C.~P.~Burgess,
  ``Phenomenological constraints on extra dimensional scalars,''
  J.\ Phys.\ G G {\bf 31}, 1 (2005)
  [hep-ph/0401125].


\bibitem{ADDgravem}
  N.~Arkani-Hamed, S.~Dimopoulos and G.~R.~Dvali,
  ``Phenomenology, astrophysics and cosmology of theories with submillimeter dimensions and TeV scale quantum gravity,''
  Phys.\ Rev.\ D {\bf 59}, 086004 (1999)
  [hep-ph/9807344];

  G.~F.~Giudice, R.~Rattazzi and J.~D.~Wells,
  ``Quantum gravity and extra dimensions at high-energy colliders,''
  Nucl.\ Phys.\ B {\bf 544}, 3 (1999)
  [hep-ph/9811291];

  T.~Han, J.~D.~Lykken and R.~-J.~Zhang,
  ``On Kaluza-Klein states from large extra dimensions,''
  Phys.\ Rev.\ D {\bf 59}, 105006 (1999)
  [hep-ph/9811350].

\bibitem{IJC}
   K.~Lanczos, Phys.\ Z.\ {\bf 23} (1922) 239--543; Ann.\ Phys.\
  {\bf 74} (1924) 518--540;

  C.W.~Misner and D.H.~Sharp,
  ``Relativistic Equations for Adiabatic, Spherically Symmetric
  Gravitational Collapse''
  Phys.\ Rev.\ {\bf 136} (1964)
  571--576;

  W.~Israel,
  ``Singular hypersurfaces and thin shells in general relativity''
  Nuov.\ Cim.\ {\bf 44B} (1966) 1--14; {\it errata}
  Nuov.\ Cim.\ {\bf 48B} 463.

\bibitem{Cod2BR}
  C.~P.~Burgess, D.~Hoover, C.~de Rham and G.~Tasinato,
  ``Effective Field Theories and Matching for Codimension-2 Branes,''
  JHEP {\bf 0903} (2009) 124
  [arXiv:0812.3820 [hep-th]];

    A.~Bayntun, C.P.~Burgess and L.~van Nierop,
  ``Codimension-2 Brane-Bulk Matching: Examples from Six and Ten Dimensions,''
  New J.\ Phys.\  {\bf 12} (2010) 075015
  [arXiv:0912.3039 [hep-th]];

\bibitem{Cod2Renorm}
   W.~D.~Goldberger and M.~B.~Wise,
  ``Renormalization group flows for brane couplings,''
  Phys.\ Rev.\ D {\bf 65} (2002) 025011
  [hep-th/0104170];

  T.~Kobayashi,
  ``UV caps, IR modification of gravity, and recovery of 4D gravity in regularized braneworlds,''
  Phys.\ Rev.\ D {\bf 78} (2008) 084018
  [arXiv:0806.0924 [hep-th]].

\bibitem{claudia}
  C.~de Rham,
  ``The Effective field theory of codimension-two branes,''
  JHEP {\bf 0801}, 060 (2008)
  [arXiv:0707.0884 [hep-th]];

 C.~de Rham,
  ``Classical renormalization of codimension-two brane couplings,''
  AIP Conf.\ Proc.\  {\bf 957} (2007) 309
  [arXiv:0710.4598 [hep-th]].

\bibitem{GW}
     W.~D.~Goldberger and M.~B.~Wise,
     ``Modulus stabilization with bulk fields'',
     Phys.\ Rev.\ Lett.\ {\bf 83} (1999) 4922-4925
     [arXiv:hep-ph/9907447]

\bibitem{6DExpStab}
  C.~P.~Burgess and L.~van Nierop,
  ``Large Dimensions and Small Curvatures from Supersymmetric Brane Back-reaction,''
  JHEP {\bf 1104} (2011) 078
  [arXiv:1101.0152 [hep-th]].

\bibitem{NS}
 H. Nishino and E. Sezgin, {\it Phys. Lett.} {\bf 144B} (1984) 187;
 ``The Complete N=2, D = 6 Supergravity With Matter And Yang-Mills
 Couplings,'' Nucl.\ Phys.\ {\bf B278} (1986) 353;

 S. Randjbar-Daemi, A. Salam, E. Sezgin and J. Strathdee,
 ``An Anomaly Free Model in Six-Dimensions''
 {\it
 Phys. Lett.} {\bf B151} (1985) 351.

\bibitem{SS}
 A.~Salam and E.~Sezgin,
 ``Chiral Compactification On Minkowski X S**2 Of N=2
 Einstein-Maxwell Supergravity In Six-Dimensions,''
 Phys.\ Lett.\ B {\bf 147} (1984) 47.

\bibitem{SLED}
 Y. Aghababaie, C.P. Burgess, S. Parameswaran and F. Quevedo,
 ``Towards a Naturally Small Cosmological Constant from Branes in 6D Supergravity''
 Nucl.\ Phys.\ {\bf B680} (2004) 389--414, [hep-th/0304256].

\bibitem{6DGW}
  C.~P.~Burgess, D.~Hoover and G.~Tasinato,
  ``UV Caps and Modulus Stabilization for 6D Gauged Chiral Supergravity,''
  JHEP {\bf 0709} (2007) 124
  [arXiv:0705.3212 [hep-th]];

  C.~P.~Burgess and L.~van Nierop,
  ``Bulk Axions, Brane Back-reaction and Fluxes,''
  JHEP {\bf 1102} (2011) 094
  [arXiv:1012.2638 [hep-th]];

\bibitem{SLEDrev}
  C.~P.~Burgess,
  ``Supersymmetric large extra dimensions and the cosmological constant: An update,''
  Annals Phys.\  {\bf 313} (2004) 283
  [arXiv:hep-th/0402200];
%
  ``Towards a natural theory of dark energy: Supersymmetric large extra dimensions,''
  AIP Conf.\ Proc.\  {\bf 743} (2005) 417
  [arXiv:hep-th/0411140].

\bibitem{6DLoops}
   C.~P.~Burgess and D.~Hoover,
  ``UV sensitivity in supersymmetric large extra dimensions: The Ricci-flat case,''
  Nucl.\ Phys.\ B {\bf 772} (2007) 175
  [hep-th/0504004];

 D.~Hoover and C.~P.~Burgess,
  ``Ultraviolet sensitivity in higher dimensions,''
  JHEP {\bf 0601} (2006) 058
  [hep-th/0507293];

  C.~P.~Burgess, D.~Hoover, G.~Tasinato,
  ``Technical Naturalness on a Codimension-2 Brane,''
  JHEP {\bf 0906 } (2009)  014.
  [arXiv:0903.0402 [hep-th]];

 M.~Williams, C.~P.~Burgess, L.~van Nierop and A.~Salvio,
  ``Running with Rugby Balls: Bulk Renormalization of Codimension-2 Branes,''
  arXiv:1210.3753 [hep-th];

 C.~P.~Burgess, L.~van Nierop, S.~Parameswaran, A.~Salvio and M.~Williams,
  ``Accidental SUSY: Enhanced Bulk Supersymmetry from Brane Back-reaction,''
  arXiv:1210.5405 [hep-th].

\bibitem{Leblond:2001ex}
  F.~Leblond,
  ``Geometry of large extra dimensions versus graviton emission,''
  Phys.\ Rev.\ D {\bf 64}, 045016 (2001)
  [hep-ph/0104273].

\bibitem{6DSolns}
    G.~W.~Gibbons, R.~Guven and C.~N.~Pope,
  ``3-branes and uniqueness of the Salam-Sezgin vacuum,''
  Phys.\ Lett.\  B {\bf 595} (2004) 498
  [hep-th/0307238];

    Y.~Aghababaie {\it et al.},
  ``Warped brane worlds in six dimensional supergravity,''
  JHEP {\bf 0309} (2003) 037
  [hep-th/0308064];

    C.~P.~Burgess, F.~Quevedo, G.~Tasinato and I.~Zavala,
  ``General axisymmetric solutions and self-tuning in 6D chiral gauged
  supergravity,''
  JHEP {\bf 0411} (2004) 069
  [hep-th/0408109];

   S.~L.~Parameswaran, G.~Tasinato and I.~Zavala,
  ``The 6D SuperSwirl,''
  Nucl.\ Phys.\  B {\bf 737} (2006) 49
  [arXiv:hep-th/0509061];

   H.~M.~Lee and C.~Ludeling,
  ``The general warped solution with conical branes in six-dimensional
  supergravity,''
  JHEP {\bf 0601} (2006) 062
  [arXiv:hep-th/0510026];

  A.~J.~Tolley, C.~P.~Burgess, D.~Hoover and Y.~Aghababaie,
  ``Bulk singularities and the effective cosmological constant for higher co-dimension branes,''
  JHEP {\bf 0603} (2006) 091
  [hep-th/0512218];

   A.~J.~Tolley, C.~P.~Burgess, C.~de Rham and D.~Hoover,
  ``Scaling solutions to 6D gauged chiral supergravity,''
  New J. Phys {\bf 8} (2006) 324
  [arXiv:0608.083 [hep-th]];

   A.~J.~Tolley, C.~P.~Burgess, C.~de Rham and D.~Hoover,
  ``Exact Wave Solutions to 6D Gauged Chiral Supergravity,''
  JHEP {\bf 0807} (2008) 075
  [arXiv:0710.3769 [hep-th]];

    M.~Minamitsuji,
  ``Instability of brane cosmological solutions with flux compactifications,''
  Class.\ Quant.\ Grav.\  {\bf 25} (2008) 075019
  [arXiv:0801.3080 [hep-th]].

\bibitem{6DKKmixing}
  C.~P.~Burgess, C.~de Rham, D.~Hoover, D.~Mason and A.~J.~Tolley,
  ``Kicking the rugby ball: Perturbations of 6D gauged chiral supergravity,''
  JCAP {\bf 0702}, 009 (2007)
  [hep-th/0610078];

  S.~L.~Parameswaran, S.~Randjbar-Daemi and A.~Salvio,
  ``General Perturbations for Braneworld Compactifications and the Six Dimensional Case,''
  JHEP {\bf 0903}, 136 (2009)
  [arXiv:0902.0375 [hep-th]];

\bibitem{brinv1}
  P.~P.~Giardino, K.~Kannike, M.~Raidal and A.~Strumia,
  ``Reconstructing Higgs boson properties from the LHC and Tevatron data,''
  JHEP {\bf 1206}, 117 (2012)
  arXiv:1203.4254 [hep-ph].

\bibitem{brinv2}
  J.~R.~Espinosa, M.~Muhlleitner, C.~Grojean and M.~Trott,
  ``Probing for Invisible Higgs Decays with Global Fits,''
  JHEP {\bf 1209}, 126 (2012)
  arXiv:1205.6790 [hep-ph].

\bibitem{brinv3}
  D.~Carmi, A.~Falkowski, E.~Kuflik, T.~Volansky and J.~Zupan,
  ``Higgs After the Discovery: A Status Report,''
  JHEP {\bf 1210}, 196 (2012)
  arXiv:1207.1718 [hep-ph].

\bibitem{brinv4}
  B.~A.~Dobrescu and J.~D.~Lykken,
  ``Coupling spans of the Higgs-like boson,''
  arXiv:1210.3342 [hep-ph].

\bibitem{raffeltBook}
  G.~G.~Raffelt,
  ``Stars as laboratories for fundamental physics: The astrophysics of neutrinos, axions, and other weakly interacting particles,''
  Chicago, USA: Univ. Pr. (1996) 664 p

\bibitem{HR}
  S.~Hannestad and G.~G.~Raffelt,
  ``Stringent neutron star limits on large extra dimensions,''
  Phys.\ Rev.\ Lett.\  {\bf 88} (2002) 071301
  [arXiv:hep-ph/0110067];
%
  ``New supernova limit on large extra dimensions,''
  Phys.\ Rev.\ Lett.\  {\bf 87} (2001) 051301
  [arXiv:hep-ph/0103201].

\bibitem{SNothers}
  S.~Cullen and M.~Perelstein,
  ``SN1987A constraints on large compact dimensions,''
  Phys.\ Rev.\ Lett.\  {\bf 83} (1999) 268
  [arXiv:hep-ph/9903422];

 V.~D.~Barger, T.~Han, C.~Kao and R.~J.~Zhang,
  ``Astrophysical constraints on large extra dimensions,''
  Phys.\ Lett.\  B {\bf 461} (1999) 34
  [arXiv:hep-ph/9905474];

\bibitem{SLEDph}
   D.~Atwood, C.~P.~Burgess, E.~Filotas, F.~Leblond, D.~London and I.~Maksymyk,
  ``Supersymmetric large extra dimensions are small and/or numerous,''
  Phys.\ Rev.\  D {\bf 63} (2001) 025007
  [arXiv:hep-ph/0007178];

\bibitem{muong2}
  J.~P.~Miller, E.~de Rafael and B.~L.~Roberts,
  ``Muon (g-2): Experiment and theory,''
  Rept.\ Prog.\ Phys.\  {\bf 70}, 795 (2007)
  [hep-ph/0703049].

\bibitem{Searches:2001ab}
  [LEP Higgs Working for Higgs boson searches and ALEPH and DELPHI and CERN-L3 and OPAL Collaborations],
  ``Searches for invisible Higgs bosons: Preliminary combined results using LEP data collected at energies up to 209-GeV,''
  hep-ex/0107032.

\bibitem{higgsphoton}
  J.~R.~Ellis, M.~K.~Gaillard and D.~V.~Nanopoulos,
  ``A Phenomenological Profile of the Higgs Boson,''
  Nucl.\ Phys.\ B {\bf 106}, 292 (1976).

\bibitem{higgsreviews}
  For general discussions see, for example:

  J.~F.~Gunion, H.~E.~Haber, G.~L.~Kane and S.~Dawson,
  ``The Higgs Hunter's Guide,''
  Front.\ Phys.\  {\bf 80} (2000) 1;

   A.~Djouadi,
  ``The Anatomy of electro-weak symmetry breaking. I: The Higgs boson in the standard model,''
  Phys.\ Rept.\  {\bf 457} (2008) 1
  [hep-ph/0503172];

  C.~P.~Burgess, J.~Matias and M.~Pospelov,
  ``A Higgs or not a Higgs? What to do if you discover a new scalar particle,''
  Int.\ J.\ Mod.\ Phys.\ A {\bf 17}, 1841 (2002)
  [hep-ph/9912459].

\bibitem{atlasMono}
  G.~Aad {\it et al.}  [ATLAS Collaboration],
  ``Search for new phenomena with the monojet and missing transverse momentum signature using the ATLAS detector in $\sqrt{s}=7$ TeV proton-proton collisions,''
  Phys.\ Lett.\ B {\bf 705}, 294 (2011)
  [arXiv:1106.5327 [hep-ex]] and ATLAS-CONF-2011-096 updates.

\bibitem{evasiveMono}
  C.~Englert, J.~Jaeckel, E.~Re and M.~Spannowsky,
  ``Evasive Higgs Maneuvers at the LHC,''
  Phys.\ Rev.\ D {\bf 85}, 035008 (2012)
  [arXiv:1111.1719 [hep-ph]].

\bibitem{cmsMono}
  S.~Chatrchyan {\it et al.}  [CMS Collaboration],
  ``Search for dark matter and large extra dimensions in monojet events in $pp$ collisions at $\sqrt{s}=7$ TeV,''
  JHEP {\bf 1209}, 094 (2012)
  [arXiv:1206.5663 [hep-ex]] and CMS-PAS-EXO-11-059 updates.

\bibitem{djouadiMono}
  A.~Djouadi, A.~Falkowski, Y.~Mambrini and J.~Quevillon,
  ``Direct Detection of Higgs-Portal Dark Matter at the LHC,''
  arXiv:1205.3169 [hep-ph].

\bibitem{baiMono}
  Y.~Bai, P.~Draper and J.~Shelton,
  ``Measuring the Invisible Higgs Width at the 7 and 8 TeV LHC,''
  JHEP {\bf 1207}, 192 (2012)
  [arXiv:1112.4496 [hep-ph]].

\bibitem{ghoshInvis}
  D.~Ghosh, R.~Godbole, M.~Guchait, K.~Mohan and D.~Sengupta,
  ``Looking for an Invisible Higgs Signal at the LHC,''
  arXiv:1211.7015 [hep-ph].

\bibitem{multScatter}
  G.~Raffelt and D.~Seckel,
  ``Multiple scattering suppression of the bremsstrahlung emission of neutrinos and axions in supernovae,''
  Phys.\ Rev.\ Lett.\  {\bf 67}, 2605 (1991).

\bibitem{yukawabounds}
  J.~A.~Grifols, E.~Masso and S.~Peris,
  ``Energy Loss From The Sun And Red Giants: Bounds On Short Range Baryonic And Leptonic Forces,''
  Mod.\ Phys.\ Lett.\ A {\bf 4}, 311 (1989).

\bibitem{higgsnucleon}
  See, for example, C.~P.~Burgess, M.~Pospelov and T.~ter Veldhuis,
  ``The Minimal model of nonbaryonic dark matter: A Singlet scalar,''
  Nucl.\ Phys.\ B {\bf 619}, 709 (2001)
  [hep-ph/0011335].

\bibitem{snYukawa}
  N.~Ishizuka and M.~Yoshimura,
  ``Axion And Dilaton Emissivity From Nascent Neutron Stars,''
  Prog.\ Theor.\ Phys.\  {\bf 84}, 233 (1990).

\bibitem{omegapole}
  C.~Hanhart, D.~R.~Phillips and S.~Reddy,
  ``Neutrino and axion emissivities of neutron stars from nucleon-nucleon scattering data,''
  Phys.\ Lett.\ B {\bf 499}, 9 (2001)
  [astro-ph/0003445];

  C.~Hanhart, D.~R.~Phillips, S.~Reddy and M.~J.~Savage,
  ``Extra dimensions, SN1987a, and nucleon-nucleon scattering data,''
  Nucl.\ Phys.\ B {\bf 595}, 335 (2001)
  [nucl-th/0007016].

\bibitem{saxion}
  D.~Arndt and P.~J.~Fox,
  ``Saxion emission from SN1987A,''
  JHEP {\bf 0302}, 036 (2003)
  [hep-ph/0207098].

\bibitem{HiggsMassBound}
 S.~Weinberg,
  ``Mass of the Higgs Boson,''
  Phys.\ Rev.\ Lett.\  {\bf 36} (1976) 294;

  A.~D.~Linde,
  ``Dynamical Symmetry Restoration and Constraints on Masses and Coupling Constants in Gauge Theories,''
  JETP Lett.\  {\bf 23}, 64 (1976)
  [Pisma Zh.\ Eksp.\ Teor.\ Fiz.\  {\bf 23}, 73 (1976)].

\bibitem{Triviality}
B.~Grzadkowski and M.~Lindner,
  ``Stability Of Triviality Mass Bounds In The Standard Model,''
  Phys.\ Lett.\ B {\bf 178} (1986) 81;

 M.~Luscher and P.~Weisz,
  ``Scaling Laws and Triviality Bounds in the Lattice phi**4 Theory. 1. One Component Model in the Symmetric Phase,''
  Nucl.\ Phys.\ B {\bf 290} (1987) 25;
%
  ``Scaling Laws and Triviality Bounds in the Lattice phi**4 Theory. 2. One Component Model in the Phase with Spontaneous Symmetry Breaking,''
  Nucl.\ Phys.\ B {\bf 295} (1988) 65.

\bibitem{sally}
  See, for example, S.~Dawson,
  ``Introduction to electroweak symmetry breaking,''
  [hep-ph/9901280];

  T.~P.~Cheng, E.~Eichten and L.~-F.~Li,
  ``Higgs Phenomena in Asymptotically Free Gauge Theories,''
  Phys.\ Rev.\ D {\bf 9}, 2259 (1974).

\bibitem{muonWidth}
  T.~Han and Z.~Liu,
  ``Direct Measurement of the Higgs Boson Total Width at a Muon Collider,''
  arXiv:1210.7803 [hep-ph].

\bibitem{ilc}
H.~Baer {\it et al.}
``Physics at the International Linear Collider.''
Physics Chapter of the ILC Detailed Baseline Design Report.
Preliminary Version: Draft of January 22, 2013.
http://lcsim.org/papers/DBDPhysics.pdf

\bibitem{6DMagic}
 J.~W.~Chen, M.~A.~Luty and E.~Ponton,
 ``A Critical cosmological constant from millimeter extra dimensions''
 JHEP {\bf 0009}(2000)012
 [arXiv:hep-th/0003067];

 F.~Leblond, R.~C.~Myers and D.~J.~Winters,
  ``Consistency conditions for brane worlds in arbitrary dimensions,''
  JHEP {\bf 0107} (2001) 031
  [arXiv:hep-th/0106140];

 S.~M.~Carroll and M.~M.~Guica,
 ``Sidestepping the cosmological constant with football-shaped extra
 dimensions,''
 [hep-th/0302067].

\bibitem{TNCC}
  C.~P.~Burgess and L.~van Nierop,
  ``Technically Natural Cosmological Constant From Supersymmetric 6D Brane Backreaction,''
  arXiv:1108.0345 [hep-th].

\end{thebibliography}
\end{document}